\documentclass[%
 reprint,
 showpacs,
 amsmath,amssymb,
 aps,
 prc,
]{revtex4-1}
\usepackage{graphicx}

\usepackage[caption=false]{subfig}
\usepackage{amsmath}
\usepackage{mathtools}
\usepackage{float}
\usepackage{dsfont}
\usepackage{dcolumn}
\usepackage{bm}
\usepackage{lipsum}
\usepackage{csquotes}
\usepackage{textcomp}
\usepackage{nicefrac} 
\usepackage{feynmp}
\usepackage{color}
\usepackage{ifpdf}
\ifpdf
\fi

\makeatletter
\def\endfmffile{%
  \fmfcmd{\p@rcent\space the end.^^J%
          end.^^J%
          endinput;}%
  \if@fmfio
    \immediate\closeout\@outfmf
  \fi
  \IfFileExists{\thefmffile.mp}{\immediate\write18{mpost \thefmffile}}{}
  \let\thefmffile\relax
}
\makeatother

\newcommand{\bra}[1]{\langle#1|}
\newcommand{\ket}[1]{|#1\rangle}

\def\cY{{\cal Y}}
\def\cX{{\cal X}}
\begin{document}
\captionsetup[subfigure]{labelformat=empty}

\preprint{APS/123-QED}

\title{Algebraic diagrammatic construction formalism with three-body interactions   }

\author{Francesco Raimondi and Carlo Barbieri}
 
\affiliation{
 Department of Physics, University of Surrey, Guildford GU2 7XH, United Kingdom
}%

\date{\today}

\begin{abstract}

\begin{description}

\item[Background] Self-consistent Green's function theory has recently been extended to the basic formalism needed to account for three-body interactions [A. Carbone, A. Cipollone, C. Barbieri, A. Rios, and A. Polls, {\color{blue}Phys. Rev. C \textbf{88}, 054326 (2013)}]. The contribution of  three-nucleon forces has so far been included in \textit{ab initio} calculations on nuclear matter and finite nuclei only as averaged two-nucleon forces.
\item[Purpose]  
We derive the working equations for all possible two- and three-nucleon terms  that enter the expansion of the self-energy up to the third order, thus including the interaction-irreducible (i.e., not averaged) diagrams with three-nucleon forces that have been previously neglected.

\item[Methods]  
We employ the algebraic diagrammatic construction up to the third order as an organization scheme for generating a \emph{non perturbative} self-energy, in which ring (particle-hole) and ladder (particle-particle) diagrams are resummed to all orders.  

\item[Results]  We derive expressions of the static and dynamic self-energy up to the third order, by taking into account the set of diagrams required when either the skeleton or nonskeleton expansions of the single-particle propagator are assumed.
A hierarchy of importance among different diagrams is revealed, and a particular emphasis is given to a third-order diagram (see Fig.~\ref{3ord_c}) that is expected to play a significant role among those featuring an interaction-irreducible three-nucleon force. 

\item[Conclusion] A consistent formalism to resum at infinite order correlations induced by three-nucleon forces in the self-consistent Green's function theory is now available and ready to be implemented in the many-body solvers.

\end{description}

\end{abstract}

\pacs{}
\maketitle


\section{\label{sec:level1}Introduction}

Three-body interactions play a prominent role in determining the behaviour of strongly interacting quantum systems~\cite{Hammer2013}.
For instance,  three-nucleon forces (3NFs) are necessary to reproduce the saturation of infinite matter as well as to determine the structure and location of the driplines in neutron rich isotopes. Hence, they have been implemented in most of the post-Hartree-Fock approaches that are currently used to study medium mass isotopes, such as self-consistent Green's function (SCGF) theory~\cite{Barb2017NLP,Cipol13}, the coupled cluster method~\cite{Hagen2014, Binder2013}, and the in-medium similarity renormalization group~\cite{Tsukiyama2011, Hergert2016}.  In all of these methods,
one typically proceeds by performing a normal ordering of the Hamiltonian, or a similar averaging, so that the dominant effect of 3NFs can be taken into account as an effective nucleon-nucleon force (2NF).
Advances of the above many-body methods,  with the concurrent introduction of chiral  two- and three-nucleon interactions,
based on the symmetries of QCD~\cite{Epelbaum2009,Machleidt20111}, have led recently to remarkable successes
in nuclear \textit{ab initio} theory~\cite{AnnPhysHebeler2015,Barbieri2014QMBT,Lapoux2016}. 
Presently,  the major sources  of error in first-principle predictions  originate from uncertainties of the nuclear Hamiltonian~\cite{AnnPhysHebeler2015,PhysRevC.94.014322}.  However, the expected progress in the next-generation realistic interactions will eventually require further developments of the many-body formalisms.

The SCGF theory is a quantum many-body method that has been extensively applied to both condensed matter and nuclear systems~\cite{Barbieri2003DRPA,Dickhoff2004,RevModPhys.74.601,Aryasetiawan1998,VONNIESSEN198457,Danovich2011,Barb2017NLP}.  This approach  relies on  the solution of the Dyson equation, which is an exact restatement of the many-body Schr\"odinger equation and it allows for a diagrammatic expansion with respect to the nuclear interaction. However, for nuclear structure and reaction  studies,  a perturbative expansion is not sufficient  due to the strong nature of the nuclear force and the importance of the long-range correlations, which affect the propagation of  nucleons inside the medium. In practice, one must resort to an efficient method in which entire classes of correlations are resummed non perturbatively.

For this purpose, a major challenge is to find a scheme capable of organizing the rapidly increasing number of Feynman diagrams entering the computation of Green's functions, especially when 3NFs and many-nucleon interactions are present. Ideally, one should include different classes of Feynman diagrams at all orders, i.e. in a nonperturbative way; at the same time, one needs to keep under control the computational resources required by the many-body problem, even for those post-Hartree-Fock approaches scaling polynomially with the number of nucleons. 
A powerful tool complying with these requirements has been devised 30 years ago in the Green's function theory applied to quantum chemistry. It is referred to as the algebraic diagrammatic construction (ADC) method~\cite{Schirmer1982,Schirmer1983}.  Born as a way to include third-order self-energy diagrams that are necessary to reproduce affinities and ionisation energies, the ADC also allows the infinite resummation of specific classes of diagrams, such as the ladder and ring series.  The general procedure is to impose the correct spectral representation of the self-energy and to require that its perturbative expansion is also consistent with the Feynman diagram series up to a given order~$n$. The spectral representation implies that diagrams up to order $n$ are actually taken as \enquote{seeds} for all-order resummations.
This generates a hierarchy of many-body truncations, labelled as ADC($n$), that contains selected relevant terms, it is non perturbative and can be systematically improved.

The aim of this paper is to derive the entire set of working equations for the  ADC(3) self-energy, when 3NFs are present. The general formalism  and diagrammatic rules for the SCGF theory with three-body interactions has been developed in Ref.~\cite{CarAr13}. 
 There, it was shown that the number of Feynman diagrams to be calculated can be reduced by 
introducing averaged effective interactions (similarly to the normal ordering strategy mentioned above), so that one needs to consider  only interaction-irreducible diagrams.  Using the resulting effective 2NFs,  a set of applications of the SCGF was put forward with  computations of binding energies~\cite{Cipol13, Soma14Ca}, spectral distributions and radii~\cite{Cipol15,Lapoux2016} for closed subshell isotopes of medium-mass. For the nuclear matter case, saturation properties of chiral forces and other thermodynamical aspects  have also been studied~\cite{Car13}. Moreover, current efforts are devoted to describing the one-nucleon scattering on a nucleus within the same framework~\cite{Idini2017}.

While Ref.~\cite{CarAr13} introduced the set of self-energy diagrams up to third order, the necessary formalism for extending them to a nonperturbative approach has not yet been derived. We fill this gap here by deriving explicit expressions of the ADC(3) based on the Feynman diagrams derived thereof. 
 For this purpose we revisit  the SCGF formalism in Sec.~\ref{sec:formalism}, with special emphasis on how 3NFs are incorporated in the self-energy expansion.  Section~\ref{ADC(N)}  discusses the ADC method at order $n$, i.e. ADC($n$), and we derive the working equations at second and third order, ADC(2) and ADC(3), in full detail. In our derivation, a hierarchy of importance among different self-energy diagrams emerges naturally: it is based on intermediate excitation energies embedded in each diagram and on the relative importance between 2N and 3NFs.  For the ADC(3), we present in Sec.~\ref{ADC(N)} diagrams that contain only two-particle-one-hole ($2p1h$) and two-hole-one-particle ($2h1p$) intermediate states with effective 2NFs and interaction-irreducible 3NFs. These diagrams are displayed in Fig.~(\ref{3ord_a_b_c}) below and are the dominant contributions to ADC(3). We also present  additional equations for a subset of diagrams with $3p2h$ and $3h2p$ intermediate states,  chosen from the different topologies appearing at third order. This will give a general overview of the formalism up to $n$=3. All the remaining parts of the ADC(3) equations, 
which complete the diagrams with $3p2h$ and $3h2p$ configurations, are given in Appendix~\ref{M_N_matrices}. 
 In Appendix~\ref{Angular} we display  the angular momentum coupled form for the leading contributions of Fig.~(\ref{3ord_a_b_c}).  The diagram in  Fig.~\ref{3ord_c} has not yet been included in calculations,  but it  could be added to current numerical implementations and it is expected to be the most important among those with an irreducible 3NF. In Appendix~\ref{General_reference_state} we derive additional nonskeleton Feynman diagrams for both the static and dynamic self-energy that need to be included in non self-consistent calculations.   The entire set of equations derived up to ADC(3) informs our conclusions, which are drawn in Sec.~\ref{sec:conclusion}.

\section{\label{sec:formalism}SCGF formalism with 3NFs}

Many-particle Green's functions, also known in the literature as propagators or correlation functions, 
are at the heart of the SCGF formalism.  The simplest Green's function is the one-body (1B) propagator describing the in-medium propagation of a particle or an hole, which are respectively created and annihilated by field operators $a_{\beta}^{\dagger}$ and $a_{\alpha}$ in the quantum states $\beta$ and $\alpha$ ~\cite{Dickhoff2005,Fetter}:
\begin{equation}
\label{Green}
 g_{\alpha \beta}(t-t') = - \frac{i}{\hbar}     \bra{\Psi^A_0} \mathcal{T}\left[ 	a_{\alpha}(t)  	a_{\beta}^{\dagger}(t') \right] \ket{\Psi^{A}_0} \, .
\end{equation} 
Here and in the following we use Greek indexes to label the states of a complete orthonormal single-particle basis $\{\alpha\}$, which could be discrete or having a continuous spectrum. 
The time interval $(t-t')$ of the propagation in Eq.~(\ref{Green}) is ordered according to the action of the time-ordering operator $\mathcal{T}$, which obeys the Fermi statistics. To describe the propagation of two particles and two holes, we introduce also the two-body (2B) Green's function,
\begin{eqnarray}
&& g^{II}_{\alpha \beta , \gamma\delta}(t_\alpha, t_\beta,t_\gamma, t_\delta) = \nonumber \\
&&- \frac{i}{\hbar}   \langle\Psi_0^A|{\cal T}[a_\beta(t_\beta)a_\alpha(t_\alpha)
a_{\gamma}^\dagger(t_{\gamma})a_{\delta}^\dagger(t_{\delta})]|\Psi_0^A\rangle \, .  
\label{g4pt}
\end{eqnarray}

For our purposes, we will concentrate on the propagator of  Eq.~(\ref{Green}), which is defined with respect to the exact $A$-body ground state $\ket{\Psi^A_0}$. The latter is the lowest eigenstate of the Schr\"odinger problem,
\begin{equation}
\label{Schro}
\hat{H} \ \ket{\Psi^A_n} = E_n^A \ \ket{\Psi^A_n} \, .
\end{equation}

The Lehmann representation of the Green's function is obtained by Fourier transforming Eq.~(\ref{Green}) to the energy plane. It contains the relevant informations on the single-particle dynamics,
%
\begin{align}
 g_{\alpha \beta}(\omega) ~={}& 
 \sum_n  \frac{ 
          \bra{\Psi^A_0}  	a_{\alpha}   \ket{\Psi^{A+1}_n}
          \bra{\Psi^{A+1}_n}  a^{\dagger}_{\beta}  \ket{\Psi^A_0}
              }{\hbar \omega - (E^{A+1}_n - E^A_0)+ \textrm{i} \eta }  \nonumber\\
 +{}& \sum_k \frac{
          \bra{\Psi^A_0}      a^{\dagger}_{\beta}    \ket{\Psi^{A-1}_k}
          \bra{\Psi^{A-1}_k}  a_{\alpha}	    \ket{\Psi^A_0}
             }{\hbar \omega - (E^A_0 - E^{A-1}_k) - \textrm{i} \eta } \; .
\label{eq:g1}
\end{align}

In the reminder of this text we will use the following shorthand notation for the quasiparticle energies, given by the poles in Eq.~(\ref{eq:g1}),
\begin{align}
\varepsilon_n^{+}\equiv{}&(E^{A+1}_n - E^A_0) \nonumber \\ 
\varepsilon_k^{-}\equiv{}&(E^A_0 - E^{A-1}_k) \, ,
\end{align}
which are the experimentally observed one-nucleon addition and removal energies. Likewise, we will use ${\cal X}^n_{\beta}$ and ${\cal Y}^k_{\alpha}$ to mark the  transition amplitudes for the addition and removal of a particle, respectively. And we will use $\mathcal{Z}_{\alpha}^i$ to collect all of them:
\begin{equation}
\label{tran_ampl}
\mathcal{Z}_{\alpha}^{i=n,k} \equiv
\begin{cases}
\left({\cal X}^n_{\alpha}\right)^* \equiv \langle\Psi_0^A |a_{\alpha}|\Psi_n^{A+1}\rangle\\
~{\cal Y}^k_{\alpha}~\equiv\langle\Psi_k^{A-1}|a_{\alpha}|\Psi_0^A\rangle \, ,
\end{cases}
\end{equation}
with the index $i$ referring to both forward-in-time ($n$) and backward-in-time ($k$) processes. The first (second) overlap integral in Eq.~\eqref{tran_ampl} is related to the probability of adding (removing) a particle to (from) a orbital $\alpha$ in a system with $A$ particles. The vectors $\mathcal{Z}^i$ (in the basis $\{\alpha\}$) form the \emph{overcomplete} set of the eigenstates of the Dyson equation. Thus, they are also referred to as `Dyson orbitals'.
 
The  1B Green's function Eq.~\eqref{eq:g1} is completely determined by solving the Dyson equation,
\begin{equation}
  \label{eq:Dy}
g_{\alpha\beta}(\omega)=g^{(0)}_{\alpha\beta}(\omega)+ \sum_{\gamma\delta} g^{(0)}_{\alpha\gamma}(\omega)\Sigma_{\gamma\delta}^{\star}(\omega) g_{\delta\beta}(\omega)  \; ,
\end{equation}
which is a nonlinear equation defining the irreducible self-energy $\Sigma_{\gamma\delta}^{\star}(\omega)$, where medium effects  on the particle propagation are encoded.
It corresponds to a set of irreducible Feynman diagrams, i.e. diagrams that cannot be divided in sub-diagrams by cutting one propagator line.
 The distinction between the unperturbed propagator $g^{(0)}_{\alpha\beta}(\omega)$ and the correlated one $ g_{\alpha\beta}(\omega)$ in Eq.~(\ref{eq:Dy}) results from the expansion with respect to the inter-particle interaction: $g^{(0)}_{\alpha\beta}(\omega)$ is then the zeroth-order term in the expansion, that is the propagator with respect to the reference state, i.e.
\begin{equation}
\label{Green_0}
 g^{(0)}_{\alpha \beta}(t-t') = - \frac{i}{\hbar}     \bra{\phi^A_0} \mathcal{T}\left[ 	a_{\alpha}(t)  	a_{\beta}^{\dagger}(t') \right] \ket{\phi^{A}_0} \, .
\end{equation}

From the derivation of the 1B propagator equation of motion, one can find the explicit separation of the irreducible self-energy in a part which is local in time $\Sigma_{\alpha\beta}^{\infty}$ (static self-energy), and a energy dependent part $\widetilde{\Sigma}_{\alpha\beta}(\omega)$ (dynamic self-energy), containing the contributions from dynamical excitations in the system:
\begin{equation}
\label{irr_SE_decomp}
\Sigma_{\alpha\beta}^{\star}(\omega) = \Sigma_{\alpha\beta}^{\infty}+ \widetilde{\Sigma}_{\alpha\beta}(\omega) \, .
\end{equation}

While $\Sigma_{\alpha\beta}^{\infty}$ accounts for the averaged (mean-field) interaction seen by a particle, $\widetilde{\Sigma}_{\alpha\beta}(\omega)$ encodes the coupling of a single-particle state with the collective configurations made by surrounding nucleons. In the continuum regime the full self-energy describes the interaction of the nucleon projectile with a target nucleus. In this respect, $\Sigma_{\alpha\beta}^{\star}(\omega)$ is investigated as  the microscopic counterpart of the dispersive optical model potentials~\cite{Waldecker2011}.

 Before proceeding with the application of the ADC formalism to the self-energy, we present in the next section the main features of the approach based on \textit{effective interactions}, which allows a simplification of the diagrammatic expansion when both two- and many-body forces are included.
 
\subsection{\label{sec:EffInts} Formalism with effective forces and interaction-irreducible terms}

Let us consider first the nuclear Hamiltonian $\hat{H}$  with a kinetic energy part $\hat{T}$ and interaction operators in the  2NF and 3NF sector, denoted with $\hat{V}$ and $\hat{W}$ respectively,
\begin{equation}
\label{plain_H}
\hat{H}  = \hat{T} + \hat{V} + \hat{W} \, .
\end{equation}
Within post Hartree-Fock approaches,  it is customary to divide the Hamiltonian into two parts, $\hat{H} = \hat{H}_0  +\hat{H}_1$, with $\hat{H}_0$ being the uncorrelated part and $\hat{H}_1$ the residual interaction. In this way, strongly interacting fermions are treated as a system of independent nucleons affected by an auxiliary mean-field potential $\hat{U}$, included in the definition of $\hat{H}_0 =  \hat{T} + \hat{U} $. The $\hat{H}_0$ defines the reference state to which the residual interaction $\hat{H}_1$ is added perturbatively. 
In the second quantization formalism the Hamiltonian reads then,
\begin{eqnarray}
\label{H}
\hat{H} &=& \! \sum_{\alpha \beta} \! h^{(0)}_{\alpha \beta} \, a^\dagger_\alpha a_\beta 
 - \! \sum_{\alpha \beta}\! U_{\alpha \beta}\, a^\dagger_\alpha a_{\beta} \! + \! \frac{1}{4} \! \sum_{\substack{\alpha\gamma\\\beta\delta}} \! V_{\alpha\gamma,\beta\delta}\, a_\alpha^\dagger a_\gamma^\dagger a_{\delta} a_{\beta}\nonumber \\
&+ &
\frac{1}{36}\sum_{\substack{\alpha\gamma\epsilon \\ \beta\delta\eta}} W_{\alpha\gamma\epsilon,\beta\delta\eta}\,
a_\alpha^\dagger a_\gamma^\dagger a_\epsilon^\dagger a_{\eta} a_{\delta} a_{\beta} \, ,
\end{eqnarray}
where $h^{(0)}_{\alpha \beta} \equiv T_{\alpha \beta} + U_{\alpha \beta}\,$ is a one-body Hamiltonian and $V_{\alpha\gamma,\beta\delta}$ and $W_{\alpha \gamma\epsilon,\beta\delta\eta}$ are the antisymmetrized matrix elements of 2NFs and 3NFs, respectively. 

The Greek indexes $\alpha$,$\beta$,$\gamma$,\ldots label a complete set of single-particle states which define the model space used in  practical computations. In most cases, one chooses this basis as the eigenstates of the unperturbed Hamiltonian $\hat{H}_0$, with the eigenvalues $\varepsilon_\alpha^0$. Then, $h^{(0)}_{\alpha \beta}= \varepsilon_\alpha^0 \delta_{\alpha,\beta}$ and the spectral representation, Eq.\eqref{eq:g1}, for unperturbed propagator $g_{\alpha,\beta}^{(0)}(\omega)$ becomes diagonal. However,  in this work we keep the most general case and the  basis $\{\alpha\}$ will be different from  the one defining the reference state.

The expansion of the self-energy $\widetilde{\Sigma}_{\alpha\beta}(\omega)$ in Eq.~(\ref{irr_SE_decomp}) involves terms with individual contributions of the 1B potential, but also of 2NFs and 3NFs  from Eq.~(\ref{H}). Of course, also terms containing combinations of different interactions are possible. The number of diagrams allowed by the Feynman diagrammatic rules is growing fast with the order of the expansion. A useful strategy is to consider only interaction-irreducible diagrams. Diagrams are considered interaction-reducible if splitting interaction vertex in two parts results in two disconnected diagrams. This happens when some (but not all) of the fermion lines leaving one interaction vertex eventually return to it.  If the interaction vertex which is cut had only one fermionic line looping over it, then all the linked diagrams can be included effectively by averaging the interaction vertex with a 1B Green's function. Alternatively, when the cut interaction vertex had two fermionic lines the averaging is performed with a 2B Green's function and so on. This process of averaging reduces the order of the interaction: for instance, a 2NF  interaction vertex averaged on a 1B Green's function gives rise to an effective 1B operator.

In Ref.~\cite{CarAr13} it is shown that diagrammatic series can be reduced to a smaller set of diagrams by excluding all interaction-reducible diagrams. The averaging procedure described above amounts to define an effective Hamiltonian up to 3NFs,
\begin{equation}
\widetilde H_1= {\widetilde U} + {\widetilde  V} + \hat{W} \, ,
\label{Heff}
\end{equation}
where $\widetilde U$ and  $\widetilde V$ represent effective interaction operators. As long as only interaction-irreducible diagrams are considered, the use of $\widetilde H_1$ is equivalent to the interaction-reducible expansion based on Eq.~(\ref{H}) (see Sec. II of Ref.~\cite{CarAr13} for details).

Explicit expressions for effective 1B and 2N  interaction operators are:
\begin{eqnarray}
\label{ueff}
\widetilde U  &=&  \sum_{\alpha \beta}  \widetilde U_{\alpha\beta} \,  a_\alpha^\dagger a_{\beta}\,, 
\end{eqnarray}
with
\begin{equation}
\label{ueff_bis}
\widetilde U_{\alpha\beta} =  -  U_{\alpha\beta} 
+  \sum_{\gamma\delta} \! V_{\alpha\gamma,\beta\delta} \, \rho_{\delta\gamma} \!
+ \!  \frac{1}{4}  \!\sum_{\substack{\gamma\epsilon \\ \delta\eta}}  W_{\alpha\gamma\epsilon,\beta\delta\eta}
\, \Gamma_{\delta\eta , \gamma\epsilon} \, ,
\end{equation}
and
\begin{eqnarray}
\label{veff}
\widetilde V \! &=& \! \frac{1}{4} \! \sum_{\substack{\alpha\gamma\\\beta\delta}}\! \left[ \! V_{\alpha\beta,\gamma\delta}
+\sum_{\epsilon \eta}W_{\alpha \beta \epsilon, \gamma \delta \eta} \, \rho_{\eta \epsilon} \!  \right] \! a_\alpha^\dagger a_\beta^\dagger a_{\delta}a_{\gamma} \, ,
\end{eqnarray}
where, in the averaging of  2NFs and 3NFs,
one- and two-body reduced density matrices of the many-body system are produced,
\begin{eqnarray}
\label{1B_densitymatrix}
\rho_{\delta\gamma} ={}& ~ \langle \Psi_0^A | \, a_{\gamma}^\dagger a_{\delta} \, | \Psi_0^A\rangle \quad &=  -i\hbar\, g_{\delta\gamma}(t-t^+) \; ,
\\
\label{2B_densitymatrix}
\Gamma_{\delta\eta , \gamma\epsilon} ={}&{} \langle \Psi_0^A| \, a_{\gamma}^\dagger a_{\epsilon}^\dagger a_{\eta} a_{\delta} \, |\Psi_0^A\rangle \; &=
~i\hbar\,g^{II}_{\delta\eta , \gamma\epsilon}(t-t^+) \, . \qquad
\end{eqnarray}
The two-body density of Eq.~(\ref{2B_densitymatrix}) is obtained when the opportune limits are taken in the time arguments of the 2B Green's function in Eq.~(\ref{g4pt}).


We note that when the irreducible self-energy is computed with the effective Hamiltonian of Eq.~(\ref{Heff}), a portion of the many-body effects is incorporated in the interactions, which become system dependent. This is done in a systematic way and the procedure  is in principle superior to the usual normal ordering approach. Here the density matrices $\rho$ and $\Gamma$ entering the contraction of the interaction vertex are obtained from the true correlated propagators, i.e. they are not computed from the reference state.

The separation of a simple unperturbed Hamiltonian $\hat{H}_0$ from Eq.~(\ref{H}) is instrumental to any approach based on perturbation theory (or on all-orders resummations): it allows us to define a reference state  upon which a perturbative series is constructed and it also leads to the expansion of the  Green's function in Feynman diagrams. Nevertheless, the auxiliary potential $\hat{U}$ eventually cancels from the SCGF formalism.
Considering the decomposition of Eq.~(\ref{irr_SE_decomp}),   the irreducible static self-energy $\Sigma_{\alpha\beta}^{\infty}$ is given exactly by the 1B effective interaction~\cite{CarAr13}:
\begin{equation}
\label{SE_eq_U}
\Sigma_{\alpha\beta}^{\infty} = \widetilde U_{\alpha\beta} \, .
\end{equation}
Since $\hat{U}$ is added to the definition of the reference propagator $g^{(0)}$ but subtracted in Eq.~(\ref{ueff_bis}), it eventually cancels out exactly from the Dyson equation (see Eq.~\eqref{eq:Dy_residual_5}).  
The dynamic self-energy $\widetilde{\Sigma}_{\alpha\beta}(\omega)$ can still depend on the auxiliary potential through the perturbative expansion in $g_{\alpha\beta}^{(0)}(\omega)$. However, in the full self-consistent approach, the perturbative series is restricted to skeleton diagrams where fully correlated propagators $g_{\alpha\beta}(\omega)$ replace the uncorrelated ones. Thus, the partition of the Hamiltonian into a uncorrelated part and residual part is completely lost in the exact SCGF formalism and one may think of the correlated propagator as playing the role of an improved reference state.

\begin{figure}[t]
  \centering
    \subfloat[(a)]{\label{2ord_2B}\includegraphics[scale=0.55]{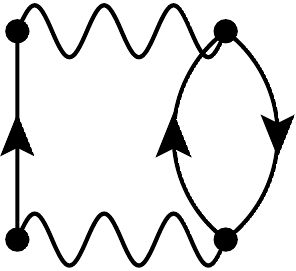}}
  \hspace{2cm}
   \subfloat[(b)]{\label{2ord_3B}\includegraphics[scale=0.55]{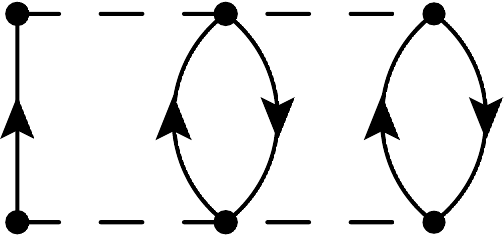}}
  \caption{ One-particle irreducible, skeleton and interaction-irreducible self-energy diagrams appearing at second order in the expansion of Eq.~(\ref{irr_SE_decomp}), using the effective Hamiltonian of Eq.~(\ref{Heff}). The wiggly lines represent the 2N effective interaction of Eq.~(\ref{veff}), while the long-dashed lines represent the interaction-irreducible 3NF $\hat{W}$.}
  \label{2ord}
\end{figure}

\begin{figure}[t]
  \centering
    \subfloat[(a)]{\label{3ord_a}\includegraphics[scale=0.55]{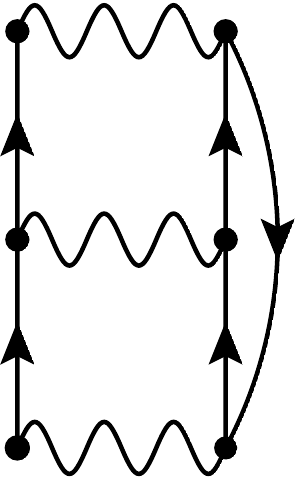}}
  \hspace{2cm}
    \subfloat[(b)]{\label{3ord_b}\includegraphics[scale=0.55]{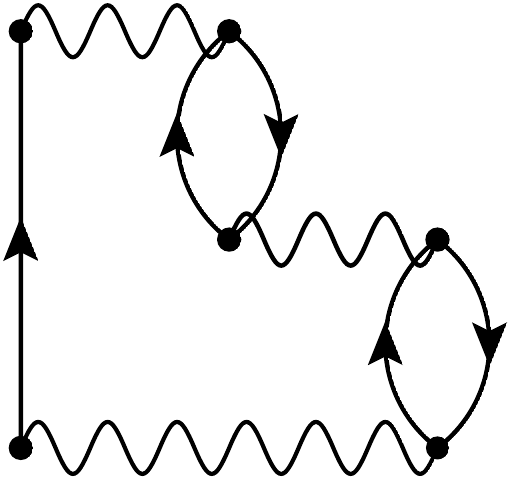}}
    \hspace{2cm}
    \subfloat[(c)]{\label{3ord_c}\includegraphics[scale=0.55]{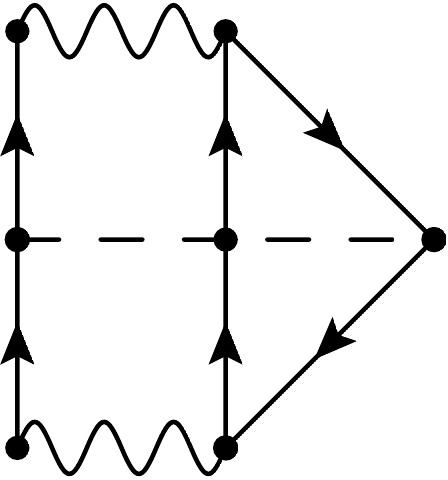}}
  \caption{As described in the caption of Fig.~\ref{2ord} but for the third-order diagrams with only $2p1h$ and $2h1p$ intermediate state configurations.}
  \label{3ord_a_b_c}
\end{figure}

For the irreducible self-energy, all one-particle irreducible, skeleton and interaction-irreducible diagrams up to third order have been derived in Ref.~\cite{CarAr13}. 
%
Within the skeleton expansion, i.e. when single-particle propagators are correlated, the irreducible self-energy up to the third order is given by the exact static part, Eq.~(\ref{SE_eq_U}), the  two second-order diagrams of Fig.~\ref{2ord}, and  the 17 third-order diagrams of Figs.~\ref{3ord_a_b_c} and~\ref{3ord_remn}. In this case, the energy-dependent part of the self-energy contains only effective 2NFs and irreducible 3NFs as interaction insertions. 
 Note that because of Eq.~\eqref{veff}, the contribution of Fig.~\ref{2ord_2B} actually corresponds to four separate diagrams if expressed in terms of the bare Hamiltonian~\eqref{plain_H}, of which three are interaction reducible~\cite{CarAr13}. Likewise, many more reducible diagrams would appear at third order.
 Without propagator renormalization, when one considers the diagrammatic expansion with reference propagators $g^{(0)}_{\alpha\beta}(\omega)$ as internal fermionic lines, other diagrams with different topologies must be included to take into account explicitly additional correlations in both the static and dynamic part of the self-energy. These terms contain also nonskeleton diagrams that include $\widetilde{U}$ and are presented in Appendix~\ref{General_reference_state}.

In Fig.~\ref{2ord} we show the only two one-particle irreducible, skeleton and interaction-irreducible diagrams at second order. These diagrams imply different sets of intermediate state configurations (ISCs), which are multiparticle-multihole, or multihole-multiparticle, excitations produced by the interaction and propagating within the nuclear medium. The diagram in Fig.~\ref{2ord_2B} involves  two-particles--one-hole ($2p1h$) and two-holes--one-particle ($2h1p$) ISCs, but it is computed with the 2N effective interaction Eq.~\eqref{veff} instead
of the original 2NF. Hence, it contains contributions form the 3NF $\hat{W}$.
 
The diagram in 
Fig.~\ref{2ord_3B} arises instead from an interaction-irreducible 3NF. 
There are two reasons to assume that this contribution is less important than the one in Fig.~\ref{2ord_2B}:
first, 3NFs are generally weaker than
corresponding  2NFs (typically, $<\widehat{W}>\approx\frac{1}{10}<\widehat{V}>$ for nuclear
interactions~\cite{Grange1989,Epelbaum2009}); second, the diagram in 
Fig.~\ref{2ord_3B} involves $3p2h$ and $3h2p$ ISCs, which involve higher excitation energies and therefore they are expected 
to play a minor role at the Fermi surface due to phase space arguments.

By the same token, we may expect that the three diagrams shown in Fig.~(\ref{3ord_a_b_c}) are the dominant ones among the 17 one-particle irreducible, skeleton, and interaction-irreducible self-energy diagrams appearing at third order in the expansion of $\widetilde{\Sigma}_{\alpha\beta}(\omega)$. While all diagrams in Fig.~(\ref{3ord_a_b_c}) involve $2p1h$ and $2h1p$ ISCs, those in Figs.~\ref{3ord_a} and \ref{3ord_b} contain only effective 2NFs. These two diagrams have been already included in actual calculations for nuclear matter and for finite nuclei~\cite{Soma08,Heb10,Heb11,Cipol13,Car13,Cipol15}. 

\subsection{\label{General} Solution of the Dyson equation as a matrix eigenvalue problem}
The ADC is a systematic  approach to calculate nonperturbative approximations to the self-energy. Once the latter is known, we still need to solve the Dyson equation to obtain the propagator, as explained below.

\begin{figure}[t]
  \centering
    \subfloat[(a)]{\label{Ex_ADC_a}\includegraphics[scale=0.55]{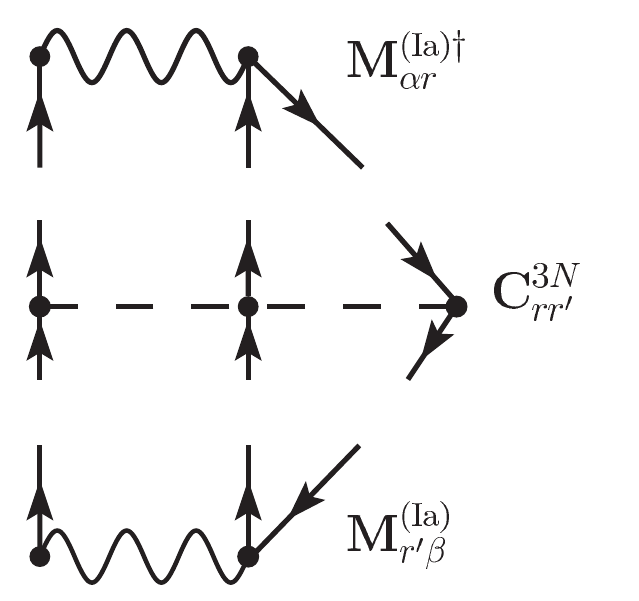}}
    \hspace{0.7cm}
    \subfloat[(b)]{\label{Ex_ADC_b}\includegraphics[scale=0.53]{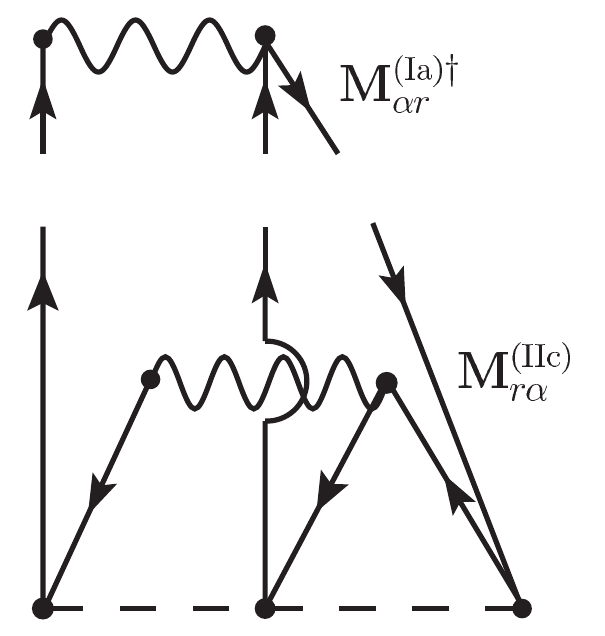}}
  \caption{Decomposition of Goldstone diagrams in terms of interaction and coupling matrices.  Two (out of the six) Goldstone diagrams that arise from the self-energy Feynman diagram of Fig.~\ref{3ord_c} are shown. Diagram (a) contains the interaction matrix $\textbf{C}^{3N}_{r  r'}  $ (see Eq.~\eqref{int_term_5c_C}) linked to two lowest order coupling matrices $\textbf{M}^{(\textrm{Ia})}_{r \alpha}$ (see Eq.~\eqref{eq:M_2a}). Diagram (b) contains only coupling matrices and includes the second-order correction $\textbf{M}^{(\textrm{IIc})}_{ r \alpha}$  reported in Eq.~(\ref{eq:M_3c}). }
  \label{Ex_ADC}
\end{figure}

Without loss of generality, the dynamical part of the self-energy $\widetilde{\Sigma}_{\alpha\beta}(\omega)$ can be written in the Lehmann representation, similar to  Eq.~(\ref{eq:g1}). Specifically, we write,
\begin{align}
\label{irr_SE_Lehmann}
\widetilde{\Sigma}_{\alpha\beta}(\omega) \! & = \! \sum_{j j'} \textbf{M}_{\alpha j}^\dagger \Bigg[ \frac{1}{ \hbar \omega \mathds{1} - (\textbf{E}^{>}   +  \textbf{C}) + \textrm{i} \eta\mathds{1}} \Bigg]_{\substack{\! \! j  j'}} \! \! \textbf{M}_{j' \beta}  \nonumber \\
 & \! + \! \sum_{k k'} \textbf{N}_{\alpha k} \Bigg[ \frac{1}{\hbar \omega \mathds{1} - (\textbf{E}^{<} +  \textbf{D}) - \textrm{i} \eta \mathds{1}} \Bigg]_{\! \!k k'} \! \! \textbf{N}_{k' \beta }^\dagger \, ,
\end{align}
with $\textbf{E}^{>}_{j j'}$ ($\textbf{E}^{<}_{k k'}$)  being energies of noninteracting ISCs, $\textbf{M}_{j \alpha} $ ($\textbf{N}_{\alpha k} $) coupling matrices, and $\textbf{C}_{\substack {j  j'}}  $ ($\textbf{D}_{\substack {k   k'}} $)  interaction matrices for the forward-in-time (backward-in-time) self-energy. 
 Coupling matrices link initial and final single-particle states of the propagator to ISCs, while interaction matrices are those parts of the self-energy diagrams that represent interactions among ISCs alone (see also Fig.~\ref{Ex_ADC}). It follows that interaction matrices contain at most one interaction vertex and are not linked to the single-particle states of the  model space.
We use Latin letters as a collective indices to label ISCs: In particular, we use $i$ for any general configuration, while  $j, j'$ ($k, k'$)  denote  forward-in-time multiparticle-multihole (backward-in-time multihole-multiparticle) configurations. In the following, we will consider explicitly $2p1h$, $3p2h$, $2h1p$, and $3h2p$ ISCs, which are included in Eq.~(\ref{irr_SE_Lehmann}). For this purpose, we set the shorthand notation,
\begin{equation}
\label{def_r}
\begin{rcases}
  r &\equiv (n_1,n_2,k_3)  \\
  r' &\equiv  (n_4,n_5,k_6)  \\
  q &\equiv  (n_1,n_2,n_3,k_4,k_5) \\
  q' &\equiv  (n_6,n_7,n_8,k_9,k_{10}) \\
  \end{rcases}
\text{ $j$, $j'$} 
\end{equation}
for forward-in-time terms, and 
\begin{equation}
\label{def_s}
\begin{rcases}
  s &\equiv (k_1,k_2,n_3) \\
  s' &\equiv (k_4,k_5,n_6) \\
  u &\equiv (k_1,k_2,k_3,n_4,n_5)  \\
  u' &\equiv  (k_6, k_7, k_8, n_9, n_{10}) \\
  \end{rcases}
\text{ $k$, $k'$} 
\end{equation}
for backward-in-time terms, where $n_i$~($k_i$) refer to the particle (hole) indices of the propagator, Eq.~\eqref{eq:g1}.
For instance, ${\textbf{M}_{r \alpha} \equiv \textbf{M}_{ (n_1,n_2,k_3)\alpha} }$ connects a single-particle state of index $\alpha$ to an intermediate state composed by a $2p1h$ configuration, 
whereas more complicated coupling matrices such as  $\textbf{M}_{q \alpha} \equiv \textbf{M}_{ (n_1,n_2,n_3,k_4,k_5) \alpha} $ involve $3p2h$ configurations. 
As one moves to higher orders beyond ADC(3) more complex multi-particle-multi-hole states appear in Eq.~(\ref{irr_SE_Lehmann}),  eventually covering the complete space of ISCs.

To better clarify how the ADC building blocks are associated to perturbation theory, we show in Fig.~\ref{Ex_ADC} the diagrammatic decomposition for two of the Goldstone contributions that arise from the self-energy Feynman diagram of Fig.~\ref{3ord_c}. The expressions for the coupling and interactions matrices can be read directly form the analytic expression of each Goldstone diagram. See Ref.~\cite{Barb2017NLP} for a detailed pedagogical  discussion.

In virtue of the Pauli principle, the expressions for coupling and interaction matrices can (and should) be made antisymmetric with respect to permutation of any two particle or any two hole indexes. This results naturally from the antisymmetry of the interaction matrix elements, Eq.~\eqref{H}, and by including a complete set of diagrams at each order, which generates all possible permutations~\cite{Schirmer1983,Soma2011}.
Using the antisymmetry, it is possible to restrict the sums in Eq.~\eqref{irr_SE_Lehmann} to sets of ordered single-particle indices, defining for instance ($n_1 < n_2,k_3$), (${n_1 < n_2 < n_3,k_4 < k_5}$), and so on. This is very important for practical implementations, owing to a substantial reduction of the dimension of the Dyson matrix, Eq.~\eqref{eq:Dy_residual_5} below.  On the other hand, manipulations such as coupling the angular momenta of ISCs is better performed in the general case of unrestricted summations.
Therefore, we will present all the working equations without assuming ordered indexes, as given by the notation of Eqs.~(\ref{def_r}) and (\ref{def_s}). One can always apply ordered summation by removing the relevant symmetry factors to the expressions given in Sec.~\ref{ADC(N)} and in the Appendices.

ISC energies are diagonal matrices in these indexes. For nucleon addition, with ${M+1}$ particles and $M$ holes, $(M+1)pMh$, we have,
\begin{eqnarray}
\label{energy_E_i}
\textbf{E}^{>}_{j j'} = \textbf{E}^{>}_{j} &=& \text{diag} (\varepsilon^{+}_{n_1} + \varepsilon^{+}_{n_2} + \cdots   + \varepsilon^{+}_{n_M}  + \varepsilon^{+}_{n_{M+1}} \nonumber \\ 
&& - \varepsilon^{-}_{k_1} - \varepsilon^{-}_{k_2} - \cdots - \varepsilon^{-}_{k_M} ) \, ,
\end{eqnarray}
while for the nucleon removal ISC,
\begin{eqnarray}
\label{energy_E_k}
\textbf{E}^{<}_{k k'} = \textbf{E}^{<}_{k} &=& \text{diag} (\varepsilon^{-}_{k_1} + \varepsilon^{-}_{k_2}  +  \cdots + \varepsilon^{-}_{k_M}  + \varepsilon^{-}_{k_{M+1}} \nonumber \\
&& - \varepsilon^{+}_{n_1} - \varepsilon^{+}_{n_2} - \cdots - \varepsilon^{+}_{n_M} ) \, .
\end{eqnarray}

To solve for the 1B propagator of Eq.~(\ref{eq:g1}), it is useful to recast the Dyson equation in matrix form, which allows a more efficient computation of 1B propagator eigenvalues. To see this point, we have to start again from the Dyson equation~(\ref{eq:Dy}) and regard the 1B propagator Eq.~\eqref{eq:g1} as a meromorphic function on the complex energy plane. This function has simple poles and residues given by  one-nucleon addition (or removal) energies and transition amplitudes, respectively. We can then find a relation among the transition amplitudes of Eq.~(\ref{tran_ampl}), by extracting them as residues of the  propagator in the Dyson equation. This gives,
\begin{equation}
  \label{eq:Dy_residual_II}
  \mathcal{Z}_{\alpha}^i (\mathcal{Z}_{\beta}^i)^{\dagger}=  \sum_{\gamma \delta} \,
  \Bigg[  \frac{1}{\hbar \omega \mathds{1} -\hat{H}_0} \Bigg]_{\alpha \gamma}  \Sigma_{\gamma \delta}^*(\omega)\mathcal{Z}_{\delta}^i (\mathcal{Z}_{\beta}^i)^{\dagger} \, \Big\rvert_{\hbar\omega =\varepsilon_i} \, .
\end{equation}

%
\begin{widetext}
By using the decomposition of Eq.~(\ref{irr_SE_Lehmann})  we obtain the relation,
\begin{eqnarray}
 \label{eq:Dy_residual_3}
\mathcal{Z}_{\alpha}^i = \sum_{\gamma \delta} \,
  \Bigg[  \frac{1}{\hbar \omega \mathds{1} -\hat{H}_0} \Bigg]_{\alpha \gamma} 
  \Sigma_{\gamma \delta}^*(\omega)\mathcal{Z}_{\delta}^i  \, \Big\rvert_{\hbar\omega =\varepsilon_i}
= \sum_{\gamma}  \Bigg[  \frac{1}{\hbar \omega \mathds{1} -\hat{H}_0} \Bigg]_{\alpha \gamma} 
 \left(\sum_{\delta} \Sigma_{\gamma \delta}^\infty \, \mathcal{Z}_{\delta}^i +\sum_{j} \textbf{M}_{\gamma j}^\dagger \mathcal{W}_j^i   + 
\sum_{k } \textbf{N}_{ \gamma k} \mathcal{W}_k^i \right)  \Bigg\rvert_{\hbar\omega =\varepsilon_i} \, ,
\end{eqnarray}
which contains both forward-in-time and backward-in-time solutions of the propagator. In the last equality of Eq.~(\ref{eq:Dy_residual_3}) we have introduced the vectors $\mathcal{W}_j^i $ and $\mathcal{W}_k^i $, defined as
\begin{equation}
  \label{eq:Dy_residual_4}
\mathcal{W}^i_j \equiv  \mathcal{W}_j(\omega) \, \Big\rvert_{\hbar \omega =\varepsilon_i}  =  \sum_{j'} \Bigg[  \frac{1}{\hbar \omega \mathds{1} -(\textbf{E}^{>} + \textbf{C} )} \Bigg]_{j j'} \, \sum_{\delta} \textbf{M}_{j' \delta} \, \mathcal{Z}_{\delta}^i \, \Bigg\rvert_{\hbar \omega =\varepsilon_i} \; ,  
\end{equation}
and
\begin{equation}
  \label{eq:Dy_residual_4_bis}
\mathcal{W}^i_k \equiv  \mathcal{W}_k(\omega) \, \Big\rvert_{\hbar \omega =\varepsilon_i}  =  \sum_{k'}  \Bigg[  \frac{1}{\hbar \omega \mathds{1} -(\textbf{E}^{<} + \textbf{D} )} \Bigg]_{k k'} \, \sum_{\delta} \textbf{N}^{\dagger}_{k' \delta} \,
\mathcal{Z}_{\delta}^i  \, \Bigg\rvert_{\hbar \omega =\varepsilon_i}\; ,
\end{equation}
respectively. These vectors are introduced to recast the Dyson equation as a large scale eigenvalue problem whose diagonalization gives the  eigenspectra of the $\ket{\Psi_{n}^{A+1}}$ and $\ket{\Psi_{k}^{A-1}}$ systems and the transition amplitudes of the 1B propagator. By recoupling  Eqs.~(\ref{eq:Dy_residual_3})-(\ref{eq:Dy_residual_4_bis}), we obtain:
\begin{equation}
  \label{eq:Dy_residual_5}
  \epsilon_i  
  \begin{pmatrix}
  \mathcal{Z}_{\alpha}^i\\
  \\
  \mathcal{W}^i_r \\
  \\
  \mathcal{W}^i_s \\
     \\
  \mathcal{W}^i_q \\
  \\
  \mathcal{W}^i_u \\ 
  \vdots
   \end{pmatrix} =
    \begin{pmatrix}
   h^{(0)}_{\alpha \delta}  \! + \! \Sigma_{\alpha \delta}^\infty    &  \textbf{M}^{\dagger}_{\alpha r'} & \textbf{N}_{\alpha s'} &  \textbf{M}^{\dagger}_{\alpha q'} & \textbf{N}_{\alpha u'} & \cdots &    \\
   \\
   \textbf{M}_{r \delta} & \textbf{ E}^{>}_{r} \delta_{r r'} \! + \! \textbf{C}_{\substack {r r'}} &  0 & \textbf{C}_{\substack {r  q'}}  &  0 & \cdots \\
   \\
   \textbf{N}^{\dagger}_{s \delta} & 0 & \textbf{E}^{<}_{s} \delta_{s s'}  \! + \! \textbf{D}_{\substack {s s'}} & 0 & \textbf{D}_{\substack { s  u'}}  & \cdots \\
   \\
     \textbf{M}_{q \delta} & \textbf{C}_{\substack {q r'}}  & 0 & \textbf{E}^{>}_{q} \delta_{q q'} \! + \! \textbf{C}_{\substack {q  q'}} &  0 &\cdots  \\
   \\
     \textbf{N}^{\dagger}_{u \delta}  & 0 & \textbf{D}_{\substack {u  s'}}  & 0 & \textbf{E}^{<}_{u} \delta_{u u'}  \! + \! \textbf{D}_{\substack {u  u'}} & \cdots\\     
     \vdots  & \vdots & \vdots &  \vdots &  \vdots
\end{pmatrix} 
   \begin{pmatrix}
  \mathcal{Z}_{\delta}^i \\
  \\
  \mathcal{W}^i_{r'} \\
  \\
  \mathcal{W}^i_{s'} \\
   \\
  \mathcal{W}^i_{q'} \\
  \\
  \mathcal{W}^i_{u'} \\ 
  \vdots
   \end{pmatrix} \, ,
\end{equation}
\end{widetext}
where the eigenmatrix is referred to as `Dyson matrix' and it is independent of $\hbar\omega$. 
The zero entries mean that the forward-in-time and backward-in-time sectors are coupled together only through single particle states, and the dots stay for the self-energy terms with ISCs beyond the $3p2h$ and $3h2p$ configurations.

Equation~(\ref{eq:Dy_residual_5}) is an energy-independent eigenvalue problem, whereas the components $\mathcal{W}^i$ of the eigenvectors are functions of the corresponding eigenvalue $\varepsilon_i$, as it is apparent from definitions~(\ref{eq:Dy_residual_4}-\ref{eq:Dy_residual_4_bis}). The diagonalization of the Dyson matrix yields all the poles of the propagator at once, while the normalization of the $i$-th eigenvector is given by
\begin{equation}
\label{normal}
\sum_{\alpha} \left|\mathcal{Z}_{\alpha}^i \right|^2  + \sum_{j } \left| \mathcal{W}^i_j\right|^2+ \sum_{k}  \left| \mathcal{W}^i_k \right|^2=1 \, .
\end{equation}
In a self-consistent calculation, the elements of the Dyson matrix Eq.~\eqref{eq:Dy_residual_5} depend on the quasiparticle energies and amplitudes, $\varepsilon_i$ and $\mathcal{Z}_{\alpha}^i$. Thus, they require an iterative solution.  The large number of poles in the dressed propagator, see Eq.~\eqref{eq:g1}, implies a severe growth in the dimension of the Dyson matrix at each iteration, as explained in Sec.~IIIA of Ref.~\cite{Soma2014}. This can be handled by projecting the space of intermediate configurations into smaller Krylov subspaces, using Lanczos-type algorithms with multiple pivots~\cite{Soma2014}.

\section{\label{ADC(N)} General outline of the ADC($n$) method}

The irreducible self-energy $\Sigma_{\alpha\beta}^{\star}(\omega)$ is the object of the ADC formalism applied in this work. Its expression  as a product of matrices, Eqs.~\eqref{irr_SE_decomp} and~\eqref{irr_SE_Lehmann}, is the most general analytic form that is consistent with the causality principle and the known Lehmann representation.

Our task is then to find expressions for  coupling and interaction matrices including the correlations due to 2NFs and 3NFs.
The ADC($n$) strategy consists in deriving explicit expressions of the coupling and interaction matrices by expanding Eq.~(\ref{irr_SE_Lehmann}) in powers of 2NFs and 3NFs and then to compare with the Goldstone-Feynman expansion up to order $n$. Formally, we have
\begin{equation}
\label{expan_M}
\textbf{M}_{j\alpha } = \textbf{M}^{(\textrm{I})}_{j \alpha} + \textbf{M}^{(\textrm{II})}_{j \alpha} + \textbf{M}^{(\textrm{III})}_{j \alpha} + \dots \; ,
\end{equation}
where the term $\textbf{M}^{(\textrm{n})}_{j \alpha} $ is of $\textrm{n}$th order in the residual Hamiltonian $H_1$, and similarly for backward-in-time coupling matrices:
\begin{equation}
\label{expan_N}
\textbf{N}_{\alpha k} = \textbf{N}^{(\textrm{I})}_{\alpha k} + \textbf{N}^{(\textrm{II})}_{\alpha k} + \textbf{N}^{(\textrm{III})}_{\alpha k} + \dots \; .
\end{equation}

By plugging Eqs.~(\ref{expan_M}) and \ref{expan_N}) into Eq.~(\ref{irr_SE_Lehmann}), we obtain the corresponding expansion for the energy-dependent irreducible self-energy up to third order (first order contributions are all included in $\Sigma_{\alpha\beta}^{\infty}$). This is

\begin{widetext}
\begin{eqnarray}
\label{irr_SE_EXPA}
\widetilde{\Sigma}_{\alpha\beta}(\omega) & = & \sum_{j} \textbf{M}^{(\textrm{I}) \dagger}_{\alpha j} \left[ \frac{1}{\hbar \omega - \textbf{E}^{>}_j + \textrm{i} \eta} \right]  \textbf{M}^{(\textrm{I})}_{j \beta  } \nonumber \\
&& + \sum_{j} \textbf{M}^{(\textrm{II}) \dagger}_{\alpha j} \left[ \frac{1}{\hbar \omega - \textbf{E}^{>}_j + \textrm{i} \eta} \right] \textbf{M}^{(\textrm{I})}_{j \beta }  + \sum_{j} \textbf{M}^{(\textrm{I}) \dagger}_{\alpha j} \left[ \frac{1}{\hbar \omega - \textbf{E}^{>}_j + \textrm{i} \eta} \right]  \textbf{M}^{(\textrm{II})}_{j \beta } \nonumber \\
&&+  \sum_{j j'} \textbf{M}^{(\textrm{I}) \dagger}_{\alpha j} \left[ \frac{1}{\hbar \omega - \textbf{E}^{>}_j  + \textrm{i} \eta} \right] \textbf{C}_{\substack {j  j'}}  \left[ \frac{1}{\hbar \omega - \textbf{E}^{>}_{j'} + \textrm{i} \eta} \right] \textbf{M}^{(\textrm{I})}_{j' \beta } + \dots \nonumber \\
&& +  \sum_{k} \textbf{N}^{(\textrm{I})}_{\alpha k} \left[ \frac{1}{\hbar \omega - \textbf{E}^{<}_k- \textrm{i} \eta} \right] \textbf{N}_{k \beta}^{(\textrm{I}) \dagger}\nonumber \\
&& +  \sum_{k} \textbf{N}^{(\textrm{II})}_{\alpha k} \left[ \frac{1}{\hbar \omega - \textbf{E}^{<}_k- \textrm{i} \eta} \right] \textbf{N}_{k \beta}^{(\textrm{I}) \dagger} +
 \sum_{k} \textbf{N}^{(\textrm{I})}_{\alpha k} \left[ \frac{1}{\hbar \omega - \textbf{E}^{<}_k- \textrm{i} \eta} \right] \textbf{N}_{k \beta }^{(\textrm{II}) \dagger} \nonumber \\
&&+  \sum_{k k'} \textbf{N}^{(\textrm{I})}_{\alpha k} \left[ \frac{1}{\hbar \omega - \textbf{E}^{<}_k  - \textrm{i} \eta} \right] \textbf{D}_{\substack {k k'}}  \left[ \frac{1}{\hbar \omega - \textbf{E}^{<}_{k'}  - \textrm{i} \eta} \right]  \textbf{N}^{(\textrm{I}) \dagger}_{k' \beta} + \dots \; ,
\end{eqnarray}
\end{widetext}
for both forward-in-time and backward-in-time self-energy parts. The comparison of the formal expansion of Eq.~(\ref{irr_SE_EXPA}) with the calculated Goldstone-type diagrams, gives the minimal expressions for interaction and coupling matrices in terms of the transition amplitudes $\mathcal{Z}_{\alpha}^{i}$ and the one-nucleon addition $\varepsilon_n^{+}$ and removal $\varepsilon_k^{-}$ energies of the 1B propagator $g_{\alpha \beta}(\omega)$ of Eq.~(\ref{eq:g1}).
By looking at the expansion in Eq.~(\ref{irr_SE_EXPA}), we see that the third-order terms containing the interaction matrices  $\textbf{C}_{\substack {j  j'}}$
and $\textbf{D}_{\substack {k k'}}$ do not retain the same analytic form as Eq.~(\ref{irr_SE_Lehmann}), which is based on the Lehmann representation of the propagator itself. To recover this analytic form in terms of self-energy poles, one must introduce higher-order terms and perform a resummation  of those diagrams up to infinite order: this resummation is implicit in Eq.~(\ref{eq:Dy_residual_5}) and it gives the nonperturbative character of the method, which takes into account at all orders several types of diagrams, particle-particle and hole-hole ladders, and  particle-hole rings, as well as other resummations induced by 3NFs.


\subsection{\label{sec:ADC(2)} ADC method at second order: ADC(2)}

In this section we present the explicit expressions of coupling  and interaction matrices entering the ADC(2) formalism. The two second order diagrams shown in Fig.~\ref{2ord} are sufficient to define the ADC(2) approximation scheme.
Coupling  and interaction matrices required to build the ADC(3) are introduced in  Sec.~\ref{sec:ADC(3)} and Appendix~\ref{M_N_matrices}. Unless otherwise stated, for coupling and interaction matrices we adopt the Einstein's convention of summing over repeated indices for both the model-space single-particle states ($\alpha$, $\beta$, $\dots$) and the particle and hole orbits ($n_1$, $n_2$, $\dots$, $k_1$, $k_2$ $\dots$). We also use collective indexes for ISCs according to the notation set in Eqs.~(\ref{def_r}) and (\ref{def_s}), where appropriate.

We have seen that for a given multiparticle-multihole or multihole-multiparticle configuration, we can have coupling matrices at different orders according to the expansions in Eqs.~(\ref{expan_M}) and~(\ref{expan_N}). Within a given order, coupling matrices can also differ with respect to the kind of interaction (2NF and/or 3NF) appearing in the term. For this reason we specify in the notation an extra superscript distinguishing different contributions at the same order.  For instance, at second order we will encounter a coupling matrix $\textbf{M}^{(\textrm{Ia})}_{ r  \alpha} $ containing a 2N interaction linked to a ${r=(n_1,n_2,k_3)}$ ISC,  $\textbf{M}^{(\textrm{Ib})}_{ q \alpha} $ containing a 3N interaction linked to  a $q=(n_1,n_2,n_3,k_4,k_5)$ ISC, and so on. The extra superscript with Latin letter corresponds to the labels of diagrams in the figures.

\begin{figure}[t]
  \centering
    \subfloat[(a)]{\label{M_2a_nnk}\includegraphics[scale=0.45]{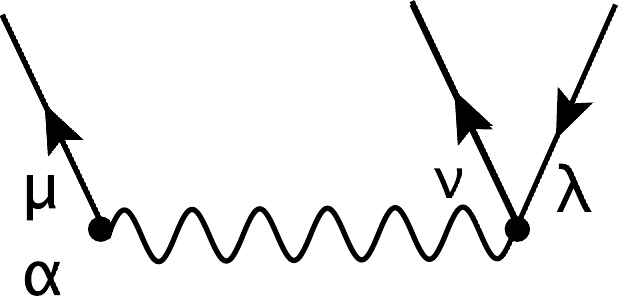}}
  \hspace{1.cm}
    \subfloat[(b)]{\label{M_2b_nnnkk}\includegraphics[scale=0.45]{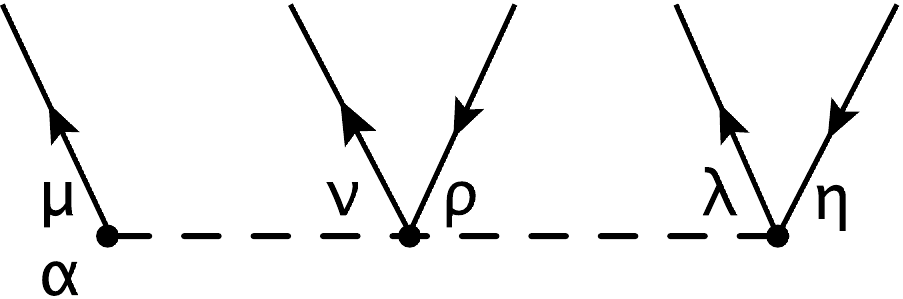}}
  \caption{ Diagrams of the self-energy coupling matrices with the effective 2NF $\widetilde{V}$ (left) and with interaction-irreducible 3NF $\hat{W}$ (right). The coupling matrix (a) connects to the $2p1h$  ISCs (see Eq.~(\ref{eq:M_2a})), while the coupling matrix (b) connects to the $3p2h$ ISCs (see Eq.~(\ref{eq:M_2b})).}
\label{M_2order}
\end{figure}

\begin{figure}[t]
  \centering
    \subfloat[(a)]{\label{N_2a_kkn}\includegraphics[scale=0.45]{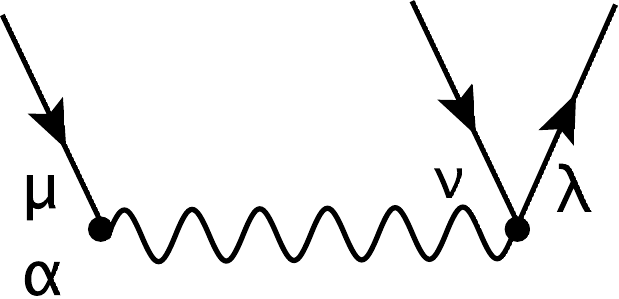}}
  \hspace{1cm}
    \subfloat[(b)]{\label{N_2b_kkknn}\includegraphics[scale=0.45]{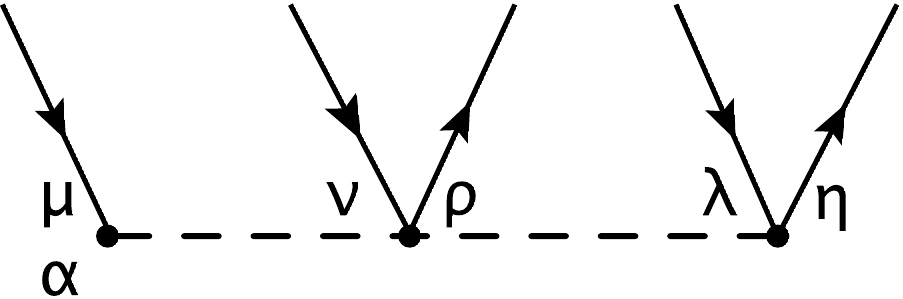}}
  \caption{ As in Fig.~\ref{M_2order} but for backward-in-time coupling matrices (see Eqs.~(\ref{eq:N_2a}) and~(\ref{eq:N_2b}) respectively).}
  \label{N_2order}
\end{figure}

\begin{widetext}
To illustrate the ADC procedure, we write first the entire expressions for all the Goldstone terms in each second order Feynman diagram of Fig.~\ref{2ord}. Then we display the formulas of the coupling matrices that can be singled out from the self-energy expressions.  The equation for the dynamic self-energy in Fig.~\ref{2ord_2B} reads,
\begin{eqnarray}
\label{1a_expr}
 \widetilde{\Sigma}^{(1a)}_{\alpha \beta}(\omega)=
 \frac{1}{2}
  \widetilde{V}_{\alpha \epsilon, \gamma \rho}
 \left(
\sum_{\substack{n_1,n_2, \\ k_3} } \frac{(\cX_{\gamma}^{n_1} \cX_{\rho}^{n_2} \cY_{\epsilon}^{k_3} )^* \cX_{\mu}^{n_1} \cX_{\nu}^{n_2}  \cY_{\lambda}^{k_3}}{\hbar \omega -\left(\varepsilon_{n_1}^{+}+\varepsilon_{n_2}^{+}-\varepsilon_{k_3}^{-} \right) + i\eta  }   + \sum_{\substack{k_1,k_2, \\ n_3} } \frac{ \cY_{\gamma}^{k_1} \cY_{\rho}^{k_2} \cX_{\epsilon}^{n_3} 
(\cY_{\mu}^{k_1} \cY_{\nu}^{k_2}  \cX_{\lambda}^{n_3})^*
 }{\hbar \omega - (\varepsilon_{k_1}^{-}+\varepsilon_{k_2}^{-}-\varepsilon_{n_3}^{+} ) - i \eta }  \right) \widetilde{V}_{\mu \nu ,  \beta \lambda} \,  . 
\end{eqnarray}

Being already in the Lehmann form of Eq.~(\ref{irr_SE_Lehmann}), we can read directly from Eq.~(\ref{1a_expr}) the forward-in-time contribution to the ADC(2) coupling matrix,
\begin{equation}
  \label{eq:M_2a}
\textbf{M}^{(\textrm{Ia})}_{ r  \alpha}  \equiv \frac{1}{\sqrt{2}}  \cX_{\mu}^{n_1}  \cX_{\nu}^{n_2} \cY_{\lambda}^{k_3}   \, \widetilde{V}_{\mu\nu,\alpha \lambda} \, ,
  \end{equation} 
  while in the backward-in-time channel we have
\begin{equation}
  \label{eq:N_2a}
\textbf{N}^{(\textrm{Ia})}_{\alpha s} \equiv \frac{1}{\sqrt{2}} \widetilde{V}_{\alpha \lambda,\mu\nu} \, \cY_{\mu}^{k_1}  \cY_{\nu}^{k_2} \cX_{\lambda}^{n_3}  \, ,
  \end{equation}
  that couples the effective  2NF with the $2h1p$ ISC. It is also clear that the interaction matrices $\textbf{C}_{\substack {j  j'}}$ and $\textbf{D}_{\substack {k k'}}$ are zero in ADC(2). The representations of Eqs.~(\ref{eq:M_2a}) and (\ref{eq:N_2a}) as fragments of Goldstone diagrams  are depicted in Figs.~\ref{M_2a_nnk} and \ref{N_2a_kkn}, respectively.

The equation for the energy-dependent self-energy with 3NFs in Fig.~\ref{2ord_3B} reads,
\begin{eqnarray}
\label{1b_expr}
 \widetilde{\Sigma}^{(1b)}_{\alpha \beta}(\omega)=
 \frac{1}{12}
  W_{\alpha \gamma \delta ,\xi \tau \sigma}
&& \left( \sum_{\substack{n_1,n_2,n_3, \\ k_4,k_5}}
  \frac{(\cX_{\xi}^{n_1}  \cX_{\tau}^{n_2} \cX_{\sigma}^{n_3} \cY_{\gamma}^{k_4} \cY_{\delta}^{k_5}  )^* \cX_{\mu}^{n_1}  \cX_{\nu}^{n_2} \cX_{\lambda}^{n_3} \cY_{\eta}^{k_4} \cY_{\rho}^{k_5}}{\hbar \omega -\left(\varepsilon_{n_1}^{+}+\varepsilon_{n_2}^{+}+\varepsilon_{n_3}^{+}-\varepsilon_{k_4}^{-}-\varepsilon_{k_5}^{-}  \right) + i\eta }  \right. \nonumber \\
&&\left. + \sum_{\substack{k_1,k_2,k_3, \\ n_4,n_5}} \frac{
 \cY_{\xi}^{k_1}  \cY_{\tau}^{k_2}  \cY_{\sigma}^{k_3}  \cX_{\gamma}^{n_4}  \cX_{\delta}^{n_5} (\cY_{\mu}^{k_1}  \cY_{\nu}^{k_2} \cY_{\lambda}^{k_3} \cX_{\eta}^{n_4} \cX_{\rho}^{n_5})^* }{\hbar \omega - (\varepsilon_{k_1}^{-}+\varepsilon_{k_2}^{-}+\varepsilon_{k_3}^{-}-\varepsilon_{n_4}^{+} -\varepsilon_{n_5}^{+}) - i\eta }  \right) W_{\mu \nu \lambda, \beta \eta \rho}  \; .
\end{eqnarray}
\end{widetext}

The coupling matrix that links the 3NF to $3p2h$ ISCs is found in the diagram of Fig.~\ref{2ord_3B} and it is read from Eq.~(\ref{1b_expr}). Its expression is,
  \begin{equation}
  \label{eq:M_2b}
\textbf{M}^{(\textrm{Ib})}_{ q \alpha} \equiv \frac{1}{\sqrt{12}}   \cX_{\mu}^{n_1} \cX_{\nu}^{n_2}  \cX_{\lambda}^{n_3}  \cY_{\rho}^{k_4} \cY_{\eta}^{k_5} \ W_{\mu\nu\lambda,\alpha \rho \eta} \, ,
  \end{equation}
  while the corresponding matrix linked to $3h2p$ ISCs is,
    \begin{equation}
  \label{eq:N_2b}
\textbf{N}^{(\textrm{Ib})}_{\alpha u} \equiv \frac{1}{\sqrt{12}}  W_{\alpha \rho \eta, \mu\nu\lambda} \, \cY_{\mu}^{k_1} \cY_{\nu}^{k_2}  \cY_{\lambda}^{k_3}  \cX_{\rho}^{n_4} \cX_{\eta}^{n_5} \, .
  \end{equation}
 Equation~(\ref{eq:N_2b}) is also found in the diagram of Fig.~\ref{2ord_3B} and in the second term of Eq.~(\ref{1b_expr}). Their representations as fragments of Goldstone diagrams are depicted in Figs.~\ref{M_2b_nnnkk} and \ref{N_2b_kkknn}.
  
The four coupling matrices in Eqs.~(\ref{eq:M_2a}), (\ref{eq:N_2a}) and~(\ref{eq:M_2b}), (\ref{eq:N_2b}), along with their complex conjugates, complete the set of matrices found in the irreducible Goldstone diagrams of the self-energy at second order, which are given by the first and fourth rows in Eq.~(\ref{irr_SE_EXPA}). 
All these matrices enter as building blocks of the ADC construction at second and third order in the expansion with respect to the nuclear interaction. To summarize, the ADC(2) approximation for Eq.~(\ref{irr_SE_Lehmann}) requires the following terms,

\begin{align}
\label{ADC2_somm}
\textbf{M}^{[\textrm{ADC(2)}]}_{j\alpha}   ={}&  
\begin{cases}
    \textbf{M}^{(\textrm{Ia})}_{r \alpha }  &\hbox{if $j$=$r$ ($2p1h$),} \\ 
    \textbf{M}^{(\textrm{Ib})}_{q \alpha }  &\hbox{if $j$=$q$ ($3p2h$),} 
 \end{cases}
 \\
\textbf{N}^{[\textrm{ADC(2)}]}_{\alpha k}  ={}& 
\begin{cases}
      \textbf{N}^{(\textrm{Ia})}_{\alpha s} &\hbox{if $k$=$s$ ($2h1p$),} \\
      \textbf{N}^{(\textrm{Ib})}_{\alpha u}   &\hbox{if $k$=$u$ ($3h2p$),}
  \end{cases}
  \\
\textbf{C}^{[\textrm{ADC(2)}]}_{j j'}  ={}&   0\, , \\
\label{ADC2_somm_end}\textbf{D}^{[\textrm{ADC(2)}]}_{k k'}  ={}&  0\, .
\end{align}
There are no interaction matrices $\textbf{C}_{j j'}$ and $\textbf{D}_{k k'}$ in the ADC(2), because  coupling matrices are linked directly without any intermediate interaction insertion. This is not true anymore in the ADC(3), where matrices $\textbf{M}^{(\textrm{Ia})}$ and $\textbf{N}^{(\textrm{Ia})}$ are linked through interaction matrices $\textbf{C}_{\substack {j  j'}}$ and $\textbf{D}_{\substack {k k'}}$ respectively, as it is the case for the third and sixth lines of Eq.~(\ref{irr_SE_EXPA}).

%
%

\subsection{\label{sec:ADC(3)} ADC method at third order: ADC(3)}

\begin{figure*}
\setcounter{subfigure}{3}
  \subfloat[(d)]{\label{3ord_223B_1}\includegraphics[scale=0.50]{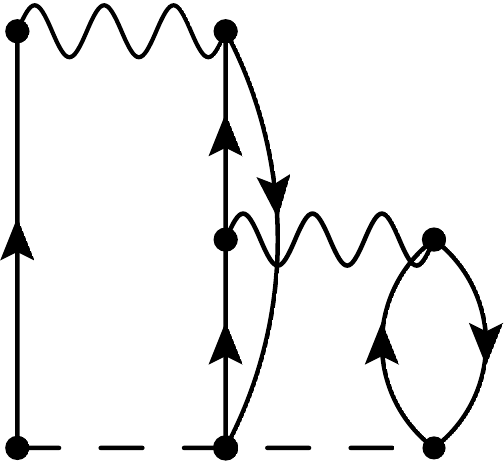}}
  \hspace{1.5cm}
  \subfloat[(e)]{\label{3ord_223B_2}\includegraphics[scale=0.50]{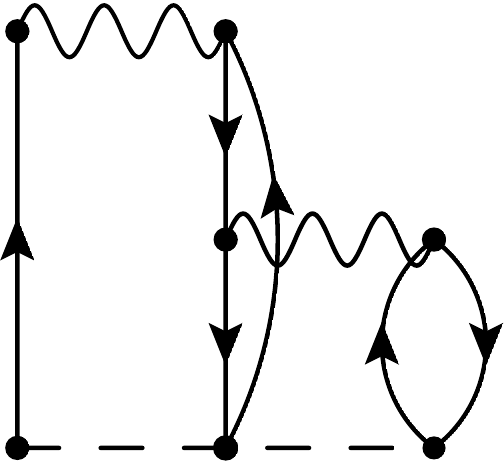}}
  \hspace{1.5cm}
   \subfloat[(f)]{\label{3ord_322B_1}\includegraphics[scale=0.50]{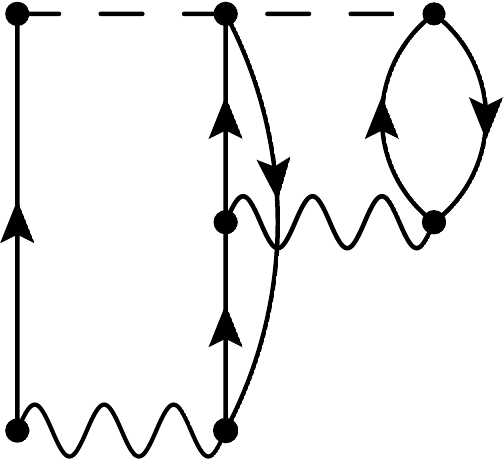}}
  \hspace{1.5cm}
  \subfloat[(g)]{\label{3ord_322B_2}\includegraphics[scale=0.50]{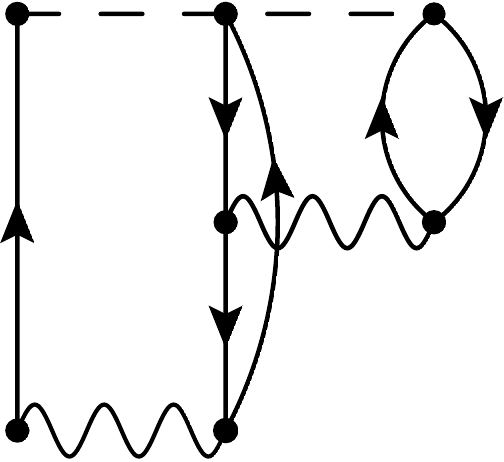}}
  \newline   \vskip .7cm
  \subfloat[(h)]{\label{3ord_233B_1}\includegraphics[scale=0.50]{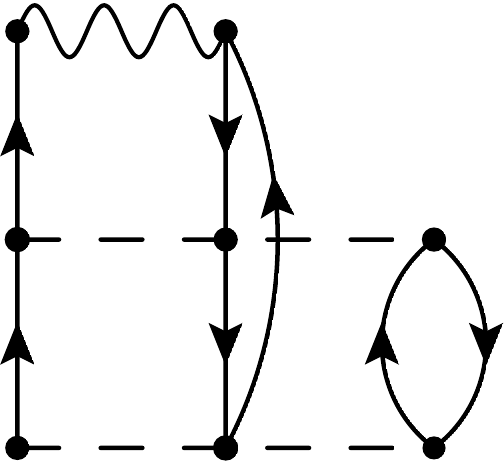}}
  \hspace{1.5cm}
  \subfloat[(i)]{\label{3ord_233B_2}\includegraphics[scale=0.50]{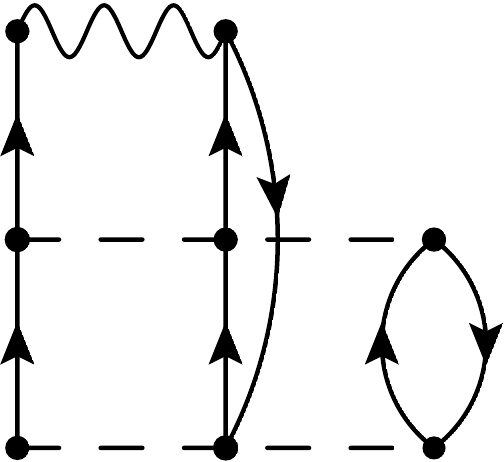}}
  \hspace{1.5cm}
  \subfloat[(j)]{\label{3ord_332B_1}\includegraphics[scale=0.50]{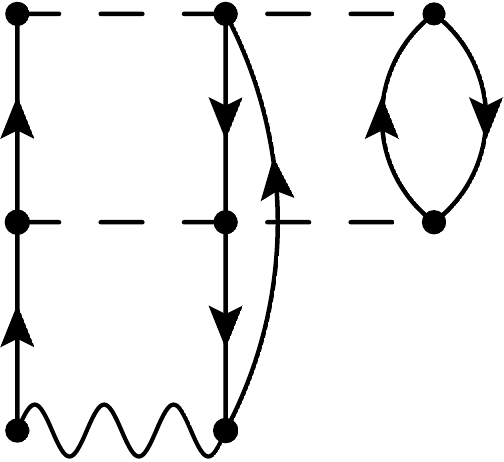}}
  \hspace{1.5cm}
  \subfloat[(k)]{\label{3ord_332B_2}\includegraphics[scale=0.50]{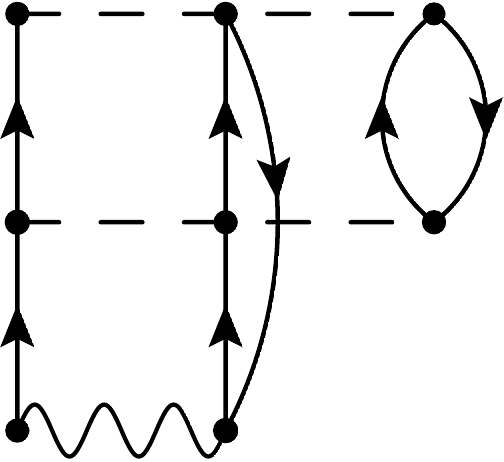}}
  \newline   \vskip .7cm
  \subfloat[(l)]{\label{3ord_323B_1}\includegraphics[scale=0.50]{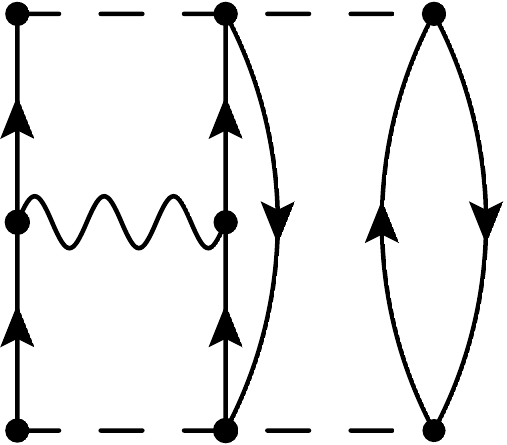}}
  \hspace{1.5cm}
  \subfloat[(m)]{\label{3ord_323B_2}\includegraphics[scale=0.50]{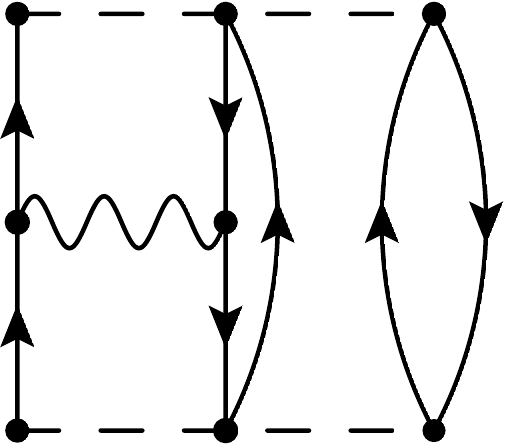}}
  \hspace{1.5cm}
  \subfloat[(n)]{\label{3ord_323B_3}\includegraphics[scale=0.50]{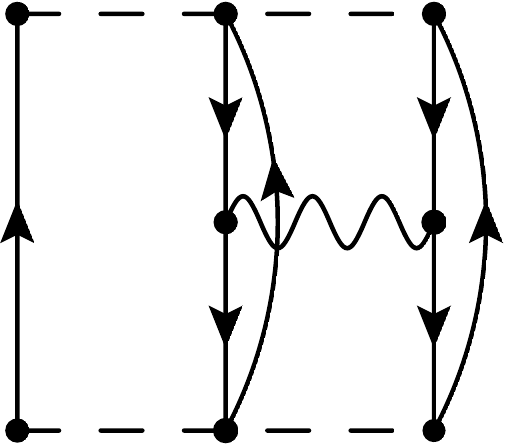}}
  \hspace{4cm}
  \newline   \vskip .7cm
  \hspace{0.4cm}
  \subfloat[(o)]{\label{3ord_333B_1}\includegraphics[scale=0.55]{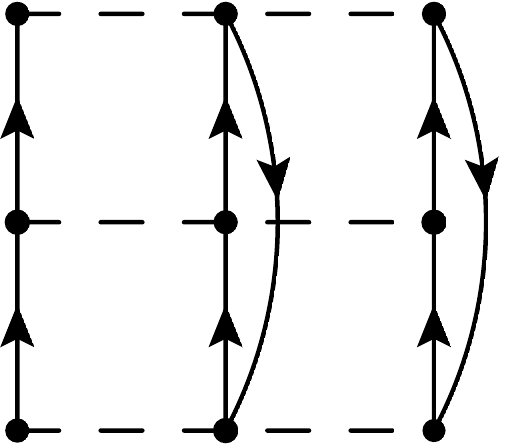}}
  \hspace{1.1cm}
  \subfloat[(p)]{\label{3ord_333B_2}\includegraphics[scale=0.55]{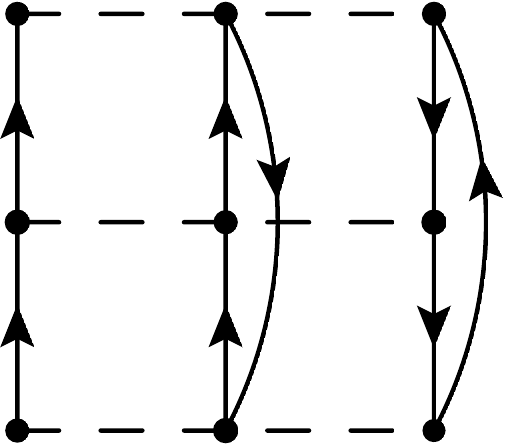}}
  \hspace{1.1cm}
  \subfloat[(q)]{\label{3ord_333B_3}\includegraphics[scale=0.55]{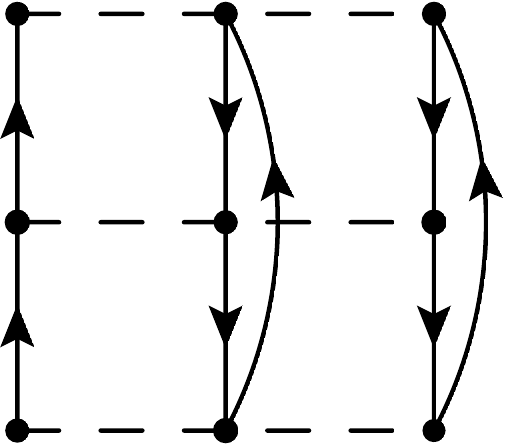}}
    \hspace{5.0cm}
    \newline
   \caption{As in Fig.~\ref{2ord} but for  third-order diagrams that include any $3p2h$ and $3h2p$ intermediate state configurations.
   Together with the diagrams (a), (b) and (c) of Fig.~\ref{3ord_a_b_c}, these are all skeleton and interaction-irreducible contributions
   present at third order. The labels in each diagram match the naming used in the text for of the corresponding coupling matrices. }
  \label{3ord_remn}
\end{figure*}

In this section we present  explicit expressions of the coupling and interaction matrices entering in the ADC formalism at third order, for the three diagrams shown in Fig.~\ref{3ord_a_b_c} and for the four diagrams appearing first in each row of Fig.~\ref{3ord_remn}. The diagrams in Fig.~\ref{3ord_a_b_c} contribute to the block-diagonal entries ($r r'$) and ($s s'$) of Eq.~(\ref{eq:Dy_residual_5}), corresponding to $2p1h$ and $2h1p$ ISCs, which are the simplest configurations to be excited in the Fock space. The diagrams depicted in Figs.~\ref{3ord_a} and \ref{3ord_b} are  the dominant ones at third order, given that only 2N interactions are present. The diagram in Fig.~\ref{3ord_c} contains instead a 3NF, but it can nonetheless play a significant role, because its Goldstone diagrams feature only $2p1h$ and $2h1p$ ISCs. 

 Each row in Fig.~\ref{3ord_remn} collects a different topology of diagrams in terms of number of effective 2NFs and interaction-irreducible 3NFs entering the diagrams.  In general, these diagrams are less important compared to the ones in Fig.~\ref{3ord_a_b_c}, because they feature at least a  $3p2h$ or a $3h2p$ ISC in all their Goldstone contributions. For forward-in-time (backward-in-time) diagrams,  topologies in the first and second row of Fig.~\ref{3ord_remn} couple $2p1h$ ($2h1p$) ISCs to $3p2h$ ($3h2p$) ISCs. They are linked by $\textbf{C}_{\substack {r q'}} $ ($\textbf{D}_{\substack {s  u'}}$) and $\textbf{C}_{\substack {q  r'}}$  ($\textbf{D}_{\substack {u s'}}$), accounting for  off-diagonal entries of Eq.~(\ref{eq:Dy_residual_5}). Within these two kinds of topologies,  diagrams in the first row contain only one 3NF, therefore they are expected to be more important than the ones in the second row, each featuring two 3NFs.  Finally, the last two rows in Fig.~\ref{3ord_remn} introduce the diagonal coupling between ISCs with five fermionic lines, $3p2h$ for forward-in-time diagrams and $3h2p$ for backward-in-time diagrams, corresponding respectively to  entries  ($q q'$) and ($u u'$) of Eq.~(\ref{eq:Dy_residual_5}). Again, there is a hierarchy between the two topologies, with those in the fourth row being less important due to the presence of three 3NFs.

The first four diagrams in each row are assumed as emblematic for each topology, and treated in the present section. The remaining coupling  and interaction matrices originating from third-order diagrams are given in Appendix~\ref{M_N_matrices}.
 

Coupling  and interaction matrices are fully antisymmetrized with respect to their particle and hole indexes. To show this explicitly, we introduce the following antisymmetrizer operators. Given a function depending on up to three particle or hole indexes, i.e.~$f(i,  j, h)\equiv f(n_i, n_j, n_h)$ or~$f(i,  j, h)\equiv f(k_i, k_j, k_h)$, the antisymmetric permutation operator of a pair of indexes is introduced,
\begin{equation}
\label{Perm_2}
\mathcal{A}_{i j} \; f(i, j, h) \equiv f(i, j, h) - f(j, i, h) \, .
\end{equation}
For  $3p2h$ and $3h2p$ configurations, it is useful to define the cyclic permutation operator as
\begin{equation}
\label{Perm_cyclic}
\mathcal{P}_{i j k} \; f(i, j, h)  \equiv  f(i, j, h) + f(h, i, j) + f(j, h, i)\, ,
\end{equation}
and the permutation operator acting on three indexes, that is
\begin{eqnarray}
\label{Perm_3}
\mathcal{A}_{i j h} \; f(i, j, h) &\equiv & f(i, j, h)+  f(h, i, j)  +  f(j, h, i) \nonumber  \\
&-&  f(i, h, j) - f(j, i, h)  -  f(h, j, i) \, .
\end{eqnarray}



Coupling matrices appearing at third order contain two interaction operators, which can be the interaction-irreducible 3NF and/or the effective 2NF. 
To simplify the equations, we write in compact form pieces of diagrams that correspond to the amplitudes appearing in the exponential ansatz of the coupled-cluster wave function~\cite{shavitt2009many}. Without assuming the Einstein's convention of summing over repeated indices, we write them as,
    \begin{equation}
  \label{eq:t_1B}
t^{n_1}_{k_2} \equiv \sum_{\substack{\alpha \beta}} \frac{ \cX_{\alpha}^{n_1}   \ \widetilde{U}_{\alpha \beta } \ \cY_{\beta}^{k_2}    }{\varepsilon_{k_2}^{-}-\varepsilon_{n_1}^{+}} \, ,
  \end{equation}
with the effective one-body potential of Eq.~(\ref{ueff}),
    \begin{equation}
  \label{eq:t_2B}
t^{n_1 n_2}_{k_3 k_4} \equiv \sum_{\substack{\alpha \beta \\  \gamma \delta}} \frac{ \cX_{\alpha}^{n_1}  \cX_{\beta}^{n_2} \ \widetilde{V}_{\alpha \beta ,\gamma \delta} \ \cY_{\gamma}^{k_3}   \cY_{\delta}^{k_4} }{\varepsilon_{k_3}^{-}+\varepsilon_{k_4}^{-}-\varepsilon_{n_1}^{+}-\varepsilon_{n_2}^{+}} \, ,
  \end{equation}
with the effective 2NF of Eq.~(\ref{veff}), and
    \begin{equation}
  \label{eq:t_3B}
t^{n_1 n_2 n_3}_{k_4 k_5 k_6} \equiv \sum_{\substack{\alpha \beta \gamma \\ \mu \nu \lambda }} \frac{ \cX_{\alpha}^{n_1}  \cX_{\beta}^{n_2} \cX_{\gamma}^{n_3}  \  W_{\alpha \beta\gamma,\mu \nu \lambda}  \ \cY_{\mu}^{k_4}   \cY_{\nu}^{k_5}  \cY_{\lambda}^{k_6}}{\varepsilon_{k_4}^{-}+\varepsilon_{k_5}^{-}+\varepsilon_{k_6}^{-}-\varepsilon_{n_1}^{+}-\varepsilon_{n_2}^{+}-\varepsilon_{n_3}^{+}} \, ,
  \end{equation}
for the terms with the interaction-irreducible 3NF.

\subsubsection{\label{sec:ADC(3)_main} ADC(3) matrices for  Feynman diagrams in Fig.~\ref{3ord_a_b_c}. }

At third order in the ADC, we consider first the subset of coupling matrices and interaction matrices that are linked to $2p1h$ and $2h1p$ ISCs. For these intermediate configurations, the ADC(3) approximation for Eq.~(\ref{irr_SE_Lehmann}) requires the following terms,
\begin{align}
\label{ADC3_somm_2_1}
&\textbf{M}^{(\textrm{II})}_{j\alpha}  =  \textbf{M}^{(\textrm{IIa})}_{r \alpha } + \textbf{M}^{(\textrm{IIb})}_{r \alpha } + \textbf{M}^{(\textrm{IIc})}_{r \alpha }\, , \\
\label{ADC3_somm_2_1_N}&\textbf{N}^{(\textrm{II})}_{\alpha k}   =  \textbf{N}^{(\textrm{IIa})}_{\alpha s} + \textbf{N}^{(\textrm{IIb})}_{\alpha s} + \textbf{N}^{(\textrm{IIc})}_{\alpha s}\, , \\
\label{C_123_456}&\textbf{C}_{j j'}  =   \textbf{C}^{pp}_{r r'} + \textbf{C}^{ph}_{r r'} +  \textbf{C}^{3N}_{r r'}\, , \\
\label{D_123_456}&\textbf{D}_{k k'}   =    \textbf{D}^{hh}_{s s'} +   \textbf{D}^{hp}_{s s'} +  \textbf{D}^{3N}_{s s'}\, ,
\end{align}
in addition to the ones already introduced by Eqs.~(\ref{ADC2_somm}-\ref{ADC2_somm_end}) at second order.

\begin{figure}[t]
  \centering
    \subfloat[(a)]{\label{M_2N_2N_nnk_a}\includegraphics[scale=0.4]{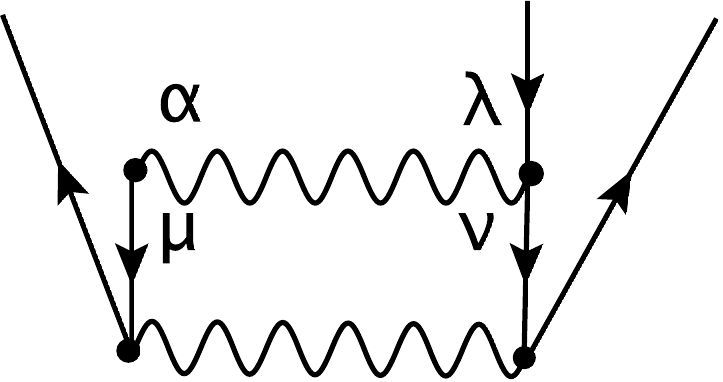}}
  \hspace{0.1cm}
    \subfloat[(b)]{\label{M_2N_2N_nnk_b}\includegraphics[scale=0.35]{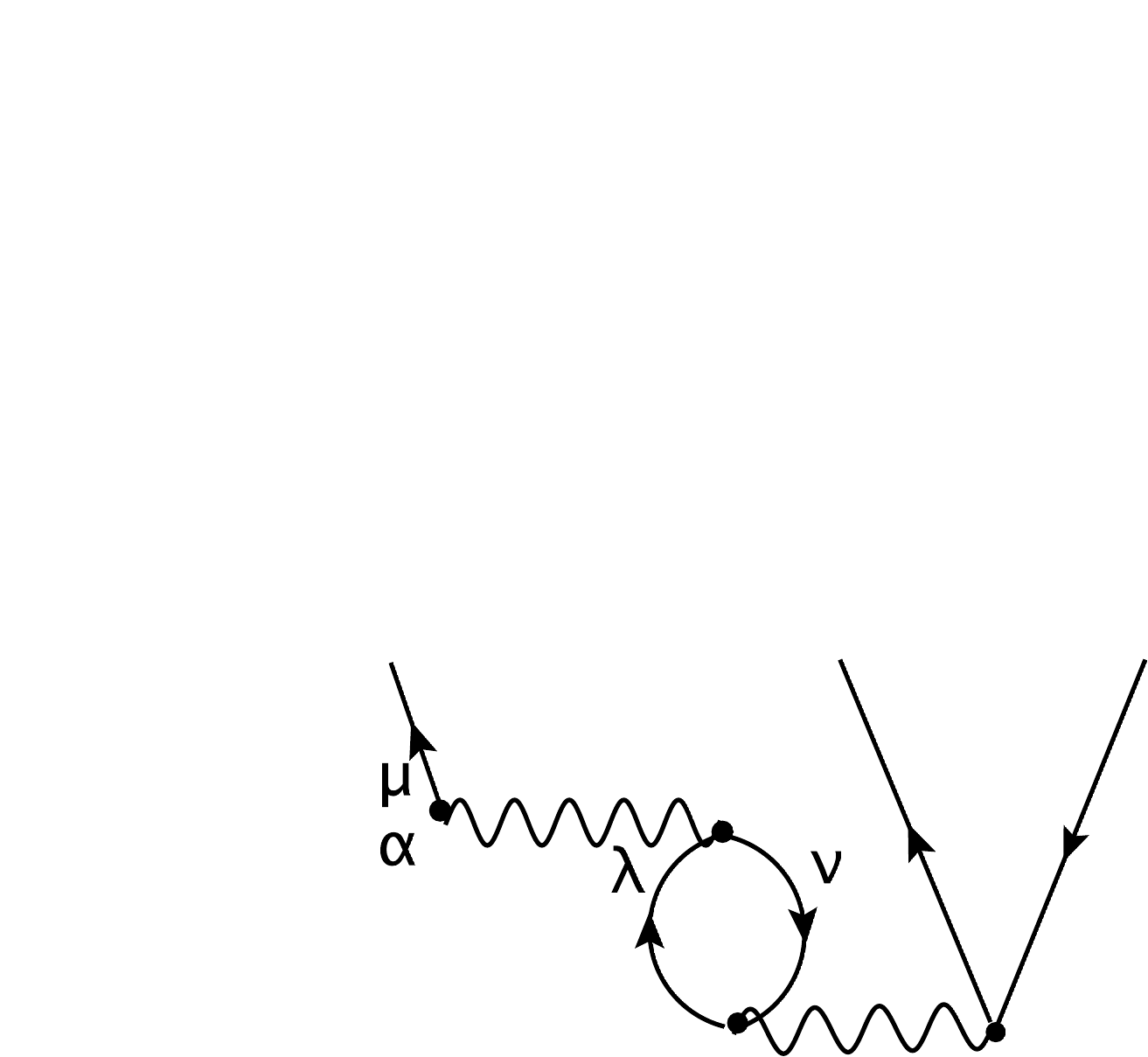}}
    \\
    \centering
    \subfloat[(c)]{\label{M_2N_3N_3c}\includegraphics[scale=0.37]{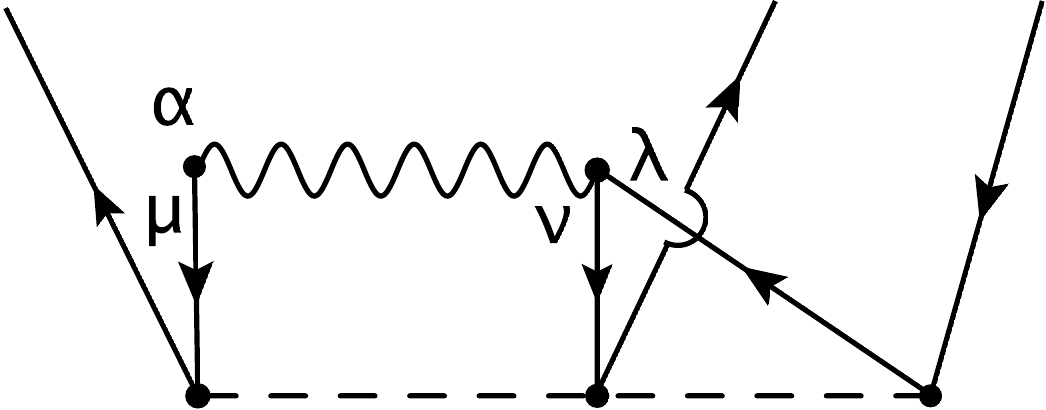}}  
  \caption{ Forward-in-time diagrams of the self-energy coupling matrices in ADC(3). Coupling matrices (a) and (b) correspond to Eqs.~(\ref{eq:M_3a}) and~(\ref{eq:M_3b}), and they feature two effective 2NFs $\widetilde{V}$ connecting $2p1h$ ISCs. The  coupling matrix (c) contains one effective 2NF $\widetilde{V}$  and one interaction-irreducible 3NF $\hat{W}$. It connects to  $2p1h$ ISCs and corresponds to Eq.~(\ref{eq:M_3c}).}
\label{M_3order_2N_2N}
\end{figure} 

 \begin{figure}[t]
  \centering
    \subfloat[(a)]{\label{N_2N_2N_kkn_a}\includegraphics[scale=0.4]{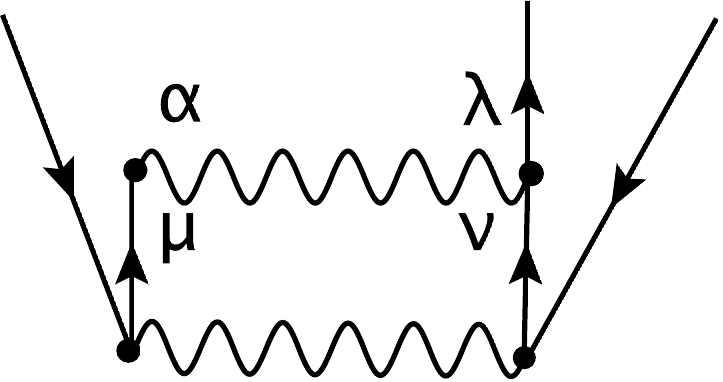}}
  \hspace{1.2cm}
    \subfloat[(b)]{\label{N_2N_2N_kkn_b}\includegraphics[scale=0.35]{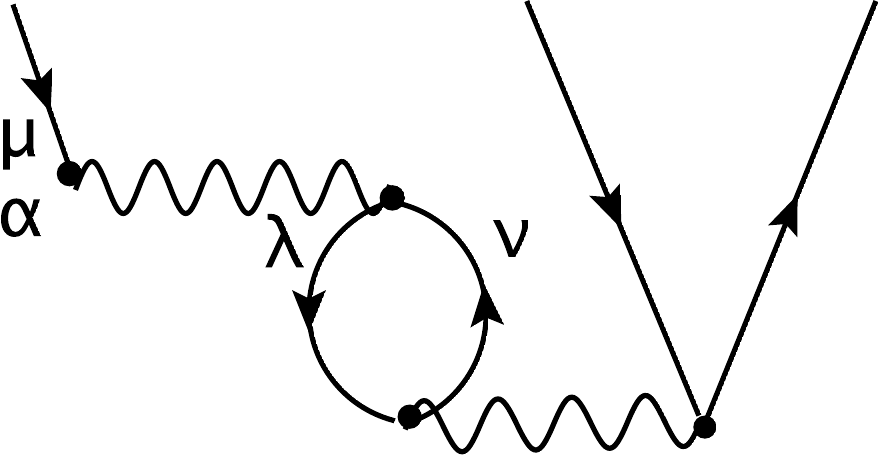}}
     \\
    \centering
      \subfloat[(c)]{\label{N_2N_3N_3c}\includegraphics[scale=0.37]{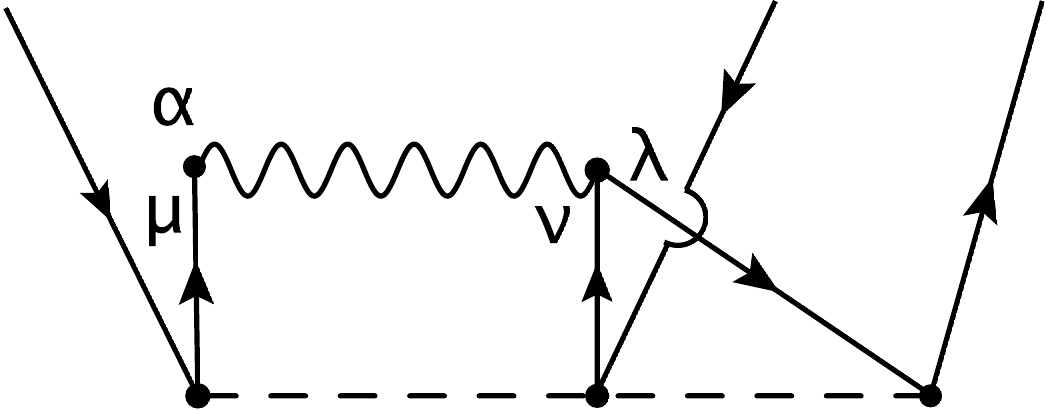}}     
  \caption{Same as in Fig.~\ref{M_3order_2N_2N} but for backward-in-time diagrams. Coupling matrices (a) and (b) correspond to Eqs.~(\ref{eq:N_3a}) and~(\ref{eq:N_3b}), while the coupling matrix (c) to  Eq.~(\ref{eq:N_3c}).}
  \label{N_3order_2N_2N}
\end{figure} 

  
We show now explicit expressions for the right-hand side of Eqs.~(\ref{ADC3_somm_2_1})-(\ref{D_123_456}), and start by presenting coupling matrices composed by two  effective 2NFs connecting to $2p1h$ ISCs.
By using the definition in Eq.~(\ref{eq:t_2B}), we have the matrices,
\begin{eqnarray}
\label{eq:M_3a}
\textbf{M}^{(\textrm{IIa})}_{r \alpha } \equiv \frac{1}{2\sqrt{2}}   t^{n_1 n_2}_{k_4 k_5}\ (\cY_{\mu}^{k_4}\cY_{\nu}^{k_5})^*\cY_{\lambda}^{k_3} \ \widetilde{V}_{\mu\nu,\alpha \lambda} \, , \nonumber \\
\, 
\end{eqnarray}
and
\begin{eqnarray}
\label{eq:M_3b}
\textbf{M}^{(\textrm{IIb})}_{r \alpha } &\equiv & \frac{1}{\sqrt{2}}   \left( t^{n_2 n_4}_{k_3 k_5}\ \cX_{\mu}^{n_1} (\cY_{\nu}^{k_5}   \cX_{\lambda}^{n_4})^* \ \widetilde{V}_{\mu\nu,\alpha \lambda} \right. \nonumber \\
&& \left.-  t^{n_1 n_4}_{k_3 k_5}\ \cX_{\mu}^{n_2} (\cY_{\nu}^{k_5}   \cX_{\lambda}^{n_4})^* \ \widetilde{V}_{\mu\nu,\alpha \lambda} \right) \, ,
\, 
\end{eqnarray}
which is explicitly antisymmetrized with respect to the $n_1, n_2$ fermion lines. The two coupling matrices in Eqs.~(\ref{eq:M_3a}) and (\ref{eq:M_3b}) are found in the Goldstone diagrams of the terms in Figs.~\ref{3ord_a} and~\ref{3ord_b}, as it is clear from their diagrammatic representations in Figs.~\ref{M_2N_2N_nnk_a} and~\ref{M_2N_2N_nnk_b} respectively.

The corresponding coupling matrices in the $2h1p$ sector read,
\begin{equation}
\label{eq:N_3a}
\textbf{N}^{(\textrm{IIa})}_{\alpha s} \equiv \frac{1}{2\sqrt{2}} \widetilde{V}_{\alpha \lambda,\mu\nu} \ \cX_{\lambda}^{n_3} (\cX_{\mu}^{n_4} \cX_{\nu}^{n_5})^* \ t^{n_4 n_5}_{k_1 k_2}  \, ,
\end{equation}
and
\begin{eqnarray}
\label{eq:N_3b}
\textbf{N}^{(\textrm{IIb})}_{\alpha s} &\equiv & \frac{1}{\sqrt{2}}   \left(\widetilde{V}_{\alpha \lambda, \mu\nu} \ \cY_{\mu}^{k_1} (\cX_{\nu}^{n_4}   \cY_{\lambda}^{k_5})^* t^{n_4 n_3}_{k_5 k_2} \right. \nonumber \\
&& \left.-  \widetilde{V}_{\alpha \lambda,\mu\nu} \  \cY_{\mu}^{k_2} (\cX_{\nu}^{n_4}   \cY_{\lambda}^{k_5})^* \ t^{n_4 n_3}_{k_5 k_1} \right) \, .
\end{eqnarray}
Both matrices in Eqs.~(\ref{eq:N_3a}) and (\ref{eq:N_3b}) are linked to $2h1p$ ISCs, as it is clear by considering backward-in-time Goldstone terms of  diagrams in Figs.~\ref{3ord_a} and~\ref{3ord_b}, respectively. Their diagrammatic representations are displayed in Figs.~\ref{N_2N_2N_kkn_a} and~\ref{N_2N_2N_kkn_b} respectively.

Other coupling matrices containing two effective 2NFs can be found in Sec.~\ref{M_N_2B_2B} of Appendix~\ref{M_N_matrices}.

Among coupling matrices containing one effective 2NF and one interaction-irreducible 3NF, we present here the ones appearing in the self-energy diagram of Fig.~\ref{3ord_c}, that is
\begin{equation}
\label{eq:M_3c}
\textbf{M}^{(\textrm{IIc})}_{ r \alpha} \equiv \frac{1}{2\sqrt{2}}  t^{ n_1 n_2 n_4}_{k_5 k_6 k_3}\ (\cY_{\mu}^{k_5}   \cY_{\nu}^{k_6} \cX_{\lambda}^{n_4})^* \ \widetilde{V}_{\mu\nu,\alpha \lambda}\, ,
\end{equation}
and
\begin{equation}
\label{eq:N_3c}
\textbf{N}^{(\textrm{IIc})}_{\alpha s} \equiv -\frac{1}{2\sqrt{2}}\widetilde{V}_{\alpha \lambda,\mu\nu} \ (\cX_{\mu}^{n_5}   \cX_{\nu}^{n_6} \cY_{\lambda}^{k_4})^* \ t^{n_5 n_6 n_3}_{ k_1 k_2 k_4} \, . 
\end{equation}
Diagrammatic representations of Eqs.~(\ref{eq:M_3c}) and~(\ref{eq:N_3c}) are displayed in Figs.~\ref{M_2N_3N_3c} and~\ref{N_2N_3N_3c} respectively.
 
All the other coupling matrices with one 2NF and one 3NF are collected in Sec.~\ref{M_N_2B_3B} of Appendix~\ref{M_N_matrices}, while matrices with two interaction-irreducible 3NFs are presented in Sec.~\ref{M_N_3B_3B}   in Appendix~\ref{M_N_matrices}.
 
 
Now we introduce expressions of the interaction matrices of Eqs.~(\ref{C_123_456}) and (\ref{D_123_456}), containing the $ \widetilde{V}_{\alpha \beta,\gamma \delta}$  and  $W_{\alpha \beta\gamma,\delta \epsilon \lambda}$ matrix elements. For the two cases, we display  interaction matrices appearing in both the forward-in-time and backward-in-time self-energy Goldstone diagrams of Fig.~\ref{3ord_a_b_c}. 
 
 
 The interaction matrix that connects $2p1h$ ISCs through a particle-particle ($pp$) interaction is
 \begin{equation}
 \label{int_term_5a_C}
 \textbf{C}^{pp}_{r r'} \equiv \frac{1}{2} \cX_{\mu}^{n_1}   \cX_{\nu}^{n_2} \widetilde{V}_{\mu\nu, \lambda \rho} (\cX_{\lambda}^{n_4}  \cX_{\rho}^{n_5})^* \delta_{k_3 k_6} \, ,
 \end{equation}
while the one connecting through a particle-hole ($ph$) interaction is
 \begin{equation}
 \label{int_term_5b_C}
  \textbf{C}^{ph}_{r r'} \! \equiv \!  \frac{1}{2} \mathcal{A}_{1 2}  \mathcal{A}_{4 5} \left(  \cX_{\nu}^{n_1}  \cY_{\rho}^{k_3} \widetilde{V}_{\nu \mu, \rho \lambda} (\cX_{\lambda}^{n_4} \cY_{\mu}^{k_6})^*   \delta_{n_2 n_5} \right) \, ,
 \end{equation}
 where the action of  two permutation operators $\mathcal{A}_{1 2} $ and $\mathcal{A}_{4 5} $ is defined in Eq.~(\ref{Perm_2}) and produces four terms.
 

 We present now the corresponding interaction matrices appearing in backward-in-time self-energy Goldstone diagrams, namely those that  are linked to propagators of hole-particle kind in  diagrams.
 
  We start by the interaction matrix that connects $2h1p$ ISCs through a hole-hole ($hh$) interaction, that  is
 \begin{equation}
 \label{int_term_5a_D}
 \textbf{D}^{hh}_{s s'} \equiv -\frac{1}{2}  (\cY_{\mu}^{k_1}   \cY_{\nu}^{k_2})^* \widetilde{V}_{\mu\nu, \lambda \rho} \cY_{\lambda}^{k_4}  \cY_{\rho}^{k_5} \delta_{n_3 n_6} \, ,
 \end{equation}
 while the one connecting through a hole-particle ($hp$) interaction is
 \begin{equation}
 \label{int_term_5b_D}
  \textbf{D}^{hp}_{s s'} \! \equiv \! - \frac{1}{2}  \mathcal{A}_{1 2}  \mathcal{A}_{4 5}   \left(
   (\cY_{\mu}^{k_1} \cX_{\rho}^{n_3})^* \widetilde{V}_{\mu\nu, \rho \lambda } \cY_{\lambda}^{k_4}   \cX_{\nu}^{n_6} \delta_{k_2 k_5} \right)\, ,
 \end{equation}
where permutation operators acting on hole states are defined in the same way as the one acting on particle states for the matrix $\textbf{C}^{ph}_{r r'}$ in Eq.~(\ref{int_term_5b_C}).
 


 The only interaction matrix that connects $2p1h$ ISCs through a 3NF is
 \begin{equation}
 \label{int_term_5c_C}
 \textbf{C}^{3N}_{r r'} \equiv  -\frac{1}{2}   \cX_{\nu}^{n_1}   \cX_{\mu}^{n_2} \cY_{\rho}^{k_3}  \, W_{ \nu\mu \lambda,   \epsilon \eta \rho}  (\cX_{\epsilon}^{n_4}  \cX_{\eta}^{n_5}  \cY_{\lambda}^{k_6})^* \, , 
 \end{equation}
which is explicitly antisymmetric in  particle indexes, while the one connecting two $2h1p$ ISCs through a 3NF is
 \begin{equation}
 \label{int_term_5c_D}
 \textbf{D}^{3N}_{s s'} \equiv  -\frac{1}{2}  (\cY_{\nu}^{k_1}   \cY_{\mu}^{k_2} \cX_{\rho}^{n_3})^* \, W_{ \nu\mu \lambda,  \epsilon \eta  \rho}   \cY_{\epsilon}^{k_4}  \cY_{\eta}^{k_5} \cX_{\lambda}^{n_6} \, , 
 \end{equation}
which is also explicitly antisymmetric in hole indexes.
 
 \subsubsection{\label{sec:ADC(3)_other}ADC(3) matrices  for selected \\ \qquad Feynman diagrams in Fig.~\ref{3ord_remn}. }

Coupling matrices presented in this section are obtained from the four Feynman diagrams in the first column of Fig.~\ref{3ord_remn}. Most of these matrices are linked to   $3p2h$ and $3h2p$ ISCs, with few exceptions derived from Goldstone diagrams where  ISCs are of $2p1h$ and $2h1p$ type. This subset of coupling matrices at the ADC(3) level is given by

\begin{eqnarray}
\label{ADC3_somm_3_2}
\textbf{M}^{(\textrm{II})}_{j\alpha }  &\! : \! & \textbf{M}^{(\textrm{IId})}_{q \alpha } ; \textbf{M}^{(\textrm{IId}')}_{r \alpha } ; \textbf{M}^{(\textrm{IIh})}_{q \alpha }; \textbf{M}^{(\textrm{IIh}')}_{r \alpha }  ;  \textbf{M}^{(\textrm{IIl})}_{q \alpha }  ; \textbf{M}^{(\textrm{IIo})}_{q \alpha }\, ,\nonumber \\  \\
\label{ADC3_somm_3_2_end}\textbf{N}^{(\textrm{II})}_{\alpha k }  & : & \textbf{N}^{(\textrm{IId})}_{\alpha u} ; \textbf{N}^{(\textrm{IId}')}_{\alpha s} ; \textbf{N}^{(\textrm{IIh})}_{\alpha u}; \textbf{N}^{(\textrm{IIh}')}_{\alpha s} ;
 \textbf{N}^{(\textrm{IIl})}_{\alpha u} ; \textbf{N}^{(\textrm{IIo})}_{\alpha u}\, .  
\end{eqnarray}

First, we present matrices containing two effective 2NFs. 
As before, we display both matrices obtained from forward-in-time  and backward-in-time Goldstone diagrams, denoted with the notation $\textbf{M}_{j \alpha }$ and $\textbf{N}_{\alpha k}$ respectively.

In Goldstone diagrams of the term in Fig.~\ref{3ord_223B_1} we have,
\begin{equation}
\label{eq:M_3dII}
\textbf{M}^{(\text{IId})}_{q \alpha }  \equiv  \frac{\sqrt{3}}{6} \mathcal{A}_{4 5}   \mathcal{P}_{1 2 3} \left(  t^{n_1 n_2}_{k_5 k_{6}} \cX_{\mu}^{n_3} (\cY_{\nu}^{k_{6}})^{\! *}   \cY_{\lambda}^{k_4}  \widetilde{V}_{\mu\nu,\alpha\lambda} \right) \, ,
\end{equation}
where the combination of permutation operators performs the antisymmetrization of the indexes \hbox{($k_4$, $k_5$)}, 
\hbox{($n_1$, $n_3$)} and \hbox{($n_2$, $n_3$)}, according to definitions in Eqs.~(\ref{Perm_2}) and (\ref{Perm_cyclic}). 

We turn now to coupling matrices containing one effective 2NF and one interaction-irreducible 3NF.
In the Goldstone diagrams of the term in Fig.~\ref{3ord_223B_1} we have also,
\begin{align}
\label{eq:M_3dI}
\textbf{M}^{(\textrm{IId}')}_{r \alpha } \equiv & -\frac{\sqrt{2}}{4}  \mathcal{A}_{1 2}  \Bigl(  \cX_{\lambda}^{n_1} t^{n_2 n_4}_{k_5 k_6} \left(\cY_{\mu}^{k_5} \cY_{\nu}^{k_6}  \cX_{\rho}^{n_4} \right)^{\! *} \,  \cY_{\eta}^{k_3}  \Bigr. \nonumber \\ 
  & \Bigl.    W_{\lambda\mu\nu, \alpha \eta \rho} \Bigr) \, .
\end{align}

In the Goldstone diagrams of the term in Fig.~\ref{3ord_233B_1} we have,
\begin{align}
\label{eq:M_5hII}
\textbf{M}^{(\textrm{IIh})}_{q \alpha } \equiv  \frac{\sqrt{3}}{6}  \mathcal{P}_{1 2 3} \left(   t^{n_1 n_3 n_6}_{k_4 k_5 k_7} \left( \cY_{\mu}^{k_{7}} \right)^{\! *} \cX_{\nu}^{n_{2}}  \left( \cX_{\lambda}^{n_{6}}  \right)^{\! *}     \widetilde{V}_{\mu\nu,\alpha\lambda}\right) \, ,
\end{align}
while in the Goldstone diagrams of the term in Fig.~\ref{3ord_323B_1} we find,
\begin{align}
\label{eq:M_5lI}
\textbf{M}^{(\textrm{IIl})}_{q \alpha } \! \equiv & \frac{\sqrt{3}}{12}   \mathcal{P}_{1 2 3}  \!   \left( t^{n_1 n_2}_{k_{6} k_{7}}  \left( \cY_{\mu}^{k_{6}}  \cY_{\nu}^{k_{7}}\right)^{\! *} \cX_{\lambda}^{n_{3}}   \cY_{\eta}^{k_{4}}  \cY_{\rho}^{k_{5}}   W_{\mu\nu\lambda, \alpha \eta \rho}  \right) \, .
\end{align}

For backward-in-time Goldstone diagrams, we can single out from the term in Fig.~\ref{3ord_223B_1} the coupling matrix,
\begin{align}
\label{eq:N_3dII}
\textbf{N}^{(\textrm{IId})}_{\alpha u} \equiv & \frac{\sqrt{3}}{6} \widetilde{V}_{\alpha\lambda, \mu\nu} \, \mathcal{A}_{4 5}  \mathcal{P}_{1 2 3}  \left(   \cY_{\mu}^{k_1} (\cX_{\nu}^{n_{6}})^{\! *}  \cX_{\lambda}^{n_4}  t^{n_6 n_5}_{k_2 k_3} \right) \, ,
\end{align}
and also the  following coupling matrix depending on 2N and 3N interactions, i.e.
\begin{align}
\label{eq:N_3dI}
\textbf{N}^{(\textrm{IId}')}_{\alpha s} \equiv & \frac{\sqrt{2}}{4}  \,  W_{\alpha \eta \rho, \lambda \mu\nu} \,  \mathcal{A}_{1 2}    \left(  \cX_{\eta}^{n_3}  (\cY_{\rho}^{k_4} )^*  \cY_{\lambda}^{k_1}  \left(\cX_{\mu}^{n_5} \cX_{\nu}^{n_6} \right)^{\! *}  \right. \nonumber \\
& \left. t^{n_5 n_6}_{k_2 k_4}  \right) \, .
\end{align}

In Fig.~\ref{3ord_233B_1}, we have the backward-in-time the coupling matrix,
\begin{align}
\label{eq:N_5hII}
\textbf{N}^{(\textrm{IIh})}_{\alpha u} \equiv & \frac{\sqrt{3}}{6} \widetilde{V}_{\alpha\lambda,\mu\nu} \, \mathcal{P}_{1 2 3}  \left( (\cY_{\lambda}^{k_{7}} \cX_{\mu}^{n_{6}})^{\! *}  \cY_{\nu}^{k_{2}}     t^{n_6 n_4 n_5}_{k_1 k_3 k_7}\right)   \, ,
\end{align}
while in Fig.~\ref{3ord_323B_1} we find the coupling matrix,
\begin{align}
\label{eq:N_5lI}
\textbf{N}^{(\textrm{IIl})}_{\alpha u} \equiv & \frac{\sqrt{3}}{12} W_{\alpha \eta \rho, \mu\nu\lambda} \, \mathcal{P}_{1 2 3}  \Bigl( \cX_{\eta}^{n_4}  \cX_{\rho}^{n_5}    \left( \cX_{\mu}^{n_{6}}\cX_{\nu}^{n_{7}} \right)^{\! *}  \cY_{\lambda}^{k_3}  \Bigr. \nonumber \\ 
  & \Bigl. 
 t^{n_{6} n_{7}}_{k_1 k_2}   \Bigr) \, .
\, 
\end{align}

Finally we introduce coupling matrices containing two interaction-irreducible 3NFs.
In Goldstone diagrams of the term  in Fig.~\ref{3ord_233B_1} we have,
\begin{align}
\label{eq:M_5hI}
\textbf{M}^{(\textrm{IIh}')}_{r \alpha} \equiv &  \frac{\sqrt{2}}{8} \mathcal{A}_{1 2}   \Bigl(t^{n_1 n_4 n_5}_{k_6 k_3 k_7} \left( \cY_{\mu}^{k_6} \cY_{\nu}^{k_7} \right)^{\!  *}  \cX_{\lambda}^{n_2} \left(\cX_{\eta}^{n_4} \cX_{\rho}^{n_5}\right)^{\! *} \Bigr. \nonumber \\ 
  & \Bigl.     W_{\lambda \mu  \nu, \alpha  \eta \rho }\Bigr) \, ,
\end{align}
while in Goldstone diagrams of the term in Fig.~\ref{3ord_333B_1} we have,
  \begin{flalign}
\label{eq:M_5oI}
\textbf{M}^{(\textrm{IIo})}_{q \alpha } & \equiv  -\frac{\sqrt{3}}{36}  t^{n_1 n_2 n_3}_{ k_{6} k_{7}  k_{8}}  \left( \cY_{\mu}^{k_{6}} \cY_{\nu}^{k_{7}}   \cY_{\lambda}^{k_{8}}  \right)^{\! \! *} \cY_{\eta}^{k_4} \cY_{\rho}^{k_5} \,  W_{\mu\nu\lambda, \alpha \eta \rho} \, , && 
\end{flalign}
that is antisymmetric in the indexes $n_1$, $n_2$, $n_3$ and $k_4$,$k_5$. 

For backward-in-time Goldstone diagrams, we can single out from the term in Fig.~\ref{3ord_233B_1} the coupling matrix
\begin{align}
\label{eq:N_5hI}
\textbf{N}^{(\textrm{IIh}')}_{\alpha s} \equiv &   \frac{\sqrt{2}}{8} \, W_{\alpha  \eta \rho , \mu \lambda \nu} \, \mathcal{A}_{1 2}    \Bigl((\cY_{\eta}^{k_4} \cY_{\rho}^{k_5} \cX_{\mu}^{n_6} \cX_{\nu}^{n_7})^{\! *}  \cY_{\lambda}^{k_2}  \Bigr. \nonumber \\ 
  & \Bigl.  t^{n_3 n_6 n_7}_{k_1 k_4 k_5}        \Bigr) \, ,
\end{align}
while the matrix
\begin{align}
\label{eq:N_5oI}
\textbf{N}^{(\textrm{IIo})}_{\alpha u} \equiv  \frac{\sqrt{3}}{36} \, W_{\alpha \eta \rho, \mu\nu\lambda} \, \cX_{\eta}^{n_4}\cX_{\rho}^{n_5} (\cX_{\mu}^{n_6}  \cX_{\nu}^{n_{7}}  \cX_{\lambda}^{n_{8}})^{\! *} \,
   t^{n_{6} n_{7} n_{8}}_{k_1 k_2 k_3} \,  ,
\end{align}
appears in the Goldstone diagrams relative to Fig.~\ref{3ord_333B_1} and it is antisymmetric in the indexes $k_1$, $k_2$, $k_3$ and $n_4$, $n_5$.

\section{\label{sec:conclusion}Conclusions}

We have calculated all possible Feynman diagrams for the self-energy up to  the third order,  for an Hamiltonian including up to three-body interactions.
Using these, we have then derived the complete set of working equations that are needed to calculate the self-energy  nonperturbatively in the ADC($n$) approach at orders $n$=2 and 3.   While the expansion of the self-energy is considered perturbatively by including diagrams featuring up to three interactions, the ADC(3) formalism  expands automatically certain classes of diagrams to infinite order. In particular, one resums series of  ladders, rings   and interaction-irreducible 3NFs diagrams. As for the usual ADC($n$) computations,  the Dyson equation for the 1B propagator can be implemented as a  large but energy independent eigenvalue problem. However, in presence of 3NFs, intermediate state configurations of $3p2h$ and $3h2p$ type contribute already at ADC(2) and  ADC(3) levels, while they would appear at ADC(4) and ADC(5) for NN-only interactions.

In showing expressions for both  coupling and interaction matrices, we have organized the equations according to their importance, using criteria based on the number of excitations implied by ISCs and the natural hierarchy of  many-nucleon forces. We started by revisiting the most relevant correlations in terms of $2p1h$ and $2h1p$ ISCs.  This sector contains the well-known ADC(3) equations for the original and effective NN interactions. A new  
contribution arises from the Feynman diagram of Fig.~\ref{3ord_c} and involves an interaction-irreducible 3NF (that is, which cannot be expressed as simpler normal ordered forces). This last term is argued to be less relevant in virtue of the hierarchy of nuclear forces. 
 Then, we have worked out the subset of ADC(3) coupling and interaction matrices that link to the $3p2h$ and $3h2p$ sector of ISCs. 
%
While this hierarchy suggests that  $3p2h$ and $3h2p$ ISCs may be necessary only for future generations of \textit{ab initio} approaches, the diagram of Fig.~\ref{3ord_c} may already have implication for present nuclear Hamiltonians.   However, these conjectures have not yet been  checked and knowing the importance of diagram~\ref{3ord_c}  would give guidance for the inclusion of further correlations~\cite{RaimondiBarbieri_in_preparation}.

To provide the ADC formalism in its most general form, we have released the assumption of a fully self-consistent expansion and considered also all the additional nonskeleton diagrams that appear in this case. The resulting corrections are important (at least conceptually) when calculations are based on standard reference propagators of mean-field type. 
 New sets of diagrams appear for both the static and dynamic self-energy and have been derived together with the corresponding contributions in the ADC framework. In total, four additional Feynman diagrams must be considered in the ADC(3) dynamic self-energy when one is working with uncorrelated propagators, while the 1B effective interaction defining the energy-independent self-energy is decomposed into 19 Feynman diagrams of different topologies. Hence, the complete ADC(3)  formalism  with 3NFs is now available for the self-energy, either self-consistent (with only skeleton diagrams) or based on an uncorrelated reference state.

The formalism presented in this work  sets the basis for future  advancements of the SCGF approach, especially useful for  studies of nuclear structure where the full inclusion of realistic three-nucleon interactions is required.
 The numerical implementation of the $3p2h$ and $3h2p$ sector is a long-term endeavour that might rely on future  
supercomputing  computing resources. At the same time, the case for such improvements in the many-body truncation will also depend on the performance and accuracy of future generations of realistic nuclear interactions.  On the other hand, calculating the diagram of Fig.~\ref{3ord_c}
involves only $2p1h$ and $2h1p$ ISCs and will not require resources beyond present day computer power~\cite{RaimondiBarbieriProc2017}. 
Thus, we plan a follow-up study  to investigate this term.

\begin{acknowledgments}
This work was supported by the United Kingdom Science and Technology Facilities Council (STFC), through Grants No. 
ST/L005743/1 and No. ST/P005314/1.
\end{acknowledgments}

\appendix

\begin{widetext}

\section{\label{M_N_matrices}ADC equations for self-energy  at third order  }

In this Appendix we give the remaining expressions of the coupling and interaction matrices that arise from Goldstone diagrams 
at third order in the expansion of the self-energy. The complete list of all terms required to build the ADC(3) formalism is 
\begin{align}
\label{ADC3_somm_final}
\textbf{M}^{[\text{ADC(3)]}}_{j\alpha}  ={}&  \textbf{M}^{[\text{ADC(2)]}}_{j\alpha} +
\begin{cases}
    \textbf{M}^{(\textrm{IIa})}_{r \alpha } + \textbf{M}^{(\textrm{IIb})}_{r \alpha } + \textbf{M}^{(\textrm{IIc})}_{r \alpha }   \\
  ~ + \textbf{M}^{(\textrm{IId}')}_{r \alpha}  + \textbf{M}^{(\textrm{IIh}')}_{r \alpha } + \textbf{M}^{(\textrm{IIe}')}_{r \alpha }  + \! \textbf{M}^{(\textrm{IIi}')}_{r \alpha } + \!  \textbf{M}^{(\textrm{IIr})}_{r \alpha } + \textbf{M}^{(\textrm{IIs})}_{r \alpha }
 &\hbox{if $j$=$r$ ($2p1h$),} \\ ~ \\
     \textbf{M}^{(\textrm{IId})}_{q \alpha } + \textbf{M}^{(\textrm{IIh})}_{q \alpha } + \textbf{M}^{(\textrm{IIl})}_{q \alpha } + \textbf{M}^{(\textrm{IIo})}_{q \alpha } + \textbf{M}^{(\textrm{IIe})}_{q \alpha } + \!  \! \textbf{M}^{(\textrm{IIi})}_{q \alpha }      \\
 ~ + \! \textbf{M}^{(\textrm{IIm})}_{q \alpha } + \!  \textbf{M}^{(\textrm{IIn})}_{q \alpha } + \! \textbf{M}^{(\textrm{IIp})}_{q \alpha }  +  \textbf{M}^{(\textrm{IIq})}_{q \alpha }  + \textbf{M}^{(\textrm{IIt})}_{q \alpha }  +  \textbf{M}^{(\textrm{IIu})}_{q \alpha } \quad ~ & \hbox{if $j$=$q$ ($3p2h$),}
\end{cases}
\\ {}& \nonumber \\
\label{ADC3_somm_final_N}
\textbf{N}^{[\text{ADC(3)]}}_{\alpha k }  ={}& \textbf{N}^{[\text{ADC(2)]}}_{\alpha k }  +
\begin{cases}
\textbf{N}^{(\textrm{IIa})}_{\alpha s} + \textbf{N}^{(\textrm{IIb})}_{\alpha s} + \textbf{N}^{(\textrm{IIc})}_{\alpha s}  \\
 ~+ \textbf{N}^{(\textrm{IId}')}_{\alpha s} + \textbf{N}^{(\textrm{IIh}')}_{\alpha s} + \textbf{N}^{(\textrm{IIe}')}_{\alpha s}
        + \textbf{N}^{(\textrm{IIi}')}_{\alpha s}  + \textbf{N}^{(\textrm{IIr})}_{\alpha s} + \textbf{N}^{(\textrm{IIs})}_{\alpha s}
              &\hbox{if $k$=$s$ ($2h1p$),} \\ ~ \\
     \textbf{N}^{(\textrm{IId})}_{\alpha u} +  \textbf{N}^{(\textrm{IIh})}_{\alpha u} +  \textbf{N}^{(\textrm{IIl})}_{\alpha u}   + \textbf{N}^{(\textrm{IIo})}_{\alpha u} +  \textbf{N}^{(\textrm{IIe})}_{\alpha u} +  \textbf{N}^{(\textrm{IIi})}_{\alpha u}  \\
  +  \textbf{N}^{(\textrm{IIm})}_{\alpha u}  + \textbf{N}^{(\textrm{IIn})}_{\alpha u}  + \textbf{N}^{(\textrm{IIp})}_{\alpha u} + \textbf{N}^{(\textrm{IIq})}_{\alpha u} +   \textbf{N}^{(\textrm{IIt})}_{\alpha u}   + \textbf{N}^{(\textrm{IIu})}_{\alpha u}    \qquad ~ ~ ~  & \hbox{if $k$=$u$ ($3h2p$),}
\end{cases}
\\ {}& \nonumber \\
\label{C_final}\textbf{C}^{[\text{ADC(3)]}}_{j j'} ={}&
\begin{cases}
   \textbf{C}^{pp}_{r r'} + \textbf{C}^{ph}_{r r'} +  \textbf{C}^{3N}_{r r'} + \textbf{C}^{\widetilde{U} p}_{r r'} + \textbf{C}^{\widetilde{U} h}_{r r'}     &\hbox{if $j$=$r$ and $j'$=$r'$ ($2p1h$),} \\ ~ \\
     \textbf{C}^{p-pph}_{r q'}    + \textbf{C}^{h-hhp}_{r q'} + \textbf{C}^{3N(I)}_{r q'} +  \textbf{C}^{3N(II)}_{r q'}   &\hbox{if $j$=$r$ ($2p1h$)  and $j'$=$q'$  ($3p2h$),} \\ ~  \\
      \textbf{C}^{pp}_{q q'} + \textbf{C}^{ph}_{q q'}  + \textbf{C}^{hh}_{q q'}\\
      ~ +  \textbf{C}^{3N(III)}_{q q'}  + \textbf{C}^{3N(IV)}_{q q'}   +       \textbf{C}^{3N(V)}_{q q'}   + \textbf{C}^{\widetilde{U} p}_{q q'} + \textbf{C}^{\widetilde{U} h}_{q q'} 
      \qquad  ~ &\hbox{if $j$=$q$ and $j'$=$q'$ ($3p2h$),} 
\end{cases}
\\ {}& \nonumber \\
\label{D_final}\textbf{D}^{[\text{ADC(3)]}}_{k k'}  ={}& 
\begin{cases}
   \textbf{D}^{hh}_{s s'} +   \textbf{D}^{hp}_{s s'} +  \textbf{D}^{3N}_{s s'}  + \textbf{D}^{\widetilde{U} p}_{s s'} +  \textbf{D}^{\widetilde{U} h}_{s s'} 
          &\hbox{if $k$=$s$ and $k'$=$s'$ ($2h1p$),} \\  ~ \\
    \textbf{D}^{h-hhp}_{s u'}  +  \textbf{D}^{p-pph}_{s u'} +   \textbf{D}^{3N(I)}_{s u'} +  \textbf{D}^{3N(II)}_{s u'} 
           &\hbox{if $k$=$s$ ($2h1p$)  and $k'$=$u'$  ($3h2p$),}\\  ~ \\
    \textbf{D}^{hh}_{u u'}   +       \textbf{D}^{hp}_{u u'}    +       \textbf{D}^{pp}_{u u'} \\ 
    ~  +  \textbf{D}^{3N(III)}_{u u'}  \ +  \textbf{D}^{3N(IV)}_{u u'}    +       \textbf{D}^{3N(V)}_{u u'}    +  \textbf{D}^{\widetilde{U} p}_{u u'}  + \textbf{D}^{\widetilde{U} h}_{u u'}  \quad ~
           &\hbox{if $k$=$u$ and $k'$=$u'$ ($3h2p$).} 
\end{cases}
\end{align}
\end{widetext}
For the coupling matrices $\textbf{M}_{j\alpha}$ and $\textbf{N}_{\alpha k }$, the list of  terms truncated at the ADC(3) level is composed by sets of ADC(2) terms, defined in Eqs.~(\ref{eq:M_2a}) and (\ref{eq:M_2b}) and in  Eqs.~(\ref{eq:N_2a}) and (\ref{eq:N_2b}) for the forward-in-time and backward-in-time self-energy, respectively; sets of terms from (IIa) to (IIc) appearing at third order of the ADC, presented in Eqs.~(\ref{eq:M_3a}), (\ref{eq:M_3b}), and (\ref{eq:M_3c}) and  in Eqs.~(\ref{eq:N_3a})), (\ref{eq:N_3b}), and (\ref{eq:N_3c}), which contain only $2p1h$ and $2h1p$ configurations; and those terms from (IId) to (IIo) with  $3p2h$ and $3h2p$ ISCs, introduced in Eqs.~(\ref{eq:M_3dII})-(\ref{eq:M_5lI}), (\ref{eq:M_5hI}), and (\ref{eq:M_5oI}) and in Eqs.~(\ref{eq:N_3dII})-(\ref{eq:N_5lI}), (\ref{eq:N_5hI}), and (\ref{eq:N_5oI}). Other terms  with  $3p2h$ and $3h2p$ ISCs, denoted with superscripts from (IIe) to (IIq), are defined  in 
Eqs.~(\ref{eq:M_3eII})-(\ref{eq:N_5qI}). Moreover, in Eqs.~(\ref{ADC3_somm_final}) and (\ref{ADC3_somm_final_N}) we find additional terms, that must be added to the ADC(3) when the single-particle propagator used to construct self-energy diagrams is uncorrelated, i.e. when one works with a nonskeleton expansion. For coupling matrices, these additional terms are denoted with superscripts ranging from $(\textrm{IIr})$ to $(\textrm{IIu})$. Their explicit expressions will be given in Appendix~\ref{Dynamic_SE}.

Interaction matrices  appear at third order in the ADC, as listed in Eqs.~(\ref{C_final}) and (\ref{D_final}). The first three terms thereof connecting to $2p1h$ and $2h1p$ configurations, are given in Eqs.~(\ref{int_term_5a_C}), (\ref{int_term_5b_C}), and (\ref{int_term_5c_C}) for  forward-in-time diagrams and in Eqs.~(\ref{int_term_5a_D}), (\ref{int_term_5b_D}), and (\ref{int_term_5c_D}) for backward-in-time ones. Other matrices required to link $3p2h$ ($3h2p$) ISCs are denoted by $\textbf{C}^{p-pph}_{r q'},...,  \textbf{C}^{3N(V)}_{q q'}  $ ($\textbf{D}^{h-hhp}_{s u'},\cdots, \textbf{D}^{3N(V)}_{u u'}$). They will be given below in Eqs.~(\ref{int_term_5d_C})-(\ref{int_term_5n_C}) and (\ref{int_term_5h_C})-(\ref{int_term_5q_C}) [Eqs.~(\ref{int_term_5d_D})-(\ref{int_term_5n_D}) and (\ref{int_term_5h_D})-(\ref{int_term_5q_D})]. Finally, additional four interaction matrices introduced in Appendix~\ref{Dynamic_SE} for the nonskeleton expansion are specified in Eqs.~(\ref{C_final}) and (\ref{D_final}) with the superscript $\widetilde{U}$. 

\subsection{\label{M_N_2B_2B}Coupling matrices with two effective 2NFs}
In  Fig.~\ref{3ord_223B_2} we find the following coupling matrices,
\begin{equation}
\label{eq:M_3eII}
\textbf{M}^{(\textrm{IIe})}_{q \alpha } \equiv  \frac{\sqrt{3}}{6} \mathcal{P}_{1 2 3} \! \left(t^{n_6 n_3}_{k_4 k_5} \,  \cX_{\mu}^{n_1} \cX_{\nu}^{n_2}  \! \left(\cX_{\lambda}^{n_{6}}\right)^{\!*} \, \widetilde{V}_{\mu \nu ,\alpha\lambda} \right) \, ,
\end{equation}
and 
\begin{equation}
\label{eq:N_3eII}
\textbf{N}^{(\textrm{IIe})}_{\alpha u} \equiv  \frac{\sqrt{3}}{6} \, \widetilde{V}_{\alpha\lambda, \mu\nu} \, \mathcal{P}_{1 2 3} \left( (\cY_{\lambda}^{k_6})^{\!*}    \cY_{\mu}^{k_1} \cY_{\nu}^{k_2} \, t^{n_4 n_5}_{k_6 k_3} \right)
\, ,
\end{equation}
for the forward-in-time and backward-in-time Goldstone diagrams, respectively.

\subsection{\label{M_N_2B_3B}Coupling matrices with one effective 2NF and one interaction-irreducible 3NF}

Diagrams  in Fig.~\ref{3ord_223B_2} contains also an interaction-irreducible 3NF, therefore another coupling matrix can be obtained from the corresponding Goldstone diagrams. For the forward-in-time and backward-in-time parts we have,
\begin{equation}
\label{eq:M_3eI}
\textbf{M}^{(\textrm{IIe}')}_{r \alpha } \equiv  \frac{\sqrt{2}}{4}  t^{n_4 n_5}_{k_3 k_6} \ \cX_{\mu}^{n_1} \cX_{\nu}^{n_2} \left(\cY_{\lambda}^{k_6} \cX_{\rho}^{n_4} \cX_{\eta}^{n_5} \right)^*    W_{\mu\nu\lambda, \alpha  \rho \eta} \, ,
\end{equation}
and
\begin{equation}
\label{eq:N_3eI}
\textbf{N}^{(\textrm{IIe}')}_{\alpha s} \equiv  \frac{\sqrt{2}}{4}  \, W_{\alpha \rho \eta , \mu\nu\lambda} \,   (\cY_{\rho}^{k_4}  \cY_{\eta}^{k_5})^* \cY_{\mu}^{k_1} \cY_{\nu}^{k_2} (\cX_{\lambda}^{n_6})^*  t^{n_3 n_6}_{k_4 k_5} \, ,
\end{equation}
respectively.

Also  diagrams in the second and third row of Fig.~\ref{3ord_remn} feature coupling matrices with 2NFs and interaction-irreducible 3NFs. We list them below considering both forward- and backward-in-time contributions. In the Goldstone diagrams of the term in Fig.~\ref{3ord_233B_2} we have,
\begin{equation}
\label{eq:M_5iII}
\textbf{M}^{(\textrm{IIi})}_{q \alpha } \equiv  \frac{\sqrt{3}}{12}  \mathcal{A}_{4 5}    \left( t^{n_1 n_2 n_{3}}_{k_5 k_6 k_7}  \left( \cY_{\mu}^{k_6}  \cY_{\nu}^{k_7}\right)^*  \cY_{\lambda}^{k_{4}}  \,   \widetilde{V}_{\mu\nu,\alpha\lambda} \right) \, ,
\end{equation}
 and
\begin{equation}
\label{eq:N_5iII}
\textbf{N}^{(\textrm{IIi})}_{\alpha u} \equiv  \frac{\sqrt{3}}{12} \mathcal{A}_{4 5} \,  \left( \widetilde{V}_{\alpha\lambda, \mu\nu} \, \cX_{\lambda}^{n_4}  (\cX_{\mu}^{n_6}\cX_{\nu}^{n_7})^* \, t^{n_5 n_6 n_7}_{k_1 k_2 k_3}    \right) \, .
\end{equation}
  
In the Goldstone diagrams of the term in Fig.~\ref{3ord_323B_2}  we can single out the coupling matrices,
\begin{eqnarray}
\label{eq:M_mI}
\textbf{M}^{(\textrm{IIm})}_{q \alpha} &\equiv & \frac{\sqrt{3}}{6} \mathcal{A}_{4 5}  \,  \mathcal{P}_{1 2 3}  \,   \left( t^{n_1 n_6}_{k_4 k_7}  \left( \cY_{\mu}^{k_7} \right)^* \cX_{\nu}^{n_2} \cX_{\lambda}^{n_3}  \right.  \nonumber \\
&& \left.  \cY_{\eta}^{k_5}   \left(\cX_{\rho}^{n_6} \right)^*   W_{\mu\nu\lambda, \alpha \rho \eta } \right) \, ,
\end{eqnarray}
and
\begin{eqnarray}
\label{eq:N_5mI}
\textbf{N}^{(\textrm{IIm})}_{\alpha u} &\equiv & \frac{\sqrt{3}}{6}  \mathcal{A}_{4 5}   \mathcal{P}_{1 2 3}  \,  \left(    W_{ \alpha  \eta \rho ,  \mu \nu \lambda} (\cY_{\eta}^{k_7})^* \cX_{\rho}^{n_5}   \right.   \nonumber \\
&& \left.  (\cX_{\mu}^{n_6})^*   \cY_{\nu}^{k_2}    \cY_{\lambda}^{k_3} t^{n_4 n_6}_{k_1 k_7}\right) \,  .
\end{eqnarray}
  
Finally,   in the Goldstone diagrams of the term  in Fig.~\ref{3ord_323B_3}  we have,
\begin{equation}
\label{eq:M_nI}
\textbf{M}^{(\textrm{IIn})}_{q \alpha } \equiv  \frac{\sqrt{3}}{12}    t^{n_6 n_7}_{k_4 k_5} \ \cX_{\mu}^{n_1}  \cX_{\nu}^{n_2} \cX_{\lambda}^{n_3}   \left( \cX_{\eta}^{n_6}   \cX_{\rho}^{n_7} \right)^*   W_{\mu\nu\lambda, \alpha \eta \rho}  \, , 
\end{equation}
and
\begin{equation}
\label{eq:N_5nI}
\textbf{N}^{(\textrm{IIn})}_{\alpha u} \equiv  \frac{\sqrt{3}}{12}  \,  W_{ \alpha \eta \rho, \mu\nu\lambda} \, \cY_{\mu}^{k_1}  \cY_{\nu}^{k_2}    \cY_{\lambda}^{k_3}     \left(\cY_{\eta}^{k_6}  \cY_{\rho}^{k_7}\right)^*  \,  t^{n_4 n_5}_{k_6 k_7} \, ,
\end{equation}
that are both antisymmetric in their particle and hole indexes.

\subsection{\label{M_N_3B_3B}Coupling matrices with two interaction-irreducible 3NFs}

All the Feynman diagrams in the second and fourth rows of Fig.~\ref{3ord_remn} can contain coupling matrices with two interaction-irreducible 3NFs. Again, for each different topology we list the expressions for  forward-in-time contributions first, followed by the backward-in-time ones.

  In the Goldstone diagrams of the term in Fig.~\ref{3ord_233B_2} we have,
  \begin{equation}
\label{eq:M_5iI}
\textbf{M}^{(\textrm{IIi}')}_{r \alpha } \equiv  \frac{\sqrt{2}}{12} \,   t^{n_1 n_2 n_4}_{k_5 k_6 k_7}  \left( \cY_{\mu}^{k_5} \cY_{\nu}^{k_6}   \cY_{\lambda}^{k_7}  \cX_{\rho}^{n_4}\right)^*  \cY_{\eta}^{k_3}    W_{\mu\nu\lambda, \alpha \eta \rho}  \, ,
\end{equation}
which is antisymmetric in the indexes $n_1$and $n_2$, and
\begin{equation}
\label{eq:N_5iI}
\textbf{N}^{(\textrm{IIi}')}_{\alpha s} \equiv  \frac{\sqrt{2}}{12} W_{\alpha \eta \rho, \lambda \mu\nu}  \,\cX_{\eta}^{n_3} (  \cY_{\rho}^{k_7}  \cX_{\lambda}^{n_4} \cX_{\mu}^{n_5} \cX_{\nu}^{n_{6}} )^*    t^{n_4 n_5 n_6}_{k_1 k_2 k_7} \,  ,
\end{equation}
which is antisymmetric in the indexes $k_1$and $k_2$. 
  
  In the Goldstone diagrams of the term in Fig.~\ref{3ord_333B_2} we have,
  \begin{align}
\label{eq:M_5pI}
\textbf{M}^{(\textrm{IIp})}_{ q \alpha} \equiv & \frac{\sqrt{3}}{12} \mathcal{A}_{4 5} \,  \mathcal{P}_{1 2 3}  \, \Bigl(t^{n_1 n_2 n_6}_{ k_7 k_8  k_5}    \cX_{\lambda}^{n_3} \cY_{\eta}^{k_4}  \left( \cY_{\mu}^{k_7} \cY_{\nu}^{k_8} \cX_{\rho}^{n_6} \right)^{\!*}   \Bigr.  \nonumber \\
& \Bigl.    W_{\mu\nu\lambda, \alpha \eta \rho  } \Bigr) \, ,
\end{align}
and
\begin{align}
\label{eq:N_5pI}
\textbf{N}^{(\textrm{IIp})}_{\alpha u} \equiv & -\frac{\sqrt{3}}{12} \,W_{\alpha \eta \rho, \mu\nu\lambda} \, \mathcal{A}_{4 5} \,  \mathcal{P}_{1 2 3} \, \Bigl( \cX_{\eta}^{n_4} (\cY_{\rho}^{k_8} \cX_{\mu}^{n_6} \cX_{\nu}^{n_7} )^* \Bigr.  \nonumber \\
& \Bigl. \cY_{\lambda}^{k_3}  \,  t^{n_6 n_7 n_5}_{k_1 k_2 k_8} \Bigr)  \, .
\end{align}

In the Goldstone diagrams of the term in Fig.~\ref{3ord_333B_3} we can also single out the following coupling matrices,
\begin{align}
\label{eq:M_5qI}
\textbf{M}^{(\textrm{IIq})}_{ q \alpha} \equiv & - \frac{\sqrt{3}}{12}  \mathcal{P}_{1 2 3}    \, \Bigl(     t^{n_1 n_6 n_7}_{ k_4  k_5  k_8}   \left( \cY_{\mu}^{k_8} \cX_{\eta}^{n_6} \cX_{\rho}^{n_7}  \right)^{\!*}  \cX_{\nu}^{n_2}  \cX_{\lambda}^{n_3} \Bigr.  \nonumber \\
& \Bigl.    W_{\mu\nu\lambda, \alpha \eta \rho}  \Bigr)    \, ,
\end{align}
and
\begin{align}
\label{eq:N_5qI}
\textbf{N}^{(\textrm{IIq})}_{ \alpha u} \equiv & \frac{\sqrt{3}}{12} \, W_{\alpha \eta \rho, \mu\nu\lambda} \, \mathcal{P}_{1 2 3}      \, \Bigl(     (\cY_{\eta}^{k_7} \cY_{\rho}^{k_8}   \cX_{\mu}^{n_6})^* \cY_{\nu}^{k_2} \cY_{\lambda}^{k_3}  
\Bigr.  \nonumber \\
& \Bigl.  t^{n_4 n_5 n_6}_{k_1 k_7 k_8}\Bigr)     \,  .
\end{align}

\subsection{\label{Interaction_T_2nd_3p2h} Formulas for 2NF interaction matrices with $3p2h$ and $3h2p$ configurations}

The interaction matrices that we have introduced in Sec.~\ref{sec:ADC(3)} do not exhaust the list of all possible terms required for the ADC(3).
A more complicated pattern in terms of ISCs is present in interaction matrices connecting  $2p1h$ and $3p2h$ ISCs, as for the forward-in-time terms in the diagrams of Figs.~\ref{3ord_223B_1}-\ref{3ord_322B_2}.

In the Feynman diagram of Fig.~\ref{3ord_223B_1} one can find,
  \begin{align}
 \label{int_term_5d_C}
  \textbf{C}^{p-pph}_{r q'}  \equiv & \frac{\sqrt{6}}{12} \mathcal{A}_{1 2} \, \mathcal{A}_{9   \, 10}  \,
 \mathcal{P}_{6 7 8} \, \Bigl(\cX_{\mu}^{n_2}   \, \widetilde{V}_{\mu\nu, \lambda \rho} (\cX_{\lambda}^{n_7} \cX_{\rho}^{n_8} \cY_{\nu}^{k_{10}})^*  \Bigr. \nonumber \\ 
  & \Bigl. \delta_{n_1 n_6} \delta_{k_3 k_9}\Bigr) \, ,
 \end{align}
  while its complex conjugate term is contained in the diagram of Fig.~\ref{3ord_322B_1}.
  
  An interaction matrix connecting the same ISCs as the one in Eq.~(\ref{int_term_5d_C}) is   
     \begin{align}
 \label{int_term_5e_C}
  \textbf{C}^{h-hhp}_{r q'}  \equiv & \frac{\sqrt{6}}{12} \mathcal{A}_{6 7 8} \, \Bigl(\!  \cY_{\lambda}^{k_3}  \, \widetilde{V}_{\mu\nu, \lambda \rho} \, (\cY_{\mu}^{k_9}   \cY_{\nu}^{k_{10}} \cX_{\rho}^{n_8})^{*} \Bigr. \nonumber \\ 
  & \Bigl. \delta_{n_1 n_6} \delta_{n_2 n_7}\Bigr) \, ,
 \end{align}
which is contained in the diagram in Fig.~\ref{3ord_223B_2}, while the diagram in Fig.~\ref{3ord_322B_2} contains the complex conjugate of Eq.~(\ref{int_term_5e_C}).

 In the self-energy diagrams represented in Figs.~\ref{3ord_323B_1}-\ref{3ord_323B_3}, when two $3p2h$ ISCs  interact through a 2NF, we find interaction matrices of the following form:
   \begin{align}
 \label{int_term_5l_C}
  \textbf{C}^{pp}_{q q'}  \equiv & \frac{1}{12} \mathcal{A}_{4 5} \,  \mathcal{P}_{1 2 3} \, \mathcal{P}_{6 7 8} \, \Bigl(\cX_{\mu}^{n_1}  \cX_{\nu}^{n_2} \, \widetilde{V}_{\mu\nu, \lambda \rho} \,  (\cX_{\lambda}^{n_6} \cX_{\rho}^{n_7})^* \Bigr. \nonumber \\ 
  & \Bigl.   \delta_{n_3 n_8} \delta_{k_4 k_9} \delta_{k_5 k_{10}} \Bigr) \, , 
 \end{align}
 which is composed by 18 terms when all the permutations indicated are taken, and the ones with a particle-hole 2NF connecting two $3p2h$ propagators, i.e.
   \begin{align}
 \label{int_term_5m_C}
  \textbf{C}^{ph}_{q q'}  \equiv & \frac{1}{12}  \mathcal{A}_{4 5} \,  \mathcal{A}_{9 \, 10} \, \mathcal{A}_{1 2 3} \, \mathcal{P}_{6 7 8} \, \Bigl(\cX_{\mu}^{n_1} \cY_{\rho}^{k_4}  \, \widetilde{V}_{\mu\nu,  \rho \lambda} \,  (\cX_{\lambda}^{n_6} \cY_{\nu}^{k_9})^{*}  \Bigr. \nonumber \\ 
  &\Bigl.  \delta_{n_2 n_7} \delta_{n_3 n_8} \delta_{k_5 k_{10}}  \Bigr)  \, , 
 \end{align}
 that contains 72 terms when the explicit antisymmetrization with respect to quasiparticle indexes is performed. The interaction matrices in Eqs.~(\ref{int_term_5l_C}) and~(\ref{int_term_5m_C}) are found in the diagrams in Figs.~\ref{3ord_323B_1} and~\ref{3ord_323B_2}, respectively.

A forward-in-time interaction matrix connecting two $3p2h$ ISCs through a hole-hole 2NF is found in the self-energy diagram in Fig.~\ref{3ord_323B_3}. This has the following expression,
     \begin{align}
 \label{int_term_5n_C}
  \textbf{C}^{hh}_{q q'}  \equiv & \frac{1}{12} \mathcal{A}_{1 2 3} \, \Bigl( \cY_{\lambda}^{k_4}  \cY_{\rho}^{k_5} \, \widetilde{V}_{\mu\nu, \lambda \rho} \, (\cY_{\mu}^{k_{9}}  \cY_{\nu}^{k_{10}})^*  \Bigr. \nonumber \\ 
  &\, \Bigl. \delta_{n_1 n_6} \delta_{n_2 n_7} \delta_{n_3 n_8}  \Bigr) \, . 
 \end{align}

 
 
  We present now the corresponding interaction matrices appearing in backward-in-time self-energy Goldstone diagrams. We remind that these interaction matrices are the ones connecting propagators of multihole-multiparticle type  in self-energy diagrams.
  
We consider first  terms contained in Figs.~\ref{3ord_223B_1}-\ref{3ord_322B_2}, namely those connecting the $2h1p$ propagator to the $3h2p$ propagator. We find,
  \begin{align}
 \label{int_term_5d_D}
  \textbf{D}^{h-hhp}_{s u'}  \equiv & \frac{\sqrt{6}}{12}  \mathcal{A}_{1 2}  \, \mathcal{A}_{9  \, 10} \,  \mathcal{P}_{6 7 8} \, \Bigl(  (\cY_{\mu}^{k_2})^* \, \widetilde{V}_{\mu\nu, \lambda \rho} \, \cY_{\lambda}^{k_7}  \cY_{\rho}^{k_8} \cX_{\nu}^{n_{10}} \Bigr. \nonumber \\ 
  &\Bigl.  \delta_{k_1 k_6} \delta_{n_3 n_9}\Bigr)  \, ,
 \end{align}
which must be combined with another interaction matrix,
\begin{align}
 \label{int_term_5e_D}
  \textbf{D}^{p-pph}_{s u'} \! \equiv &\! -\frac{\sqrt{6}}{12} \mathcal{A}_{6 7 8} \Bigl(\! (\cX_{\lambda}^{n_3})^{\!*}  \, \widetilde{V}_{\mu\nu, \lambda \rho}  \cX_{\mu}^{n_9}   \cX_{\nu}^{n_{10}}  \cY_{\rho}^{k_8}   \delta_{k_1 k_6} \delta_{k_2 k_7}\!\Bigr)  \, .
 \end{align}
Interaction matrices in Eqs.~(\ref{int_term_5d_D}) and~(\ref{int_term_5e_D}) are found in the self-energy diagrams of Figs.~\ref{3ord_223B_1} and~\ref{3ord_223B_2} respectively, while their complex conjugates are contained in the diagrams of Figs.~\ref{3ord_322B_1} and~\ref{3ord_322B_2}.

 When two $3h2p$ propagators  in a self-energy diagram are connected via ISCs linked to 2NFs, as in diagrams of Figs.~\ref{3ord_323B_1}-\ref{3ord_323B_3}, we have an interaction matrix of the following form,
\begin{align}
 \label{int_term_5l_D}
  \textbf{D}^{hh}_{u u'}  \equiv & -\frac{1}{12} \mathcal{A}_{4 5} \,  \mathcal{P}_{1 2 3} \, \mathcal{P}_{6 7 8} \, \Bigl(   (\cY_{\mu}^{k_1} \cY_{\nu}^{k_2})^*  \,  \widetilde{V}_{\mu\nu, \lambda \rho} \, \cY_{\lambda}^{k_6}  \cY_{\rho}^{k_7}    \Bigr. \nonumber \\ 
  &  \Bigl.\delta_{k_3 k_8} \delta_{n_4 n_9} \delta_{n_5 n_{10}} \Bigr) \, ,
 \end{align}
 contained in the diagram of Fig.~\ref{3ord_323B_1}, and another one found when a particle-hole 2NF connects two $3h2p$ ISCs, i.e.
 \begin{align}
 \label{int_term_5m_D}
  \textbf{D}^{hp}_{u u'}  \equiv & -\frac{1}{12}  \mathcal{A}_{4 5} \,  \mathcal{A}_{9 \,10} \, \mathcal{A}_{1 2 3} \, \mathcal{P}_{6 7 8} \, \Bigl(  (\cY_{\mu}^{k_1} \cX_{\rho}^{n_4})^*  \, \widetilde{V}_{\mu\nu,  \rho \lambda} \Bigr. \nonumber \\ 
  & \Bigl. \cY_{\lambda}^{k_6} \cX_{\nu}^{n_9}  \delta_{k_2 k_7} \delta_{k_3 k_8} \delta_{n_5 n_{10}} \Bigr)\, ,
 \end{align}
 which appears in the Feynman diagram of Fig.~\ref{3ord_323B_2}.
 
 Finally, also a backward-in-time interaction matrix connecting two $3h2p$ ISCs through a particle-particle 2NF is found in the self-energy diagram of Fig.~\ref{3ord_323B_3}, that is
    \begin{align}
 \label{int_term_5n_D}
  \textbf{D}^{pp}_{u u'}  \equiv & -\frac{1}{12} \mathcal{A}_{1 2 3}  \, \Bigl(  (\cX_{\lambda}^{n_4}  \cX_{\rho}^{n_5})^* \,  \widetilde{V}_{\mu\nu, \lambda \rho} \,  \cX_{\mu}^{n_{9}}  \cX_{\nu}^{n_{10}}  \Bigr. \nonumber \\ 
  & \Bigl.   \delta_{k_1 k_6} \delta_{k_2 k_7} \delta_{k_3 k_8}  \Bigr) \, . 
 \end{align}


\subsection{\label{Interaction_T_3nd_3p2h} Formulas for 3NF interaction matrices with $3p2h$ and $3h2p$ configurations}

The last set of interaction matrices is required to complete the ADC(3), which is given by those terms containing the interaction-irreducible 3NF. 

First we consider interaction matrices in which the 3NF connects  $2p1h$ ISCs to $3p2h$ ISCs, as in the diagrams in Figs.~\ref{3ord_233B_1}-\ref{3ord_332B_2}. We find in the diagram of Fig.~\ref{3ord_233B_1},
  \begin{align}
 \label{int_term_5h_C}
  \textbf{C}^{3N(I)}_{r q'}  \equiv & \frac{\sqrt{6}}{12} \mathcal{A}_{1 2} \, \mathcal{P}_{6 7 8}  \, \Bigl( \delta_{n_2 n_7} \, \cX_{\lambda}^{n_1} \cY_{\eta}^{k_3}   \, W_{\lambda \mu\nu,  \rho \eta \epsilon }  \Bigr. \nonumber \\ 
  & \Bigl.    (\cX_{\rho}^{n_6} \cX_{\epsilon}^{n_8}  \cY_{\mu}^{k_9} \cY_{\nu}^{k_{10}})^*  \Bigr) \, ,
 \end{align}
and a second 3NF interaction matrix connecting the $2p1h$ propagator to the $3p2h$ propagator, requiring an explicit antisymmetrization only with respect to two particle indexes. Its expression from a Goldstone diagram of Fig.~\ref{3ord_233B_2} reads,
  \begin{align}
 \label{int_term_5i_C}
  \textbf{C}^{3N(II)}_{r q'}  \equiv & \frac{\sqrt{6}}{12} \mathcal{A}_{9 \, 10}  \, \Bigl(   \cX_{\lambda}^{n_1} \cX_{\nu}^{n_2}\, W_{\lambda \nu\mu,  \rho \epsilon \eta}   \Bigr. \nonumber \\ 
  & \Bigl.    (\cX_{\rho}^{n_6} \cX_{\epsilon}^{n_7}  \cX_{\eta}^{n_8} \cY_{\mu}^{k_{10}})^*  \,  \delta_{k_3 k_9} \Bigr)\, .
 \end{align}
Complex conjugate interaction matrices corresponding to Eqs.~(\ref{int_term_5h_C}) and ~(\ref{int_term_5i_C}) can be found in the Goldstone diagrams of Figs.~\ref{3ord_332B_1} and~\ref{3ord_332B_2}.

Two $3p2h$  propagators can be connected in a self-energy diagram via ISCs containing a 3NF. They are contained in the three self-energy diagrams represented in Figs.~\ref{3ord_333B_1}-\ref{3ord_333B_3}.  We have then an interaction matrix of the following form in Fig.~\ref{3ord_333B_1},
   \begin{align}
 \label{int_term_5o_C}
  \textbf{C}^{3N(III)}_{q q'}  \! \equiv & \frac{1}{12} \mathcal{A}_{4 5}   \, \Bigl(  \cX_{\lambda}^{n_1} \cX_{\nu}^{n_2}   \cX_{\mu}^{n_3} \, W_{\lambda \nu\mu,  \rho \epsilon \eta} \, (\cX_{\rho}^{n_6} \cX_{\epsilon}^{n_7}  \cX_{\eta}^{n_8})^*\Bigr. \nonumber \\ 
  & \Bigl.  \delta_{k_4 k_9} \delta_{k_5 k_{10}} \Bigr) \, . 
 \end{align}

 Other 3NF interaction matrices  connecting two $3p2h$ ISCs are found in Fig.~\ref{3ord_333B_2}, 
 \begin{align}
 \label{int_term_5p_C}
  \textbf{C}^{3N(IV)}_{q q'}  \equiv & -\frac{1}{12}  \mathcal{A}_{4 5} \,  \mathcal{A}_{9 \,10} \, \mathcal{P}_{1 2 3} \, \mathcal{P}_{6 7 8}  \, \Bigl(  \cX_{\lambda}^{n_1} \cX_{\nu}^{n_2}  \cY_{\eta}^{k_5}  \Bigr. \nonumber \\ 
  & \Bigl.  W_{\lambda \nu\mu,  \rho \epsilon \eta} \,  (\cX_{\rho}^{n_6} \cX_{\epsilon}^{n_7} \cY_{\mu}^{k_{10}})^* \, \delta_{k_4 k_9} \delta_{n_3 n_8} \Bigr)  \, ,
 \end{align}
and in Fig.~\ref{3ord_333B_3}:
 \begin{align}
 \label{int_term_5q_C}
  \textbf{C}^{3N(V)}_{q q'}  \equiv & \frac{1}{12}  \, \mathcal{A}_{1 2 3} \,  \mathcal{P}_{6 7 8}  \, \Bigl( \cX_{\mu}^{n_1} \cY_{\epsilon}^{k_4}  \cY_{\eta}^{k_5} \, W_{\mu \nu\lambda,  \rho \epsilon \eta}  \Bigr. \nonumber \\ 
  & \Bigl.  (\cX_{\rho}^{n_6} \cY_{\nu}^{k_9}   \cY_{\lambda}^{k_{10}})^*  \, \delta_{n_2 n_7} \delta_{n_3 n_8} \Bigr) \, .
 \end{align}

%
 
 We can present now 3NF interaction matrices appearing in backward-in-time self-energy Goldstone diagrams. These interaction matrices connect  hole-particle ISCs in the diagrams, i.e. $2h1p$ and/or $3h2p$ propagators. As the corresponding interaction matrices for the forward-in-time part shown above, they are found in the diagrams of Figs.~\ref{3ord_233B_1}-\ref{3ord_332B_2} and Figs.~\ref{3ord_333B_1}-\ref{3ord_333B_3}.
 
First we have the terms,
  \begin{align}
 \label{int_term_5h_D}
  \textbf{D}^{3N(I)}_{s u'}  \equiv & - \frac{\sqrt{6}}{12}  \mathcal{P}_{1 2} \mathcal{P}_{6 7 8} \, \Bigl(  \delta_{k_2 k_7} \,  (\cY_{\lambda}^{k_1}  \cX_{\eta}^{n_3})^*    \, W_{\lambda \nu\mu,  \rho \eta \epsilon } \Bigr. \nonumber \\ 
  & \Bigl.  \cY_{\rho}^{k_6}  \cY_{\epsilon}^{k_8}    \cX_{\nu}^{n_9}  \cX_{\mu}^{n_{10}} \Bigr) \, ,
 \end{align}
 and 
 \begin{align}
 \label{int_term_5i_D}
  \textbf{D}^{3N(II)}_{s u'}  \equiv & -\frac{\sqrt{6}}{12} \mathcal{A}_{9 \, 10} \, \Bigl(  (\cY_{\lambda}^{k_1} \cY_{\nu}^{k_2} )^* \, W_{\lambda \nu\mu,  \rho \epsilon \eta}\Bigr. \nonumber \\ 
  & \Bigl.  \cY_{\rho}^{k_6} \cY_{\epsilon}^{k_7}  \cY_{\eta}^{k_8}  \cX_{\mu}^{n_{10}} \, \delta_{n_3 n_9}  \Bigr) \, ,
 \end{align}
 which are found in Figs.~\ref{3ord_233B_1} and~\ref{3ord_233B_2},  respectively.
 
 Finally, 3NF interaction matrices can connect  two $3h2p$  ISCs within a self-energy diagram, as in diagrams of Figs.~\ref{3ord_333B_1}-\ref{3ord_333B_3}. Specifically, in Figs.~\ref{3ord_333B_1} and~\ref{3ord_333B_2} we can single out the interaction matrices,
\begin{align}
 \label{int_term_5o_D}
  \textbf{D}^{3N(III)}_{u u'}  \equiv & \frac{1}{12} \mathcal{A}_{4 5}  \, \Bigl(  (\cY_{\lambda}^{k_1} \cY_{\nu}^{k_2}   \cY_{\mu}^{k_3})^* \, W_{\lambda \nu\mu,  \rho \epsilon \eta} \, \cY_{\rho}^{k_6} \cY_{\epsilon}^{k_7}  \cY_{\eta}^{k_8}  \Bigr. \nonumber \\ 
  & \Bigl.  \delta_{n_4 n_9} \delta_{n_5 n_{10}} \Bigr) \, ,
 \end{align}
and
 \begin{align}
 \label{int_term_5p_D}
  \textbf{D}^{3N(IV)}_{u u'}  \equiv & -\frac{1}{12} \mathcal{A}_{4 5} \,  \mathcal{A}_{9 \,10} \,  \mathcal{P}_{1 2 3} \,  \mathcal{P}_{6 7 8} \, \Bigl( (\cY_{\lambda}^{k_1} \cY_{\nu}^{k_2} \cX_{\eta}^{n_5})^* \Bigr. \nonumber \\ 
  & \Bigl. W_{\lambda \nu\mu,  \rho \epsilon \eta} \, \cY_{\rho}^{k_6} \cY_{\epsilon}^{k_7}  \cX_{\mu}^{n_{10}}   \, \delta_{n_4 n_9} \delta_{k_3 k_8} \Bigr)  \, ,
 \end{align}
 respectively.
Matrices in Eqs.~(\ref{int_term_5o_D}) and~(\ref{int_term_5p_D}) must be complemented with another backward-in-time interaction matrix that also connects two $3h2p$ ISCs. This is given by the expression,
\begin{align}
 \label{int_term_5q_D}
  \textbf{D}^{3N(V)}_{u u'}  \equiv & \frac{1}{12} \mathcal{A}_{1 2 3} \, \mathcal{P}_{6 7 8} \, \Bigl(  (\cY_{\mu}^{k_1} \cX_{\epsilon}^{n_4}  \cX_{\eta}^{n_5})^*  \, W_{\mu \nu \lambda,  \rho \epsilon \eta} \Bigr. \nonumber \\ 
  & \Bigl.
  \cY_{\rho}^{k_6} \cX_{\nu}^{n_9}   \cX_{\lambda}^{n_{10}} \, \delta_{k_2 k_7} \delta_{k_3 k_8} \Bigr) \,  , 
 \end{align}
as it is singled out from the Goldstone diagram corresponding to Fig.~\ref{3ord_333B_3}.
 
%

\section{\label{Angular}Irreducible self-energy in angular momentum coupling formalism}

Most implementations of \textit{ab initio} approaches in nuclear physics are based on the assumption of a spherical ground state, so that one can exploit a spherical single-particle basis and the angular momentum formalism to decouple different channels. The diagonalization of Eq.~(\ref{eq:Dy_residual_5}) can then be performed separately for each partial wave, and the required computational resources are significantly reduced.
In this Appendix we follow this path and derive the corresponding working equations for ADC(3) and 3NFs  including up to $2p1h$  and $2h1p$ ISCs. 
The ADC(3) matrix elements for 2NFs have been discussed in the past~\cite{Walter1981JPB} and used in applications for several years~\cite{Barbieri2006plbO16,Barbieri2007,Barbieri2014QMBT}. However, they were never published in detail for the case of fully dressed propagators. Moreover, we complement them with the new terms arising from 3NFs. We present them here in the hope that they might turn out to be useful to practitioners.

In general, a spherical single-particle state with isospin function $\chi_q$, is given by coupling the spherical harmonic $Y_l(\hat{r})$ to $\chi_{\frac{1}{2}}$, the function of the intrinsic spin of the nucleon,
\begin{equation}
\label{eq:phy}
\phi_\beta(\vec{r},\sigma, \tau)=f_{n_\beta l_\beta}(r) [Y_{l_\beta}(\hat{r}) \otimes \chi_{\frac{1}{2}}(\sigma)]_{j_\beta}^{m_\beta} \chi_{q_\beta}(\tau) \, ,
\end{equation}
with $\sigma$ and $\tau$ being  spin and isospin coordinates, respectively.

The collective index $\beta$  denotes the set of quantum numbers ($n_\beta,\pi_\beta,j_\beta,m_\beta,q_\beta$), where $n_\beta$ is the principal quantum number, $\pi_\beta$ is the parity corresponding to the orbital angular momentum $l_\beta$, $j_\beta$ and $m_\beta$ are  the total angular momentum and its projection along the $z$ axis, and $q_\beta$ represents the isospin projection. In this basis, the creation operator $a^{\dagger}_{\beta}$ of the single-particle basis is the $m^{th}_\beta$ component of an irreducible tensor of rank $j_\beta$:
\begin{equation}
  \label{eq:a_alpha}
a^{\dagger}_{\beta} \equiv a^{\dagger}_{n_\beta, \pi_\beta, j_\beta, m_\beta, q_\beta} \equiv a^{\dagger}_{b, m_\beta} \, ,
\end{equation}
where we made use of the notation $\beta = (b, m_\beta)$, i.e.  
\begin{equation}
  \label{beta}
b \equiv (n_\beta, \pi_\beta, j_\beta, q_\beta) \, .
\end{equation}
 The destruction operators are dealt with in the same fashion but we add a phase factor and invert the quantum number $m$, which is needed to obtain an irreducible tensor  $\bar{a}_\beta\equiv(-1)^{j_\beta+m_\beta}a_{b, -m_\beta}$. For particle and hole Dyson orbits, corresponding to the eigenstates $\ket{\Psi^{A+1}_n}$ and $\ket{\Psi^{A-1}_k}$ of the (A+1)- and (A-1)-body systems, we use  a compact notation analogous to Eq.~(\ref{eq:a_alpha}) and write \hbox{$n = (\tilde{n}, m_n)$} and \hbox{$k = (\tilde{k}, m_k)$}, with \hbox{$\tilde{n} \equiv (n_n, \pi_n, j_n, q_n)$} and \hbox{$\tilde{k} \equiv (n_k, \pi_k, j_k, q_k)$}. Using these definitions, the shorthand notations of Eqs.~(\ref{def_r}) and (\ref{def_s}) can be coupled to total angular momentum and the overall quantum number $M$ is separated. This leads us to define \hbox{$\tilde{r} \equiv  [(\tilde{n}_1, \tilde{n}_2, J_{1 2}, \tilde{k}_3), J_r]$}, \hbox{$ \tilde{s} \equiv [(\tilde{k}_1, \tilde{k}_2, J_{1 2}, \tilde{n}_3), J_s]$} and so on, where we follow the coupling conventions of Eqs.~\eqref{choice_Mppk} and~\eqref{choice_Nkkp} below. 

We now revisit the angular momentum coupling of the self-energy, when the ground state $\ket{\Psi^{A}_0}$ in Eq.~(\ref{Green}) has total angular momentum and parity $J^\pi$=0$^+$. For these systems, the formalism is considerably simplified because the total angular momentum  $j_n$ ($j_k$), its projection  along the $z$ axis $m_n$ ($m_k$) and the parity $\pi_n$ ($\pi_k$) of excited states $\Psi_{n}^{A+1}$ ($\Psi_{k}^{A-1}$) are the same as the corresponding quantum numbers of the particle creation (annihilation) tensor operators entering the definition of the Green's function, Eq.~(\ref{Green}). The total isospin is also uniquely determined by isospin projections of the reference state and tensor operators.
With our assumption that $\ket{\Psi^{A}_0}$  has  $J^\pi$=0$^+$, the irreducible self-energy in the Dyson equation~(\ref{eq:Dy}) becomes diagonal in the quantum numbers ($\pi, j,m,q$), and it is independent on $m$:
\begin{equation}
  \label{eq:self_diag}
\Sigma_{\alpha\beta}^{\star}(\omega)=
\delta_{a b}^{(\pi j q)} \delta_{m_\alpha m_\beta} \Sigma_{a b}^{\star}(\omega) \, ,
\end{equation}
where we have introduced a compact notation for multiple Kronecker deltas:
\begin{equation}
\delta_{a b}^{(\pi j q)} \equiv \delta_{\pi_\alpha \pi_\beta} \delta_{j_\alpha j_\beta}  \delta_{q_\alpha q_\beta} \, .
\end{equation}

\begin{widetext}
By applying the Wigner-Eckart theorem to the transition amplitudes  from Eq.~\eqref{tran_ampl}, one finds:
\begin{eqnarray}
\label{tran_ampl_WE}
{\cal X}^{\tilde{n},m_n}_{\alpha}&=&(-1)^{2 j_n}\frac{\delta_{a \tilde{n}}^{(\pi j q)} \delta_{m_\alpha m_n}}{\sqrt{2 j_\alpha + 1}}\langle\Psi_{\tilde{n}}^{A+1}||a_{a}^{\dagger}||\Psi_0^A\rangle 
\equiv  {\cal X}^{\tilde{n}}_{a}\delta_{a \tilde{n}}^{(\pi j q)} \delta_{m_\alpha m_n}
\end{eqnarray}
and
\begin{eqnarray}
\label{tran_ampl_Y_WE}
{\cal Y}^{\tilde{k},m_k}_{\alpha}&=&(-1)^{2 j_k}(-1)^{j_\alpha-m_\alpha}\frac{\delta_{a \tilde{k}}^{(\pi j q)}  \delta_{-m_\alpha m_k}}{\sqrt{2 j_\alpha + 1}}\langle\Psi_{\tilde{k}}^{A-1}||\bar{a}_{a}||\Psi_0^A\rangle 
\equiv  {\cal Y}^{\tilde{k}}_{a}(-1)^{j_\alpha-m_\alpha} \delta_{a \tilde{k}}^{(\pi j q)} \delta_{-m_\alpha m_k} \, ,
\end{eqnarray}
which define the $m$-independent spectroscopic amplitudes $ {\cal X}^{\tilde{n}}_{a}$ and $ {\cal Y}^{\tilde{k}}_{a}$.

Because of rotational invariance of the Hamiltonian, the 2NFs and 3NFs are coupled as $m$-independent matrix elements according to
\begin{eqnarray}
\label{V_2B_ren}
\bar{V}_{a b, g d}^J & = &  \frac{1}{\sqrt{1+\delta_{a b}}\sqrt{1+\delta_{g d}}} \sum_{\substack{m_\alpha m_\beta \\ m_\gamma m_\delta}}  (j_\alpha j_\beta m_\alpha m_\beta | J M)  (j_\gamma j_\delta m_\gamma m_\delta | J M) \tilde{V}_{\alpha \beta, \gamma \delta} 
\end{eqnarray}
and 
\begin{eqnarray}
\label{V_3B_J}
W_{ a b l, g d v}^{J_1 J_2 J} &=& \sum_{\substack{ m_\alpha m_\beta \\ m_\lambda }} \sum_{\substack{ m_\gamma m_\delta \\ m_\nu }}(j_\alpha j_\beta m_\alpha m_\beta | J_1 M_1)  (j_\gamma j_\delta m_\gamma m_\delta | J_2 M_2)   ( J_1 j_\lambda M_1 m_\lambda | J M) ( J_2 j_\nu M_2 m_\nu | J M)  W_{ \alpha \beta \lambda, \gamma \delta \nu} \, .
\end{eqnarray}
Note that we chose to properly normalize the matrix elements of $\bar{V}$ in Eq.~\eqref{V_2B_ren} but not those of $W.$
When considering  the Goldstone diagrams that result from Fig.~\ref{3ord_c}, it is also convenient to recouple the angular momenta of the 3NF in the  particle-particle-hole channel:
\begin{eqnarray}
\label{V_3B_Jpph}
W_{ a b l^{-1}, g d v^{-1}}^{(pph) J_1 J_2 J} &=& 
 - \sum_{\substack{ m_\alpha m_\beta \\ m_\lambda }} \sum_{\substack{ m_\gamma m_\delta \\ m_\nu }}
  (j_\alpha j_\beta m_\alpha m_\beta | J_1 M_1)  (j_\gamma j_\delta m_\gamma m_\delta | J_2 M_2) 
  \nonumber \\
  & & \qquad \times ( J_1 j_\lambda M_1 -m_\lambda | J M) (J_2 j_\nu M_2 -m_\nu | J M) 
  (-)^{j_\nu - m_\nu}  (-)^{j_\lambda - m_\lambda} 
  W_{ \alpha \beta \nu , \gamma \delta \lambda } \, .
\end{eqnarray}
The  matrix elements $W$ can be transformed into the $W^{(pph)}$ form by using  the Pandya relation~\cite{Pandya1956}:
\begin{eqnarray}
\label{V_3B_J_pandya}
W_{ a b l^{-1}, g d v^{-1}}^{(pph) J_1 J_2 J} &=& 
 - (-)^{2j_\nu + 2 j_\lambda + 2J}  \,  \sum_{J'} (2J'+1) \left\{ \begin{array}{@{\!~}c@{\!~}c@{\!~}c@{\!~}} 
J_1 & j_{\lambda} & J\\[2mm] 
J_2 & j_{\nu} & J'  
\end{array}\right\}
W_{ a b v, g d l}^{ J_1 J_2 J'} \, .
\end{eqnarray}

\subsection{\label{choices} General J-coupling conventions for ISCs}

The ADC coupling and interaction matrices derived in earlier sections already separate naturally in terms of the parity and isospin (i.e., charge differences with respect to the ground state $|\Psi^A_0\rangle$).  To decouple the Dyson eigenvalue problem according to Eq.~(\ref{eq:self_diag}), we  then need to recouple the ISCs to total
angular momentum. In this Appendix we are going to consider only the ADC contributions arising from Feynman diagrams of Fig.~\ref{3ord_a_b_c}. 
Thus, we need to specify an angular momentum coupling convention for $2p1h$ and $2h1p$ ISCs. 
As usual, the particular choice of coupling may introduce particular phases and factors in the equations but the final result of the Dyson diagonalization, Eq.~\eqref{eq:Dy_residual_5}, does not depend on these.

For $2p1h$ coupling matrices we define the coupled elements of $\textbf{M}_{\tilde{r} a } $ as
\begin{eqnarray}
\label{choice_Mppk}
 \sum_{\substack{ m_{n_1} m_{n_2} \\ m_{k_3}}} (j_{n_1} j_{n_2} m_{n_1} m_{n_2} | J_{1 2} M_{1 2})  (J_{1 2} j_{k_3} M_{1 2} m_{k_3} | J_r M_r) \textbf{M}_{r \alpha }  & \equiv & \delta_{a r}^{(\pi j q)} \delta_{m_\alpha M_r} \, \sqrt { \frac{1+\delta_{\tilde{n}_1 \tilde{n}_2}}  2} \, \textbf{M}_{\tilde{r} a } \, ,
\end{eqnarray}
with the parity $\pi_r=(-1)^{l_1+l_2+l_3}$, the charge $q_r=q_1+q_2-q_3$  and the total angular momentum $J_r$ of the $2p1h$ ISC $\tilde{r} =  [(\tilde{n}_1, \tilde{n}_2, J_{1 2}, \tilde{k}_3), J_r]$. 
For the backward-in-time $2h1p$ matrix $\textbf{N}_{\alpha \tilde{s}} $, we write: 
\begin{eqnarray}
\label{choice_Nkkp}
 \sum_{\substack{ m_{k_1} m_{k_2} \\ m_{n_3}}} (j_{k_1} j_{k_2} m_{k_1} m_{k_2} | J_{1 2} M_{1 2})  (J_{1 2} j_{n_3} M_{1 2} m_{n_3} | J_s M_s) \textbf{N}_{\alpha s} & \equiv &  \delta_{a s}^{(\pi j q)} \delta_{m_\alpha -M_s} \, \sqrt { \frac{1+\delta_{\tilde{k}_1 \tilde{k}_2}}  2} \, \textbf{N}_{a \tilde{s}} \, ,
\end{eqnarray}
with the charge $q_s=q_1+q_2-q_3$  and the total angular momentum $J_s$ of the $2h1p$ ISC $\tilde{s} =  [(\tilde{k}_1, \tilde{k}_2, J_{1 2}, \tilde{n}_3), J_s]$. 
The interaction matrices $\textbf{C}_{\substack {j j'}}$ and $\textbf{D}_{\substack {k k'}}$ are coupled in the same way.
For  $2p1h$  matrix elements $\textbf{C}_{\substack{r  r'}}$, we have
\begin{eqnarray}
\label{choice_Cppk}
 \sum_{\substack{ m_{n_1} m_{n_2} \\ m_{k_3}}}  \sum_{\substack{ m_{n_4} m_{n_5} \\ m_{k_6}}} && (j_{n_1} j_{n_2} m_{n_1} m_{n_2} | J_{1 2} M_{1 2})  (J_{1 2} j_{k_3} M_{1 2} m_{k_3} | J_r M_r) (j_{n_4} j_{n_5} m_{n_4} m_{n_5} | J_{4 5} M_{4 5})  (J_{4 5} j_{k_6} M_{4 5} m_{k_6} | J_{r'} M_{r'}) \textbf{C}_{\substack{r  r'}} \nonumber \\
 &&= \delta_{r r'}^{(\pi j q)} \delta_{M_r M_{r'}} \, \sqrt { \frac{1+\delta_{\tilde{n}_1 \tilde{n}_2}}  2} \, \textbf{C}_{\substack{\tilde{r}  \tilde{r}'}} \, \sqrt { \frac{1+\delta_{\tilde{n}_4 \tilde{n}_5}}  2} \; .
\end{eqnarray}
And for the $2h1p$ matrix elements $ \textbf{D}_{\substack{s  s'}} $:
\begin{eqnarray}
\label{choice_Dkkp}
 \sum_{\substack{ m_{k_1} m_{k_2} \\ m_{n_3}}}  \sum_{\substack{ m_{k_4} m_{k_5} \\ m_{n_6}}} && (j_{k_1} j_{k_2} m_{k_1} m_{k_2} | J_{1 2} M_{1 2})  (J_{1 2} j_{n_3} M_{1 2} m_{n_3} | J_s M_s) (j_{k_4} j_{k_5} m_{k_4} m_{k_5} | J_{4 5} M_{4 5})  (J_{4 5} j_{n_6} M_{4 5} m_{n_6} | J_{s'} M_{s'}) \textbf{D}_{\substack{s  s'}} \nonumber \\
 &&= \delta_{s s'}^{(\pi j q)} \delta_{M_s M_{s'}}  \, \sqrt { \frac{1+\delta_{\tilde{k}_1 \tilde{k}_2}}  2} \, \textbf{D}_{\substack{\tilde{s}  \tilde{s}'}} \, \sqrt { \frac{1+\delta_{\tilde{k}_4 \tilde{k}_5}}  2} \; .
\end{eqnarray}
Equations~(\ref{choice_Cppk}) and (\ref{choice_Dkkp}) define the coupled  $\textbf{C}_{\substack{\tilde{r}  \tilde{r}'}}$ and  $\textbf{D}_{\substack{\tilde{s}  \tilde{s}'}}$ that are independent of $M_r$ and $M_s$, respectively.
With the above definitions, the unperturbed energies for each ISC do not depend on the angular momentum coupling and their expressions, Eqs.~\eqref{energy_E_i} and~\eqref{energy_E_k}, remain unchanged. For example:
\begin{equation}
\label{Ene_conf}
\textbf{E}_{\tilde{r} \, \tilde{r}'}  = \text{diag}\{\varepsilon^+_{\tilde{n}_{1} }+\varepsilon^+_{\tilde{n}_{2} } -\varepsilon^-_{\tilde{k}_{3}} \} \, .
\end{equation}
Note that the coupling and interaction matrices introduced in Secs.~\ref{sec:ADC(2)} and~\ref{sec:ADC(3)} are explicitly antisymmetric for the exchanges in any of the particle ($n_1$, $n_2$, $n_3$, ...) or any of the hole ($k_1$, $k_2$, $k_3$, ...) indices. They are nevertheless defined assuming unrestricted sum over all indexes. In practical calculations, one can drop the symmetry factors of $1/\sqrt{2}$ for $2p1h$ ($2h1p$) ISCs and restrict the sums over the fist two indexes, $n_1<n_2$~($k_1<k_2$).  In angular momentum coupling formalism one  has similarly ordered sums---$\tilde{n}_1\leq\tilde{n}_2$ and $\tilde{k}_1\leq\tilde{k}_2$ in this case---and factors such as $\sqrt{(1+\delta_{\tilde{n}_1 \, \tilde{n}_2})/2}$ are to be dropped. Thus, we separated these factors already in the definitions~(\ref{choice_Mppk})-(\ref{choice_Dkkp}).

%
%
%

With the above definitions, and assuming ordered sums, the  dynamic self-energy of Eq.~(\ref{irr_SE_Lehmann}) can be decoupled as 
\begin{eqnarray}
\label{irr_SE_couplings}
\widetilde{\Sigma}_{\alpha\beta}(\omega)  &=&  \sum_{r r'} \textbf{M}_{\alpha r}^*  \frac{1}{\hbar \omega - (\textbf{E}_r \delta_{r r'} +  \textbf{C}_{\substack{r  r'}}) + \textrm{i} \eta}   \textbf{M}_{r' \beta} 
 +  \sum_{s s'} \textbf{N}_{\alpha s} \frac{1}{\hbar \omega - (\textbf{E}_s \delta_{s s'}  +  \textbf{D}_{\substack{s  s'}}) - \textrm{i} \eta} \textbf{N}_{s' \beta}^*  \nonumber \\
 &=&  \sum_{\tilde{r} \, \tilde{r}'\, M_r \, M_{r'}} 
 \delta_{a r}^{(\pi j q)} \delta_{m_\alpha M_r} \textbf{M}_{a \tilde{r}}^*  \frac{1}{\hbar \omega - (\textbf{E}_{\tilde{r}} \delta_{\tilde{r}  \tilde{r}'}  +   \textbf{C}_{\substack{\tilde{r}  \tilde{r}'}} ) \delta_{r r'}^{(\pi j q)} \delta_{M_r M_{r'}} + \textrm{i} \eta}  \delta_{b r'}^{(\pi j q)} \delta_{m_\beta M_{r'}} \textbf{M}_{ \tilde{r}' b} \nonumber \\
 &&+  \sum_{\tilde{s} \, \tilde{s}' \, M_s \, M_{s'} } \delta_{a s}^{(\pi j q)} \delta_{m_\alpha -M_s} \textbf{N}_{a \tilde{s}} \frac{1}{\hbar \omega - (\textbf{E}_{\tilde{s}} \delta_{\tilde{s}  \tilde{s}'}  +   \textbf{D}_{\substack{\tilde{s}  \tilde{s}'}} ) \delta_{s s'}^{(\pi j q)} \delta_{M_s M_{s'}} - \textrm{i} \eta}
  \delta_{b s'}^{(\pi j q)} \delta_{m_\beta -M_{s'}}\textbf{N}^*_{b \tilde{s}'}  \nonumber \\
 &=& \delta_{a b}^{(\pi j q)} \delta_{m_\alpha m_\beta}  \left( \sum_{\tilde{r} \tilde{r}'} 
 \delta_{a r}^{(\pi j q)} \textbf{M}_{a \tilde{r}}^*  \frac{1}{\hbar \omega - (\textbf{E}_{\tilde{r}} \delta_{\tilde{r}  \tilde{r}'}  +   \textbf{C}_{\substack{\tilde{r}  \tilde{r}'}} )  + \textrm{i} \eta}  \delta_{b r'}^{(\pi j q)}  \textbf{M}_{ \tilde{r}' b} \right. \nonumber \\
 &&+  \left. \sum_{\tilde{s} \tilde{s}'} \delta_{a s}^{(\pi j q)} \textbf{N}_{a \tilde{s}} \frac{1}{\hbar \omega - (\textbf{E}_{\tilde{s}} \delta_{\tilde{s}  \tilde{s}'}   +   \textbf{D}_{\substack{\tilde{s}  \tilde{s}'}} )  - \textrm{i} \eta}
  \delta_{b s'}^{(\pi j q)} \textbf{N}^*_{b \tilde{s}'}  \right) \nonumber \\
 &\equiv & 
\delta_{a b}^{(\pi j q)} \delta_{m_\alpha m_\beta} \widetilde\Sigma_{a b}(\omega) \, ,
\end{eqnarray}
 which proves the energy dependent part of Eq.~(\ref{eq:self_diag}) and defines the $m$-independent irreducible self-energy $\Sigma_{a b}^{\star}(\omega)$.

\subsection{\label{2p1h} Angular momentum coupling of the ADC(3) equations}
We are now in the position to give  expressions for the coupling  and interaction matrices resulting from the diagrams of Figs.~\ref{2ord} and~\ref{3ord_a_b_c} in the angular momentum formalism.

 \subsubsection{\label{2p1h_only} Coupling matrices with $2p1h$ ISCs}
 
 The simplest $2p1h$ coupling matrix is  $\textbf{M}^{(\textrm{Ia})}_{r \alpha} $,  Eq.~(\ref{eq:M_2a}), and it appears already at the ADC(2) level.  Using the definition~(\ref{choice_Mppk}), we obtain
%
%
\begin{eqnarray}
\label{eq:M_2a_ANG}
 \textbf{M}^{(\textrm{Ia})}_{\tilde{r} a } &\equiv & \Delta(j_{n_1},j_{n_2},J_{1 2}) \Delta(J_{1 2},j_{k_3}, J_r)(-1)^{j_\alpha+j_{k_3}-J_{1 2}}\frac{\hat{J}_{1 2}}{\hat{j}_\alpha} \nonumber \\ 
 && \times
 \sum_{\substack{ m \leq v \\ l}} 
  \frac{
     \cX_{m}^{\tilde{n}_1} \delta_{m n_1}^{(\pi j q)}   \cX_{v}^{\tilde{n}_2}  \delta_{v n_2}^{(\pi j q)}
       - (-1)^{j_{n_1}+j_{n_2}-J_{1 2}} \cX_{v}^{\tilde{n}_1} \delta_{v n_1}^{(\pi j q)}  \cX_{m}^{\tilde{n}_2}   \delta_{m n_2}^{(\pi j q)}
     }{\sqrt{1+\delta_{v m}} \sqrt{1+\delta_{\tilde{n}_1 \tilde{n}_2}}}
  \, \bar{V}^{J_{1 2}}_{m v,a l} \,  \sqrt{1+\delta_{a l}} \,  \cY_{l}^{\tilde{k}_3}  \delta_{l k_3}^{(\pi j q)} \, ,  \qquad
\end{eqnarray}
with the usual \enquote{hat } notation
\begin{equation}
\label{hat}
\hat{j} \equiv \sqrt{2j+1} 
\end{equation}
and the triangular condition
\begin{equation}
\label{triang}
\Delta(j, j', J) = \left\{ \begin{array}{ccl}
1  &\quad\quad\quad&   \hbox{if~} |j-j'| \leq J \leq j+j' \, , \\
~\\
0 & & \hbox{otherwise.}
\end{array}  \right.
\end{equation}
The $\textbf{M}^{(\textrm{IIa})}_{r \alpha }$ contribution from Eq.~(\ref{eq:M_3a}) becomes
\begin{eqnarray}
\label{eq:M_3a_ANG}
 \textbf{M}^{(\textrm{IIa})}_{\tilde{r} a } &\equiv & \Delta(j_{n_1},j_{n_2},J_{1 2}) \Delta(J_{1 2},j_{k_3}, J_r) \frac {(-1)^{j_\alpha+j_{k_3}-J_{1 2}} }{\sqrt{1+\delta_{\tilde{n}_1 \tilde{n}_2} } } \frac{\hat{J}_{1 2}}{\hat{j}_\alpha} \sum_{\substack{ \tilde{k}_{4} \leq \tilde{k}_{5} }} \sum_{\substack{ g \leq d \\ l}} \Delta(j_{k_4},j_{k_5},J_{1 2})   \,
  t^{\tilde{n}_1 \tilde{n}_2, J_{1 2}}_{\tilde{k}_4 \tilde{k}_5}  \qquad
   \nonumber \\
&& \quad \times \frac{
   \left(\cY_{g}^{\tilde{k}_4} \delta_{g  k_4}^{(\pi j q)} \,  \cY_{d}^{\tilde{k}_5}  \delta_{d   k_5}^{(\pi j q)}   
      - (-1)^{j_{k_4}+j_{k_5}-J_{1 2}} \,  \cY_{d}^{\tilde{k}_4}   \delta_{d   k_4}^{(\pi j q)} \,  \cY_{g}^{\tilde{k}_5} \delta_{g  k_5}^{(\pi j q)}  \right)^*
     }{ (1+\delta_{\tilde{k}_4 \tilde{k}_5})  \, \sqrt{1+\delta_{g d}} }  \, \bar{V}^{J_{1 2}}_{g d, a l} \sqrt{1 + \delta_{a l}} \, 
 \cY_{l}^{\tilde{k}_3} 
    \delta_{l   k_3}^{(\pi j q)} \, , 
\end{eqnarray}
where we have coupled the angular momenta of the  $2p2h$ amplitude, Eq.~(\ref{eq:t_2B}), as follows:
\begin{eqnarray}
\label{eq:t_2B_ANG}
t^{\tilde{n}_1 \tilde{n}_2 , J}_{\tilde{k}_3 \tilde{k}_3} &\equiv &
 \sum_{\substack{ m_{n_1} m_{n_2} \\ m_{k_3} m_{k_4}}}  (j_{n_1} j_{n_2} m_{n_1} m_{n_2} | J M)  ( j_{k_3} j_{k_4}  m_{k_3} m_{k_4} | J \, -M) \, (-1)^{J - M} \,  t^{n_1 n_2}_{k_3 k_4} 
\nonumber \\
&=&\sum_{\substack{ m \leq v \\ g \leq d}}  
\frac{
   \cX_{m}^{\tilde{n}_1}  \delta_{m  n_1}^{(\pi j q)} \, \cX_{v}^{\tilde{n}_2}  \delta_{v   n_2}^{(\pi j q)}
       - (-1)^{j_{n_1}+j_{n_2}-J} \,  \cX_{v}^{\tilde{n}_1}  \delta_{v  n_1}^{(\pi j q)} \,  \cX_{m}^{\tilde{n}_2}  \delta_{m   n_2}^{(\pi j q)} }{\sqrt{1+\delta_{m v}}  } 
\nonumber \\ 
&& \qquad \times   
\frac{  \bar{V}^{J}_{m v, g d}}
{\varepsilon_{\tilde{k}_4}^{-}+\varepsilon_{\tilde{k}_3}^{-}-\varepsilon_{\tilde{n}_1}^{+} -\varepsilon_{\tilde{n}_2}^{+}}
\frac{
   \cY_{g}^{\tilde{k}_3} \delta_{g   k_3}^{(\pi j q)} \,  \cY_{d}^{\tilde{k}_4} \delta_{d   k_4}^{(\pi j q)} 
    - (-1)^{j_{k_3}+j_{k_4}-J} \, \cY_{d}^{\tilde{k}_3}  \delta_{d   k_3}^{(\pi j q)} \, \cY_{g}^{\tilde{k}_4} \delta_{g   k_4}^{(\pi j q)}  }{\sqrt{1+\delta_{r s}} }  
  \, .
\end{eqnarray}

The $\textbf{M}^{(\textrm{IIb})}_{r \alpha}$ of Eq.~(\ref{eq:M_3b}) in the angular momentum coupled representation is given by
\begin{eqnarray}
\label{eq:M_3b_ANG}
 \textbf{M}^{(\textrm{IIb})}_{\tilde{r}  a } &\equiv & \Delta(j_{n_1},j_{n_2},J_{1 2}) \Delta(J_{1 2},j_{k_3}, J_r)  (-1)^{j_\alpha+j_{k_3}-J_{1 2}} \frac{\hat{J}_{1 2}}{\hat{j}_\alpha} \frac{1}{\sqrt{1+ \delta_{\tilde{n}_1 \tilde{n}_2}}} \;  \sum_{\substack{ \tilde{k}_{4}  \, \tilde{n}_{5} }} \; \sum_{\substack{  m \, v \\ l  }} \; \sum_{J_2 J_3} \, 
   (2J_2+1) (2J_3+1) 
   \nonumber \\
    && \times \left( (-1)^{-j_{n_1}-j_{n_2}+J_{1 2}} 
 \left\{ \begin{array}{@{\!~}c@{\!~}c@{\!~}c@{\!~}} 
j_{n_1} & j_{k_4} & J_2 \\[2mm] 
j_{n_2} & J_3 & j_{n_5}  \\[2mm] 
J_{1 2} &  j_{k_3}  & J_r
\end{array}\right\}  
 \cX_{m}^{\tilde{n}_1} \delta_{m  n_1}^{(\pi j q)}  \, t^{\tilde{n}_2 \tilde{n}_5, J_3}_{ \tilde{k}_4 \tilde{k}_3} \, (\cY_{v}^{\tilde{k}_4} 
      \delta_{v   k_4}^{(\pi j q)}  \, \cX_{l}^{\tilde{n}_5}  \delta_{l   n_5}^{(\pi j q)})^*  \,
  \sqrt{1 + \delta_{g m}}  \bar{V}^{J_2}_{v m, a l}  \sqrt{1 + \delta_{a d}}    \;
   \right.
 \nonumber \\
&& \qquad \quad \left. -  
 \left\{ \begin{array}{@{\!~}c@{\!~}c@{\!~}c@{\!~}} 
j_{n_2} & j_{k_4} & J_2 \\[2mm] 
j_{n_1} & J_3 & j_{n_5}  \\[2mm] 
J_{1 2} &  j_{k_3}  & J_r
\end{array}\right\}  \cX_{m}^{\tilde{n}_2} \delta_{m  n_2}^{(\pi j q)}  \,  t^{\tilde{n}_1 \tilde{n}_5, J_3}_{ \tilde{k}_4 \tilde{k}_3}  \, 
  (\cY_{v}^{\tilde{k}_4}  \delta_{v   k_4}^{(\pi j q)} \, \cX_{l}^{\tilde{n}_5}  \delta_{l   n_5}^{(\pi j q)})^* \,
 \sqrt{1 + \delta_{g m}}   \bar{V}^{J_2}_{v m, a l}    \sqrt{1 + \delta_{a d}}   \;
\right) \, ,
\end{eqnarray}
which is explicitly antisymmetrized with respect to the $\tilde{n}_1$, $\tilde{n}_2$ indexes.
%
%
%

The coupling matrix  $\textbf{M}^{(\textrm{IIc})}_{ r \alpha}$ of Eq.~(\ref{eq:M_3c}) is implied by the diagram of Fig.~\ref{3ord_c}. This is the first term that contains an interaction-irreducible 3NF and it has the following form:
\begin{eqnarray}
\label{eq:M_3c_ANG}
 \textbf{M}^{(\textrm{IIc})}_{\tilde{r} a } &\equiv &
 \Delta(j_{n_1},j_{n_2},J_{1 2}) \Delta(J_{1 2},j_{k_3}, J_r)  \, 
  \sum_{J_{56} \,  J'} \sum_{\substack{ \tilde{k}_{5} \leq \tilde{k}_{6} \\ \tilde{n}_{4} }} \sum_{\substack{ v \leq m \\ l}} \, 
  \Delta(j_{k_5},j_{k_6},J_{56}) \Delta(j_{n_4},J_{56}, j_\alpha) \,  
  \frac{ (-1)^{j_{n_4} + j_\alpha - J_{56}  + 2J' } }{\sqrt{1+\delta_{\tilde{n}_1 \tilde{n}_2}}}
   \frac{\hat{J}_{56}}{\hat{j}_{\alpha}}  (2J'+1)
  \nonumber \\ 
&&  \qquad \times  \left\{ \begin{array}{@{\!~}c@{\!~}c@{\!~}c@{\!~}} 
 j_{k_3} &  J_{1 2}  & j_\alpha \\[2mm] 
 j_{n_4} & J_{5 6} & J'  
\end{array}\right\}  t^{\tilde{n}_1 \tilde{n}_2 \tilde{n}_4, J_{1 2} J'}_{\tilde{k}_5 \tilde{k}_6 \tilde{k}_3, J_{56}} 
\frac{  \left( 
   \cY_{m}^{\tilde{k}_5}  \delta_{n  k_5}^{(\pi j q)}  \,  \cY_{v}^{\tilde{k}_6} \delta_{v  k_6}^{(\pi j q)} 
     - (-1)^{j_{k_5}+j_{k_6}-J_{56}} \,  \cY_{v}^{\tilde{k}_5} \delta_{v  k_5}^{(\pi j q)}  \,  \cY_{m}^{\tilde{k}_6}  \delta_{m  k_6}^{(\pi j q)}
     \right)^* }{(1+\delta_{\tilde{k}_5 \tilde{k}_6}) \, \sqrt{1+\delta_{m v}}  } 
 \nonumber \\ 
&& \qquad \quad \times (\cX_{l}^{\tilde{n}_4} \delta_{l n_4}^{(\pi j q)})^*    \bar{V}^{J_{56}}_{m v,a l} \sqrt{1+\delta_{a l}}   \, ,
\end{eqnarray}
which is expressed in terms of the $3p3h$ amplitude Eq.~(\ref{eq:t_3B}). In the angular momentum form, this is
\begin{eqnarray}
\label{eq:t_3B_ANG}
t^{\tilde{n}_1 \tilde{n}_2 \tilde{n}_4, J_{1 2} J'}_{\tilde{k}_5 \tilde{k}_6 \tilde{k}_3, J_{56}} 
&\equiv &
 \sum_{\substack{ g \leq d \\ r s \\ t \leq p}}  
 \frac{ 
   \cX_{g}^{\tilde{n}_1} \delta_{g  n_1}^{(\pi j q)}  \,  \cX_{d}^{\tilde{n}_2} \delta_{d    n_2}^{(\pi j q)} 
    - (-1)^{j_{n_1}+j_{n_2}-J_{1 2}} \cX_{d}^{\tilde{n}_1} \delta_{d  n_1}^{(\pi j q)}  \,   \cX_{g}^{\tilde{n}_2}  \delta_{g    n_2}^{(\pi j q)}
    }{(1+\delta_{g d}) } \,  
\frac{ 
 \cX_r^{\tilde{n}_4}   \delta_{r   n_4}^{(\pi j q)} \; W^{J_{1 2} J_{56} J'}_{g d r, t p s}   \;   \cY_{s}^{\tilde{k}_3}  \delta_{s  k_3}^{(\pi j q)} 
}
    {\varepsilon_{\tilde{k}_3}^{-}+\varepsilon_{\tilde{k}_5}^{-}+\varepsilon_{\tilde{k}_6}^{-}-\varepsilon_{\tilde{n}_1}^{+} -\varepsilon_{\tilde{n}_2}^{+}-\varepsilon_{\tilde{n}_4}^{+}} 
\nonumber \\ 
&& \qquad  \quad \times \frac{
  \cY_{t}^{\tilde{k}_5} \delta_{t   k_5}^{(\pi j q)} \, \cY_{p}^{\tilde{k}_6}  \delta_{p   k_6}^{(\pi j q)}
    - (-1)^{j_{k_5}+j_{k_6}-J_{56}} \cY_{p}^{\tilde{k}_5} \delta_{p   k_5}^{(\pi j q)} \, \cY_{t}^{\tilde{k}_6}   \delta_{t   k_6}^{(\pi j q)}
    }{(1+\delta_{t p}) } 
     \, .
\end{eqnarray}

 \subsubsection{\label{2h1p} Coupling matrices with $2h1p$ ISCs}

The backward-in-time contributions to the self-energy involve ISCs with $2h1p$. The angular momentum representation for the ADC(2) coupling matrix, Eq.~(\ref{eq:N_2a}), is:
\begin{eqnarray}
\label{eq:N_2a_ANG}
 \textbf{N}^{ (\textrm{Ia})}_{a \tilde{s}} &\equiv& \Delta(j_{k_1},j_{k_2},J_{1 2}) \Delta(J_{1 2}, j_{n_3}, J_s)  \,
  (-1)^{j_\alpha - m_\alpha} (-1)^{j_\alpha - j_{n_3} - J_{1 2}} \,
  \frac{\hat{J}_{1 2}}{\hat{j}_\alpha} 
  \nonumber \\ 
&& \times \sum_{\substack{ m \leq v \\ l}}  
      \sqrt{1+\delta_{a l}} \bar{V}^{J_{1 2}}_{a l, m v}     \, 
   \frac{    \cY_{m}^{\tilde{k}_1} \delta_{m k_1}^{(\pi j q)} \,  \cY_{v}^{\tilde{k}_2}  \delta_{v k_2}^{(\pi j q)} 
      - (-1)^{j_{k_1}+j_{k_2}-J_{1 2}}  \, \cY_{v}^{\tilde{k}_1}  \delta_{v k_1}^{(\pi j q)} \, \cY_{m}^{\tilde{k}_2}  \delta_{m k_2}^{(\pi j q)}  }
   {\sqrt{1+\delta_{v m}} \sqrt{1+\delta_{\tilde{k}_1 \tilde{k}_2}} } \,    \cX_{l}^{\tilde{n}_3} \delta_{l n_3}^{(\pi j q)} \,  .
\end{eqnarray}
The ADC(3) ladder term coupling matrix of Eq.~(\ref{eq:N_3a}) becomes
\begin{eqnarray}
\label{eq:N_3a_ANG}
 \textbf{N}^{(\textrm{IIa})}_{a \tilde{s}} &\equiv & \Delta(j_{k_1},j_{k_2},J_{1 2}) \Delta(J_{1 2}, j_{n_3}, J_s)  \, 
 (-1)^{j_\alpha - m_\alpha}   \frac{ (-1)^{j_\alpha-j_{n_3}-J_{1 2}} }{\sqrt{1+\delta_{\tilde{k}_1 \tilde{k}_2}}} \, 
 \frac{\hat{J}_{1 2}}{\hat{j}_\alpha}  \, 
 \sum_{\substack{ \tilde{n}_{4} \leq \tilde{n}_{5} }} \sum_{\substack{ g \leq d \\ l}} \Delta(j_{n_4},j_{n_5},J_1)  \, 
 \sqrt{1 + \delta_{a l} } \bar{V}^{J_{1 2}}_{a l, g d} 
 \nonumber \\
&&
\qquad \times \cX_{l}^{\tilde{n}_3}   \delta_{l   n_3}^{(\pi j q)}  \, 
 \frac{  \left(
     \cX_{g}^{\tilde{n}_4} \delta_{g  n_4}^{(\pi j q)}  \,  \cX_{d}^{\tilde{n}_5}  \delta_{d   n_5}^{(\pi j q)}
     - (-1)^{j_{\gamma}+j_{\delta}-J_{1 2}} \,  \cX_{d}^{\tilde{n}_4} \delta_{d  n_4}^{(\pi j q)} \,  \cX_{g}^{\tilde{n}_5}  \delta_{g   n_5}^{(\pi j q)} 
     \right)^* }{\sqrt{1+\delta_{g d}} (1+\delta_{\tilde{n}_4 \tilde{n}_5}) } 
 t^{\tilde{n}_4 \tilde{n}_5, J_{12}}_{\tilde{k}_1 \tilde{k}_2}    \, ,
\end{eqnarray}
and the corresponding particle-hole channel, Eq.~(\ref{eq:N_3b}), is 
\begin{eqnarray}
\label{eq:N_3b_ANG}
 \textbf{N}^{(\textrm{IIb})}_{a \tilde{s}} &\equiv &  \Delta(j_{k_1},j_{k_2},J_{1 2}) \Delta(J_{1 2}, j_{n_3}, J_s) \, 
   (-1)^{j_\alpha - m_\alpha}    \frac { (-1)^{j_\alpha-j_{n_3}-J_{1 2}}  } {\sqrt{1+ \delta_{\tilde{k}_1 \tilde{k}_2}}}  \,
   \frac{\hat{J}_{1 2}}{\hat{j}_\alpha}  \, 
   \sum_{\substack{ \tilde{n}_{4}  \, \tilde{k}_{5} }} \, \sum_{\substack{  m \, v \\ l }} \,  \sum_{J_2 J_3} \, (2J_2+1) (2J_3+1) 
\nonumber \\
&& \times \left( (-1)^{-j_{k_1}-j_{k_2}+J_{1 2}}   \left\{ \begin{array}{@{\!~}c@{\!~}c@{\!~}c@{\!~}} 
j_{k_1} & j_{n_4} & J_2 \\[2mm] 
j_{k_2} & J_3 & j_{k_5}  \\[2mm] 
J_{1 2} &  j_{n_3}  & J_s 
\end{array}\right\} 
 \sqrt{1+ \delta_{a l}} \bar{V}^{J_2}_{a l, v  m}  \sqrt{1+ \delta_{v m}} \, 
(\cX_{v}^{\tilde{n}_4}\delta_{v  n_4}^{(\pi j q)}  \cY_{l}^{\tilde{k}_5} \delta_{l  k_5}^{(\pi j q)})^* \,  \cY_{m}^{\tilde{k}_1} \delta_{m k_1}^{(\pi j q)}  \,  
  t^{\tilde{n}_4 \tilde{n}_3, J_3}_{\tilde{k}_2 \tilde{k}_5} 
\right. \nonumber \\
&& \left. - 
  \left\{ \begin{array}{@{\!~}c@{\!~}c@{\!~}c@{\!~}} 
j_{k_2} & j_{n_4} & J_2 \\[2mm] 
j_{k_1} & J_3 & j_{k_5}  \\[2mm] 
J_{1 2} &  j_{n_3}  & J_s
\end{array}\right\} 
 \sqrt{1+ \delta_{a l}} \bar{V}^{J_2}_{a l, v  m}  \sqrt{1+ \delta_{v m}} \, 
(\cX_{v}^{\tilde{n}_4}\delta_{v  n_4}^{(\pi j q)}  \cY_{l}^{\tilde{k}_5} \delta_{l  k_5}^{(\pi j q)})^* \,  \cY_{m}^{\tilde{k}_2} \delta_{m k_2}^{(\pi j q)}  \,  
  t^{\tilde{n}_4 \tilde{n}_3, J_3}_{\tilde{k}_1 \tilde{k}_5} 
\right) \, ,
\end{eqnarray}
which is explicitly antisymmetrized with respect to the $\tilde{k}_1$, $\tilde{k}_2$ indices. 

Finally, the coupling matrix  $\textbf{N}^{(\textrm{IIc})}_{\alpha s}$ of Eq.~(\ref{eq:N_3c}) is found in the backward-in-time diagram of Fig.~\ref{3ord_c} and contains a 3NF. It has the following form in the angular momentum coupling representation:
\begin{eqnarray}
\label{eq:N_3c_ANG}
 \textbf{N}^{(\textrm{IIc})}_{a \tilde{s}} &\equiv & 
 \Delta(j_{k_1},j_{k_2},J_{1 2}) \Delta(J_{1 2},j_{n_3},J_s)  \, 
 \sum_{J_{4 5} J'} \sum_{\substack{ \tilde{n}_{4} \leq \tilde{n}_{5} \\ \tilde{k}_{6} }} \sum_{\substack{ v \leq m \\ l}} \, 
  \Delta(j_{n_5},j_{n_6},J_{5 6}) \Delta(J_{5 6},j_{k_4}, j_\alpha) 
\nonumber \\ 
&& \quad  \times     (-1)^{ j_\alpha - m_\alpha}  \frac{(-1)^{ j_\alpha + 2 j_{n_3} + j_{k_4} - J_{56} + 2 J'} }{ \sqrt{1+\delta_{\tilde{k}_1 \tilde{k}_2}} }
    \frac{\hat{J}_{1 2}}{\hat{j}_{\alpha}} 
   (2J'+1) 
 \left\{ \begin{array}{@{\!~}c@{\!~}c@{\!~}c@{\!~}} 
 j_{n_3} & J_{4 5} & J' \\[2mm] 
 j_{k_4} & J_{1 2}  &  j_\alpha  
\end{array}\right\}  \,
 \sqrt{1+\delta_{a l}}  \bar{V}^{J_{5 6}}_{a l, m v}  
 \nonumber \\ 
&& \qquad  \times 
    \frac{ \left(
      \cX_{m}^{\tilde{n}_5} \delta_{m  n_5}^{(\pi j q)}    \,  \cX_{v}^{\tilde{n}_6}  \delta_{v  n_6}^{(\pi j q)}
        - (-1)^{j_{n_5}+j_{n_6}-J_{5 6}}  \, \cX_{v}^{\tilde{n}_5} \delta_{v  n_5}^{(\pi j q)}  \,  \cX_{m}^{\tilde{n}_6}  \delta_{m  n_6}^{(\pi j q)}  \right)^*  
       }{(1+\delta_{\tilde{n}_5 \tilde{n}_6}) \, \sqrt{1+\delta_{m v}}   }  
  \left( \cY_{l}^{\tilde{k}_4} \delta_{l k_4}^{(\pi j q)}  \right)^*  \,
  t^{\tilde{n}_5 \tilde{n}_6 \tilde{n}_3, J_{5 6} J'}_{\tilde{k}_1 \tilde{k}_2 \tilde{k}_4, J_{1 2}} \,  .  \qquad
\end{eqnarray}

 \subsubsection{\label{2p1h_2h1p_int} Interaction matrices with $2p1h$ and $2h1p$ ISCs}
 
 The interaction matrix $\textbf{C}_{\substack{\tilde{r}  \tilde{r}'}}$  can connect $2p1h$ propagators through particle-particle, particle-hole and 3NFs, according to the terms
  \begin{equation}
 \label{C_123_456_bis}
  \textbf{C}_{\substack{\tilde{r}  \tilde{r}'}} \equiv  \textbf{C}^{pp}_{\substack{\tilde{r}  \tilde{r}'}} + \textbf{C}^{ph}_{\substack{\tilde{r}  \tilde{r}'}}  + \textbf{C}^{3N}_{\substack{\tilde{r}  \tilde{r}'}} \, ,
 \end{equation}
which have been introduced in Eqs.~(\ref{int_term_5a_C}),~(\ref{int_term_5b_C}), and~(\ref{int_term_5c_C}), respectively.

The particle-particle interaction matrix results from the diagram in Fig.~\ref{3ord_a}. Using the coupling convention of Eq.~(\ref{choice_Cppk}), we have:
\begin{eqnarray}
\label{eq:C_5a_ANG}
\textbf{C}^{pp}_{\substack{\tilde{r}  \tilde{r}'}} &\equiv & 
 \Delta(j_{n_1},j_{n_2},J_{1 2}) \Delta(J_{1 2},j_{k_3}, J_r) \Delta(j_{n_4},j_{n_5},J_{1 2}) \Delta(J_{1 2}, j_{k_6}, J_{r})  \,  
 \nonumber \\
 && \quad \times  \delta_{J_{12} J_{45}} \delta_{\tilde{k}_3 \tilde{k}_6}   \sum_{\substack{ m \leq v \\  l \leq p}}   \left.
\frac{\cX_{m}^{\tilde{n}_1} \delta_{m  n_1}^{(\pi j q)} \,  \cX_{v}^{\tilde{n}_2}  \delta_{v    n_2}^{(\pi j q)} 
    - (-1)^{j_{n_1}+j_{n_2}-J_{1 2}} \cX_{v}^{\tilde{n}_1} \delta_{v  n_1}^{(\pi j q)} \,  \cX_{m}^{\tilde{n}_2} \delta_{m    n_2}^{(\pi j q)}  }
    {  \sqrt{1+\delta_{\tilde{n}_1 \tilde{n}_2} }   {\sqrt{1+\delta_{m v}}  }  }  \right.
   \nonumber \\ 
&& \qquad \qquad \qquad \times  \bar{V}^{J_{1 2}}_{m v, l p}  \frac{  \left(
  \cX_{l}^{\tilde{n}_4}  \delta_{l   n_4}^{(\pi j q)}  \,  \cX_{p}^{\tilde{n}_5}  \delta_{p   n_5}^{(\pi j q)} 
     - (-1)^{j_{n_4}+j_{n_5}-J_{1 2}} \,  \cX_{p}^{\tilde{n}_4}  \delta_{p  n_4}^{(\pi j q)}  \, \cX_{l}^{\tilde{n}_5}  \delta_{l   n_5}^{(\pi j q)} 
     \right)^*  }{\sqrt{1+\delta_{l p}} \sqrt{1+\delta_{\tilde{n}_4 \tilde{n}_5}} }    \, . \qquad\qquad
\end{eqnarray}
The particle-hole $\textbf{C}^{ph}_{\substack{\tilde{r}  \tilde{r}'}} $ comes from the ring diagram in Fig.~\ref{3ord_b}, which contains four terms owing to the antisymmetrization specified in Eq.~(\ref{int_term_5b_C}),
\begin{eqnarray}
\label{eq:C_5b_ANG}
\textbf{C}^{ph}_{\substack{\tilde{r}  \tilde{r}'}} &\equiv & \Delta(j_{n_1},j_{n_2},J_{1 2}) \Delta(J_{1 2},j_{k_3}, J_r) \Delta(j_{n_4},j_{n_5},J_{4 5}) \Delta(J_{4 5}, j_{k_6}, J_r) \frac{1 }{2} \sum_{\substack{ m \, v \\  \, p \; l}} 
 \sum_J  \frac{ \hat{J}_{1 2} \; \hat{J}_{4 5} \; (2J+1)}{ \sqrt{1 + \delta_{\tilde{n}_1 \tilde{n}_2}} \, \sqrt{1 + \delta_{\tilde{n}_4 \tilde{n}_5}} }
   \nonumber \\ 
&&\times \Bigg( 
\left\{ \begin{array}{@{\!~}c@{\!~}c@{\!~}c@{\!~}} 
 j_{n_1}  & j_{n_2}   &  J_{1 2}   \\[2mm]
 j_{n_5}  &    J          &  j_{k_3}   \\[2mm] 
 J_{4 5}  & j_{k_6}    &  J_{r}
\end{array}\right\} 
\cX_{v}^{\tilde{n}_2} \delta_{v  n_2}^{(\pi j q)}  \,  \cY_{p}^{\tilde{k}_6}   \delta_{p    k_6}^{(\pi j q)}  
\sqrt{1 + \delta_{m v} } V^{J}_{m v, p l} \sqrt{1 + \delta_{p l} }
 \left( \cX_{l}^{\tilde{n}_5}  \delta_{l n_5}^{(\pi j q)} \,  \cY_{m}^{\tilde{k}_3}   \delta_{m   k_3}^{(\pi j q)} \right)^*   
\delta_{\tilde{n}_1 \tilde{n}_4}  
\Bigg. \nonumber \\ 
&& \Bigg. -(-1)^{j_{n_1}+j_{n_2}-J_{1 2}}
\left\{ \begin{array}{@{\!~}c@{\!~}c@{\!~}c@{\!~}} 
 j_{n_2}  & j_{n_1}   &  J_{1 2}   \\[2mm]
 j_{n_5}  &    J          &  j_{k_3}   \\[2mm] 
 J_{4 5}  & j_{k_6}    &  J_{r}
\end{array}\right\} 
\cX_{v}^{\tilde{n}_1} \delta_{v  n_1}^{(\pi j q)}  \,  \cY_{p}^{\tilde{k}_6}   \delta_{p    k_6}^{(\pi j q)}  
\sqrt{1 + \delta_{m v} } V^{J}_{m v, p l} \sqrt{1 + \delta_{p l} }
 \left( \cX_{l}^{\tilde{n}_5}  \delta_{l n_5}^{(\pi j q)} \,  \cY_{m}^{\tilde{k}_3}   \delta_{m   k_3}^{(\pi j q)} \right)^*   
\delta_{\tilde{n}_2 \tilde{n}_4}  
\Bigg. \nonumber \\ 
&& -(-1)^{j_{n_4}+j_{n_5}-J_{4 5}}
\left\{ \begin{array}{@{\!~}c@{\!~}c@{\!~}c@{\!~}} 
 j_{n_1}  & j_{n_2}   &  J_{1 2}   \\[2mm]
 j_{n_4}  &    J          &  j_{k_3}   \\[2mm] 
 J_{4 5}  & j_{k_6}    &  J_{r}
\end{array}\right\} 
\cX_{v}^{\tilde{n}_2} \delta_{v  n_2}^{(\pi j q)}  \,  \cY_{p}^{\tilde{k}_6}   \delta_{p    k_6}^{(\pi j q)}  
\sqrt{1 + \delta_{m v} } V^{J}_{m v, p l} \sqrt{1 + \delta_{p l} }
 \left( \cX_{l}^{\tilde{n}_4}  \delta_{l n_4}^{(\pi j q)} \,  \cY_{m}^{\tilde{k}_3}   \delta_{m   k_3}^{(\pi j q)} \right)^*   
\delta_{\tilde{n}_1 \tilde{n}_5}  
\Bigg. \nonumber \\ 
&&  +(-1)^{j_{n_1}+j_{n_2}-J_{1 2}} (-1)^{j_{n_4}+j_{n_5}-J_{4 5}} \nonumber \\
&& \Bigg. \qquad \quad \times
\left\{ \begin{array}{@{\!~}c@{\!~}c@{\!~}c@{\!~}} 
 j_{n_2}  & j_{n_1}   &  J_{1 2}   \\[2mm]
 j_{n_4}  &    J          &  j_{k_3}   \\[2mm] 
 J_{4 5}  & j_{k_6}    &  J_{r}
\end{array}\right\} 
\cX_{v}^{\tilde{n}_1} \delta_{v  n_1}^{(\pi j q)}  \,  \cY_{p}^{\tilde{k}_6}   \delta_{p    k_6}^{(\pi j q)}  
\sqrt{1 + \delta_{m v} } V^{J}_{m v, p l} \sqrt{1 + \delta_{p l} }
 \left( \cX_{l}^{\tilde{n}_4}  \delta_{l n_4}^{(\pi j q)} \,  \cY_{m}^{\tilde{k}_3}   \delta_{m   k_3}^{(\pi j q)} \right)^*   
\delta_{\tilde{n}_2 \tilde{n}_5}  
\Bigg).
\end{eqnarray}
The 3NF interaction matrix in Eq.~(\ref{C_123_456_bis}) reads,
\begin{eqnarray}
\label{eq:C_5c_ANG_3N}
\textbf{C}^{3N}_{\substack{\tilde{r}  \tilde{r}'}} &\equiv & 
 \Delta(j_{n_1},j_{n_2},J_{1 2}) \Delta(J_{1 2},j_{k_3}, J_r)  \Delta(j_{n_4},j_{n_5}, J_{4 5}) \Delta(J_{4 5}, j_{k_6}, J_r)  
 \nonumber \\ 
&& \times
 \sum_{\substack{ m \leq v \\ l }}   \sum_{\substack{ g \leq d \\ p}} 
\frac{ 
   \cX_{m}^{\tilde{n}_1}\delta_{m  n_1}^{(\pi j q)} \,  \cX_{v}^{\tilde{n}_2}  \delta_{v    n_2}^{(\pi j q)} 
   - (-1)^{j_{n_1}+j_{n_2}-J_{1 2}} \cX_{v}^{\tilde{n}_1} \delta_{v  n_1}^{(\pi j q)} \,  \cX_{m}^{\tilde{n}_2}   \delta_{m    n_2}^{(\pi j q)}
   }{ \sqrt{1+\delta_{\tilde{n}_1 \tilde{n}_2}}  (1+\delta_{m v})} 
   \cY_l^{\tilde{k}_3} \delta_{l  k_3}^{(\pi j q)} \,
    \nonumber \\
 && \quad \times
       W^{(pph) \, J_{1 2} J_{4 5} J'}_{m v l^{-1}, g d p^{-1}} 
\frac{
   (\cX_{g}^{\tilde{n}_4} \delta_{g   n_4}^{(\pi j q)} \,  \cX_{d}^{\tilde{n}_5}  \delta_{d   n_5}^{(\pi j q)}
    - (-1)^{j_{n_4}+j_{n_5}-J_{4 5}} \cX_{d}^{\tilde{n}_4} \delta_{d   n_4}^{(\pi j q)} \,  \cX_{g}^{\tilde{n}_5}  \delta_{g   n_5}^{(\pi j q)} )^*  
    }{ (1+\delta_{g d}) \sqrt{1+\delta_{\tilde{n}_4 \tilde{n}_5}}  } 
   (\cY_{p}^{\tilde{k}_6} \delta_{p   k_6}^{(\pi j q)})^*     \, ,
\end{eqnarray}
where we have the 3NF coupled in the $pph$ channel from Eq.~\eqref{V_3B_Jpph}.

The backward-in-time interaction matrix $\textbf{D}_{\substack{\tilde{s}  \tilde{s}'}}$  can connect the $2h1p$ propagators through hole-hole, hole-particle and backward-in-time 3NFs, according to,
 
  \begin{equation}
 \label{D_123_456_bis}
  \textbf{D}_{\substack{\tilde{s}  \tilde{s}'}} \equiv  \textbf{D}^{hh}_{\substack{\tilde{s}  \tilde{s}'}} + \textbf{D}^{hp}_{\substack{\tilde{s}  \tilde{s}'}}  + \textbf{D}^{3N}_{\substack{\tilde{s}  \tilde{s}'}} \, ,
 \end{equation}
with the three matrices on the right-hand side introduced in Eqs.~(\ref{int_term_5a_D}),~(\ref{int_term_5b_D}), and~(\ref{int_term_5c_D}), respectively.

The hole-hole interaction matrix resulting from the diagram in Fig.~\ref{3ord_a} reads,
\begin{eqnarray}
\label{eq:D_5a_ANG}
\textbf{D}^{hh}_{\substack{\tilde{s}   \tilde{s}'}} &\equiv & 
- \Delta(j_{k_1},j_{k_2},J_{1 2}) \Delta(J_{1 2}, j_{n_3}, J_s) \Delta(j_{k_4},j_{k_5},J_{1 2}) \Delta(J_{1 2}, j_{n_6}, J_{s})
\nonumber \\
& & \quad \times    \delta_{J_{12} J_{45}}   \delta_{\tilde{n}_3 \tilde{n}_6} \, 
\sum_{\substack{ m \leq v \\  l \leq p}}  
 \frac{  \left(
     \cY_{m}^{\tilde{k}_1} \delta_{m  k_1}^{(\pi j q)} \,  \cY_{v}^{\tilde{k}_2} \delta_{v    k_2}^{(\pi j q)}
      - (-1)^{j_{k_1}+j_{k_2}-J_{1 2}} \cY_{v}^{\tilde{k}_1} \delta_{v  k_1}^{(\pi j q) } \,  \cY_{m}^{\tilde{k}_2} \delta_{n    k_2}^{(\pi j q)} 
      \right)^* }{\sqrt{1+\delta_{g d}}  \sqrt{1+\delta_{\tilde{k}_1 \tilde{k}_2}}  } 
  \bar{V}^{J_{1 2}}_{m v, l p} \nonumber \\ 
&&   \qquad \qquad \qquad \times  
  \frac{
  \cY_{l}^{\tilde{k}_4}  \delta_{l   k_4}^{(\pi j q)} \,  \cY_{p}^{\tilde{k}_5}    \delta_{p   k_5}^{(\pi j q)} 
    - (-1)^{j_{k_4}+j_{k_5}-J_{1 2}} \cY_{p}^{\tilde{k}_4}  \delta_{p   k_4}^{(\pi j q)} \,   \cY_{l}^{\tilde{k}_5}    \delta_{l   k_5}^{(\pi j q)} 
     }{\sqrt{1+\delta_{t p}} \sqrt{1+\delta_{\tilde{k}_4 \tilde{k}_5}} }   \, ,
\end{eqnarray}
while $\textbf{D}^{hp}_{\substack{\tilde{s}  \tilde{s}'}} $ results from the ring diagram in Fig.~\ref{3ord_b} and contains four different terms owing to the antisymmetrization specified in Eq.~(\ref{int_term_5b_D}):
\begin{eqnarray}
\label{eq:D_5b_ANG}
\textbf{D}^{hp}_{\substack{\tilde{s}  \tilde{s}'}} &\equiv & 
 \Delta(j_{k_1},j_{k_2},J_{1 2}) \Delta(J_{1 2}, j_{n_3}, J_s) \Delta(j_{k_4},j_{k_5},J_{4 5}) \Delta(J_{4 5}, j_{n_6}, J_s) 
   \frac{1}{2}  \sum_{\substack{ g  d \\  t  p}} \sum_J 
   (-1)^{2j_{k_6}-2j_{k_3}} 
   \frac{  \hat{J}_{1 2} \, \hat{J}_{4 5} \, (2J+1) }{ \sqrt{1+\delta_{\tilde{k}_1 \tilde{k}_2}}  \sqrt{1+\delta_{\tilde{k}_4 \tilde{k}_5}} } 
    \nonumber \\ 
&& \times \Bigg(  \left\{ \begin{array}{@{\!~}c@{\!~}c@{\!~}c@{\!~}} 
  j_{k_1}  & j_{k_2}  &  J_{1 2} \\[2mm] 
  j_{k_5} & J   &  j_{n_3}  \\[2mm] 
 J_{4 5} & j_{n_6}  & J_{s}
\end{array}\right\} 
  \left( \cY_{m}^{\tilde{k}_2}   \delta_{m   k_2}^{(\pi j q)} \,  \cX_{p}^{\tilde{n}_3}  \delta_{p  n_3}^{(\pi j q)} \right)^* 
    \sqrt{1+\delta_{m v}} V^{J}_{v m, p l}  \sqrt{1+\delta_{p l}}   \, 
         \cY_{l}^{\tilde{k}_5}  \delta_{l    k_5}^{(\pi j q)} \,   \cX_{v}^{\tilde{n}_6} \delta_{v   n_6}^{(\pi j q)}   \, \delta_{\tilde{k}_1 \tilde{k}_4} 
           \Bigg. \nonumber \\ 
 & & - (-1)^{j_{k_1}+j_{k_2}-J_{1 2}}
\left\{ \begin{array}{@{\!~}c@{\!~}c@{\!~}c@{\!~}} 
  j_{k_2}  & j_{k_1}  &  J_{1 2} \\[2mm] 
  j_{k_5} & J   &  j_{n_3}  \\[2mm] 
 J_{4 5} & j_{n_6}  & J_{s}
\end{array}\right\} 
  \left( \cY_{m}^{\tilde{k}_1}   \delta_{m   k_1}^{(\pi j q)} \,  \cX_{p}^{\tilde{n}_3}  \delta_{p  n_3}^{(\pi j q)} \right)^* 
    \sqrt{1+\delta_{m v}} V^{J}_{v m, p l}  \sqrt{1+\delta_{p l}}   \, 
         \cY_{l}^{\tilde{k}_5}  \delta_{l    k_5}^{(\pi j q)} \,   \cX_{v}^{\tilde{n}_6} \delta_{v   n_6}^{(\pi j q)}   \, \delta_{\tilde{k}_2 \tilde{k}_4} 
            \nonumber \\ 
& & - (-1)^{j_{k_4}+j_{k_5}-J_{4 5}}
\left\{ \begin{array}{@{\!~}c@{\!~}c@{\!~}c@{\!~}} 
  j_{k_1}  & j_{k_2}  &  J_{1 2} \\[2mm] 
  j_{k_4} & J   &  j_{n_3}  \\[2mm] 
 J_{4 5} & j_{n_6}  & J_{s}
\end{array}\right\} 
  \left( \cY_{m}^{\tilde{k}_2}   \delta_{m   k_2}^{(\pi j q)} \,  \cX_{p}^{\tilde{n}_3}  \delta_{p  n_3}^{(\pi j q)} \right)^* 
    \sqrt{1+\delta_{m v}} V^{J}_{v m, p l}  \sqrt{1+\delta_{p l}}   \, 
         \cY_{l}^{\tilde{k}_4}  \delta_{l    k_4}^{(\pi j q)} \,   \cX_{v}^{\tilde{n}_6} \delta_{v   n_6}^{(\pi j q)}   \, \delta_{\tilde{k}_1 \tilde{k}_5} 
            \nonumber \\ 
 && +  (-1)^{j_{k_1}+j_{k_2}-J_{1 2} } (-1)^{j_{k_4}+j_{k_5}-J_{4 5}}  \nonumber \\
&& \Bigg. \qquad \quad \times
\left\{ \begin{array}{@{\!~}c@{\!~}c@{\!~}c@{\!~}} 
  j_{k_2}  & j_{k_1}  &  J_{1 2} \\[2mm] 
  j_{k_4} & J   &  j_{n_3}  \\[2mm] 
 J_{4 5} & j_{n_6}  & J_{s}
\end{array}\right\} 
  \left( \cY_{m}^{\tilde{k}_1}   \delta_{m   k_1}^{(\pi j q)} \,  \cX_{p}^{\tilde{n}_3}  \delta_{p  n_3}^{(\pi j q)} \right)^* 
    \sqrt{1+\delta_{m v}} V^{J}_{v m, p l}  \sqrt{1+\delta_{p l}}   \, 
         \cY_{l}^{\tilde{k}_4}  \delta_{l    k_4}^{(\pi j q)} \,   \cX_{v}^{\tilde{n}_6} \delta_{v   n_6}^{(\pi j q)}   \, \delta_{\tilde{k}_2 \tilde{k}_5} 
  \Bigg) \, .
\end{eqnarray}
Finally, the backward-in-time 3NF interaction matrix in Eq.~(\ref{D_123_456_bis}) is given by
\begin{eqnarray}
\label{eq:D_5c_ANG_3N}
\textbf{D}^{3N}_{\substack{\tilde{s} \tilde{s}'}} &\equiv & - 
  \Delta(j_{k_1},j_{k_2},J_{1 2}) \Delta(J_{1 2}, j_{n_3}, J_s) \Delta(j_{k_4},j_{k_5},J_{4 5}) \Delta(J_{4 5}, j_{n_6}, J_s) 
\nonumber \\
& & \times  \sum_{\substack{ m \leq v \\ l }}   \sum_{\substack{ g \leq d \\ p}} 
\frac{
     (\cY_{m}^{\tilde{k}_1} \delta_{m  k_1}^{(\pi j q)} \,  \cY_{v}^{\tilde{k}_2} \delta_{v k_2}^{(\pi j q)}
     - (-1)^{j_{k_1}+j_{k_2}-J_{1 2}} \cY_{v}^{\tilde{k}_1} \delta_{v  k_1}^{(\pi j q)} \,  \cY_{m}^{\tilde{k}_2}  \delta_{m k_2}^{(\pi j q)} )^* 
     }{\sqrt{1+\delta_{\tilde{k}_1 \tilde{k}_2}} (1+ \delta_{g d}) }
     (\cX_l^{\tilde{n}_3} \delta_{l  n_3}^{(\pi j q)}  )^*
\nonumber \\
& & \qquad \times
 W^{(pph) \, J_{1 2} J_{4 5} J'}_{m v l^{-1},g d p^{-1}}
\frac{
    \cY_{g}^{\tilde{k}_4}    \delta_{g   k_4}^{(\pi j q)} \, \cY_{d}^{\tilde{k}_5} \delta_{d   k_5}^{(\pi j q)}
    - (-1)^{j_{k_4}+j_{k_5}-J_{4 5}} \cY_{d}^{\tilde{k}_4}   \delta_{g   k_4}^{(\pi j q)} \,  \cY_{g}^{\tilde{k}_5}   \delta_{g   k_5}^{(\pi j q)}
      }{ \sqrt{1+\delta_{\tilde{k}_4 \tilde{k}_5}} (1+ \delta_{ t p})}
       \cX_{p}^{\tilde{n}_6} \delta_{p   n_6}^{(\pi j q)}   \, .
\end{eqnarray}

\end{widetext}
\section{\label{General_reference_state} Self-energy without renormalization of the propagators}

This Appendix  discusses how the ADC(3)  equations  need to be modified when one
releases the assumption of full self-consistency.
This is the case for most of applications in quantum chemistry and also for state-of-the-art nuclear
structure studies,  where fully dressed propagators become too complex to be able to expand
the dynamic self-energy~$\widetilde{\Sigma}_{\alpha\beta}(\omega)$. In the latter case, one is forced
to implement self-consistency only at the level of the static self-energy, Eq.~\eqref{SE_eq_U}, 
while  Eqs.~(\ref{ADC3_somm_final})-(\ref{D_final}) that generate the dynamic part are based on
an uncorrelated (bare) propagator (this is the so--called `$sc0$' approximation introduced and discussed
in Ref.~\cite{Soma2014}).
Without self-consistency, one needs to follow the standard perturbation approach and to expand
the self-energy in terms of reference mean-field propagators. This means that nonskeleton
diagrams also need to be added to the perturbative expansion that is used to constrain
the ADC($n$) interaction and coupling matrices.   
%
%
For calculations up to order $n$=3, there are substantially two consequences. 
First, Eq.~\eqref{ueff_bis} for the static self-energy must be re-expressed in terms of the reference propagator as shown in Appendix~\ref{Static_SE}.
One does this by expanding both the (correlated) 1B and 2B density matrices, $\rho_{\alpha\beta}$ and $\Gamma_{\alpha\beta,\gamma\delta}$, with the inclusion of nonskeleton terms.  
The three terms of Eq.~\eqref{ueff_bis} still generate the skeleton diagrams of Fig.~\ref{Utilda_1} but now there are 16 additional higher order contributions in terms of the effective forces~$\widetilde U$ and~$\widetilde V$, as shown in Figs.~\ref{Utilda_2_V} and~\ref{Utilda_3} below.  
Note that a few of these diagrams are of skeleton type and they should be included \emph{also} when $\widetilde{U}$ is calculated self-consistently. They result from the skeleton expansion of the 2B Green's function and density matrix (see Eqs.~\eqref{ueff_bis} and~\eqref{2B_densitymatrix}) and will be identified further below.
Second, the dynamic self-energy receives four third-order nonskeleton diagrams that are obtained by inserting the (first-order term of the) 1B operator into the uncorrelated fermionic lines that form the diagrams of the  dynamic self-energy  at second order. These are derived in Section~\ref{Dynamic_SE} and they generate additional contributions to the  ADC(3) equations. It is useful to note that these additional terms cancel exactly when one chooses a Hartree-Fock (HF) state as the reference (but not in the general case). 

In the following, we will consider the expansion with respect to the uncorrelated propagator $g^{(0)}(\omega)$ that is associated with a mean-field reference state~$|\phi_0^A\rangle$. Hence, we  redefine  the transition amplitudes of the unperturbed ($A\pm 1$)-body systems, denoted by $|\phi^{(A\pm 1)}\rangle$, as follows:
\begin{equation}
\label{tran_ampl_ref}
Z_{\alpha}^{i=n,k} \equiv
\begin{cases}
\left( X^n_{\alpha}\right)^* \equiv \langle\phi_0^A |a_{\alpha}|\phi_n^{A+1}\rangle \\
Y^k_{\alpha}\equiv\langle\phi_k^{A-1}|a_{\alpha}|\phi_0^A\rangle \, ,
\end{cases}
\end{equation}
which build the reference propagator similarly to Eq.~\eqref{eq:g1}.  In general, to obtain the ADC(3) approximation for such reference state, one only needs to substitute the amplitudes from Eq.~\eqref{tran_ampl} in the expressions for Eqs.~(\ref{ADC3_somm_final})-(\ref{D_final}) with the corresponding ones from Eq.~\eqref{tran_ampl_ref}.
  Note that, differently from the true Dyson eigenstates of Eq.~\eqref{tran_ampl}, the orbitals~$n$ and~$k$ of Eq.~\eqref{tran_ampl_ref} form a complete orthonormal set. Thus, the great advantage in using a mean-field reference state is that the numbers of particle and hole states is drastically reduced and, in fact, tractable.  In the following we will consider the most general case in which these orbits are different from the model space basis $\{\alpha\}$, then the $Z_{\alpha}^i$ give the unitary transformation between the two sets. In standard applications of perturbation theory  it is customary to identify the basis $\{\alpha\}$ with the unperturbed orbits of the reference state. One can always reduce to this particular case by substituting $X^n_{\alpha}\rightarrow\delta_{n  \alpha}$ and~$Y^k_{\alpha}\rightarrow\delta_{k  \alpha}$,  which recovers the expressions  reported in Refs.~\cite{Schirmer1982,Schirmer1983}.

\subsection{\label{Static_SE} Static self-energy}

\begin{figure}[t!]
\vspace{0.5cm}
  \centering
    \subfloat{\includegraphics[scale=0.47]{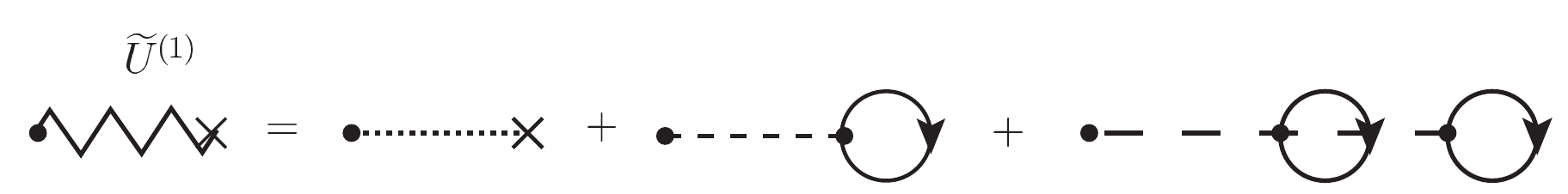}}
  \caption{ Diagrammatic representation of the first-order part $\widetilde U^{(1)}$ (zigzag line) of the effective 1B interaction of Eq.~(\ref{ueff}). Fermionic lines here denote  uncorrelated propagators. Dotted lines denote the 1B potential $U_{\alpha \beta}$, while short (long) dashed lines denote 2N (3N) interactions.
   This is the fist term of the expansion of Eq.~\eqref{U_exp} and it is given in full by Eq.~\eqref{eq:U_1st_me}. }
  \label{Utilda_1}
\end{figure}

\begin{figure}[t]
  \centering
    \subfloat{\includegraphics[scale=0.55]{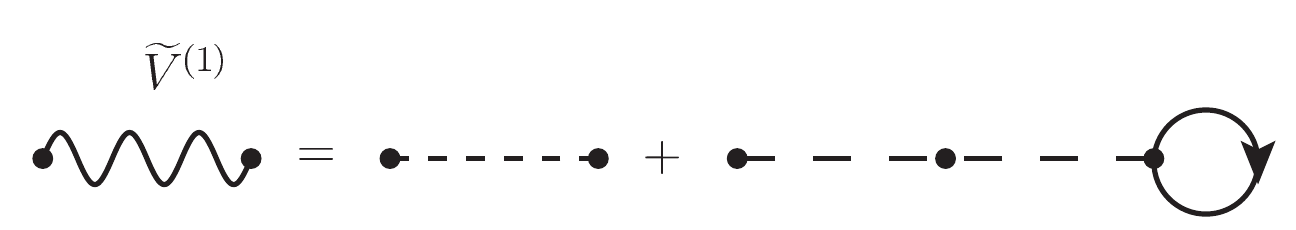}}
  \caption{ Diagrammatic representation of $\widetilde V^{(1)}$ given by  Eq.~(\ref{eq:V_1st_me}). This is the first-order term in the expansion of Eq.~(\ref{V_exp}) of the effective 2N  interaction.}
  \label{Vtilda_1}
\end{figure}

 Equation~(\ref{ueff_bis}) for the 1B effective interaction is exact and it is given in terms of correlated 1B and 2B propagators: for this reason, practical calculations (such as the above mentioned \emph{sc0} approach) may follow an iterative procedure to \enquote{dress} the propagator and evaluate self-consistently the first-order static irreducible self-energy. Alternatively, one can  consider  the explicit expansion of $\Sigma_{\alpha\beta}^{\infty}$  in terms of uncorrelated propagators. This should be done up to the same order $n$ that matches the ADC($n$) truncation for~$\widetilde{\Sigma}_{\alpha\beta}(\omega)$. Thus, we expand the effective interaction $\widetilde U$ as follows:
\begin{equation}
\label{U_exp}
 \widetilde U = \sum_{\alpha \beta}  \Big( \widetilde U^{(1)}_{\alpha \beta}  +  \widetilde U^{(2)}_{\alpha \beta}    +   \widetilde U^{(3)}_{\alpha \beta}   + \dots \Big) a_\alpha^\dagger a_{\beta} \, .
\end{equation} 

In a similar fashion, one considers the expansion (up to second order) of $\widetilde V$, that is
\begin{equation}
\label{V_exp}
 \widetilde V =  \sum_{\alpha \beta,  \gamma\delta}  \Big(  \widetilde V^{(1)}_{\alpha \beta,  \gamma\delta}  +  \widetilde V^{(2)}_{\alpha \beta,  \gamma\delta}    + \dots \Big) a_\alpha^\dagger a_\beta^\dagger a_ \delta a_\gamma \, .
\end{equation} 


The first-order term in Eq.~(\ref{U_exp}) is given by the expression in Eq.~(\ref{ueff_bis}) once correlated propagators are substituted with bare ones. It is composed by the three diagrams represented in Fig.~\ref{Utilda_1} and can be written in terms of  one-body  reduced density matrix, $\rho^{(0)}_{\alpha\beta}=-i\hbar g^{(0)}_{\alpha\beta}(t-t^+)$,  which is the  uncorrelated version  of Eq.~(\ref{1B_densitymatrix}).
We have
\begin{equation}
\label{eq:U_1st_me}
  \widetilde U^{(1)}_{\alpha \beta} = - U_{\alpha\beta} 
+ \sum_{\gamma\delta}V_{\alpha\gamma,\beta\delta} \, \rho^{(0)}_{\delta\gamma} 
+ \frac{1}{4} \sum_{\substack{\gamma\delta \\ \epsilon\eta}} W_{\alpha\gamma\epsilon,\beta\delta\eta}
\, \rho^{(0)}_{\delta \gamma} \rho^{(0)}_{\eta \epsilon} \,.
\end{equation}
Similarly, the explicit expression for the matrix element $\widetilde V^{(1)}_{\alpha \beta,  \gamma\delta}$ is depicted in Fig.~\ref{Vtilda_1}. It can be directly read from Eq.~(\ref{veff}) once the correlated fermionic loop is substituted with an uncorrelated one:
\begin{equation}
\label{eq:V_1st_me}
  \widetilde V^{(1)}_{\alpha \beta,\gamma\delta} = V_{\alpha\beta,\gamma\delta} 
+  \sum_{\epsilon \eta} W_{\alpha\beta\epsilon,\gamma\delta \eta}
\, \rho^{(0)}_{\eta , \epsilon}  \,.
\end{equation}

\begin{widetext}

\begin{figure}
  \centering
    \subfloat{\includegraphics[scale=0.6]{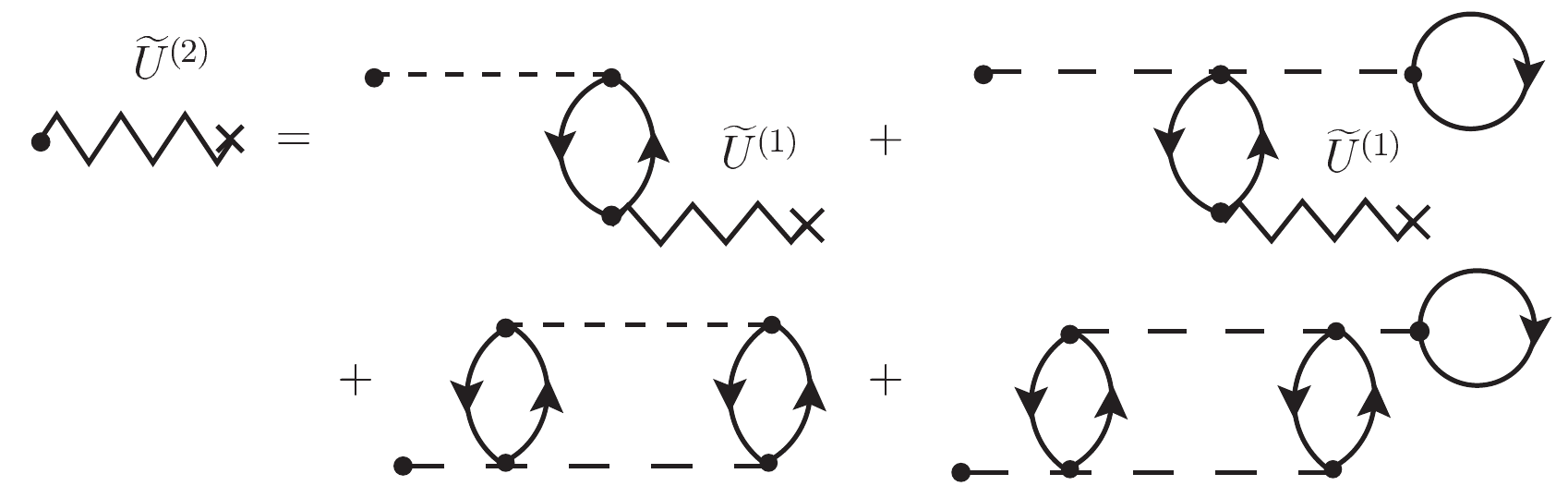}}
  \caption{ As described in the caption of Fig.~\ref{Utilda_1} but for the second-order term $\widetilde U^{(2)}$.}
  \label{Utilda_2}
\end{figure}

The second-order term $ \widetilde U^{(2)}$ is composed by eight different Feynman diagrams, that can be grouped into four by using the effective interaction $ \widetilde U^{(1)}$:
\begin{eqnarray}
\label{eq:U_2nd_me}
  \widetilde U^{(2)}_{\alpha  \beta } &=& 
-i \hbar \! \int \!\frac{d \omega}{2 \pi} \sum_{\substack{\epsilon \eta \\ \gamma \delta }} \left[V_{\alpha\gamma,\beta\delta}
+\sum_{\epsilon \eta}W_{\alpha \gamma \epsilon,\beta \delta \eta} \, \rho^{(0)}_{\eta \epsilon} \right]  g^{(0)}_{\delta \epsilon}(\omega) g^{(0)}_{\eta \gamma}(\omega) \widetilde  U^{(1)}_{\epsilon \eta}   \nonumber \\
&& +\frac{(i\hbar)^3}{4} \! \! \int  \! \frac{\mathrm d \omega_1}{2 \pi} \! \int  \! \frac{\mathrm d \omega_2}{2 \pi}  \! \int  \! \frac{\mathrm d \omega_3}{2 \pi} \! \sum_{\substack{\epsilon \eta \nu \delta \\ \gamma \lambda \mu \rho} } \! W_{\alpha\gamma \epsilon,\beta\delta \eta} \, g^{(0)}_{\nu \gamma}(\omega_1) g^{(0)}_{\delta \rho}(\omega_2)  g^{(0)}_{\lambda \epsilon}(\omega_3) g^{(0)}_{\eta \mu}(\omega_1+\omega_3-\omega_2) \left[V_{\rho \mu, \nu \lambda}
+\sum_{\epsilon \eta}W_{\rho \mu \epsilon, \nu \lambda \eta} \, \rho^{(0)}_{\eta \epsilon} \right]\,, \nonumber \\
\end{eqnarray}
with the corresponding diagrams displayed in Fig.~\ref{Utilda_2}.
The terms in Eq.~(\ref{eq:U_2nd_me}) can be further reduced by using the effective 2NF at the first order,
\begin{eqnarray}
\label{eq:U_2nd_me_recast}
  \widetilde U^{(2)}_{\alpha  \beta } &=& 
-i \hbar \! \int \!\frac{d \omega}{2 \pi} \sum_{\substack{\epsilon \eta \\ \gamma \delta }} \widetilde V^{(1)}_{\alpha \gamma, \beta \delta}  \, g^{(0)}_{\delta \epsilon}(\omega) g^{(0)}_{\eta \gamma}(\omega) \widetilde  U^{(1)}_{\epsilon \eta}   \nonumber \\
&& +\frac{(i\hbar)^3}{4} \! \! \int  \! \frac{\mathrm d \omega_1}{2 \pi} \! \int  \! \frac{\mathrm d \omega_2}{2 \pi}  \! \int  \! \frac{\mathrm d \omega_3}{2 \pi} \! \sum_{\substack{\epsilon \eta \nu \delta \\ \gamma \lambda \mu \rho} } \! W_{\alpha\gamma \epsilon,\beta\delta \eta} \, g^{(0)}_{\nu \gamma}(\omega_1)  g^{(0)}_{\delta \rho}(\omega_2)  g^{(0)}_{\lambda \epsilon}(\omega_3) g^{(0)}_{\eta \mu}(\omega_1+\omega_3-\omega_2) \widetilde V^{(1)}_{\rho \mu, \nu \lambda}\,, 
\end{eqnarray}
as depicted in Fig.~\ref{Utilda_2_V}.
\begin{figure}
  \centering
    \subfloat{\includegraphics[scale=0.6]{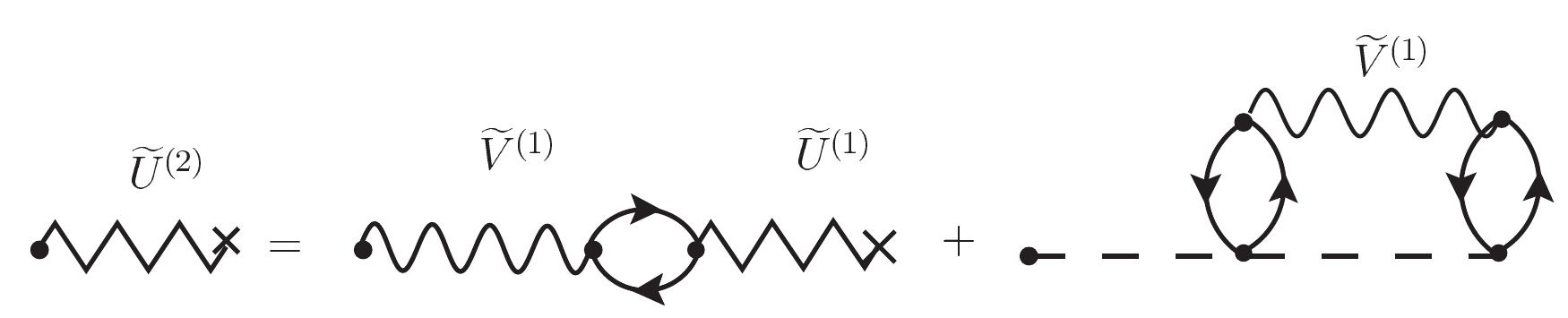}}
  \caption{ As described in the caption of Fig.~\ref{Utilda_1} but for the second-order term $\widetilde U^{(2)}$ rewritten in terms of $\widetilde V^{(1)}$.}
  \label{Utilda_2_V}
\end{figure}

When the integration over the frequencies is performed, the second order term $ \widetilde U^{(2)}$ becomes 
\begin{eqnarray}
\label{eq:U_2nd_me_XY}
  \widetilde U^{(2)}_{\alpha  \beta } &=& 
\sum_{\substack{\epsilon \eta \\ \gamma \delta }} \widetilde V^{(1)}_{\alpha \gamma, \beta \delta}  \, \widetilde  U^{(1)}_{\epsilon \eta} \left(\sum_{n_1 k_2 } \frac{ (X_{\delta}^{n_1})^* X_{\epsilon}^{n_1}  Y_{\eta}^{k_2} (Y_{\gamma}^{k_2})^* }{- (\varepsilon_{n_1}^{+}-\varepsilon_{k_2}^{-}) + i\eta} -\sum_{k_1 n_2 } \frac{ Y_{\delta}^{k_1} (Y_{\epsilon}^{k_1})^* (X_{\eta}^{n_2})^* X_{\gamma}^{n_2}   }{- (\varepsilon_{k_1}^{-}-\varepsilon_{n_2}^{+}) - i\eta} \right) \\
& +&\frac{1}{4} \sum_{\substack{\epsilon \eta \nu \delta \\ \gamma \lambda \mu \rho} } \! W_{\alpha\gamma \epsilon,\beta\delta \eta} \,     \left(
\sum_{\substack{n_1 n_2 \\ k_3 k_4}} \frac{(X_{\nu}^{n_1} X_{\lambda}^{n_2} Y_{\rho}^{k_3} Y_{\mu}^{k_4})^* X_{\gamma}^{n_1} X_{\epsilon}^{n_2}  Y_{\delta}^{k_3}  Y_{\eta}^{k_4}}{ -\left(\varepsilon_{n_1}^{+}+\varepsilon_{n_2}^{+}-\varepsilon_{k_3}^{-}-\varepsilon_{k_4}^{-} \right) + i\eta  }    - \! \sum_{ \substack{k_1 k_2 \\ n_3 n_4}} \! \frac{ Y_{\nu}^{k_1} Y_{\lambda}^{k_2} X_{\rho}^{n_3} X_{\mu}^{n_4} 
(Y_{\gamma}^{k_1} Y_{\epsilon}^{k_2}  X_{\delta}^{n_3}  X_{\eta}^{n_4})^*
 }{- (\varepsilon_{k_1}^{-}+\varepsilon_{k_2}^{-}-\varepsilon_{n_3}^{+} -\varepsilon_{n_4}^{+} ) - i \eta }  \right) 
 \widetilde V^{(1)}_{\rho \mu, \nu \lambda}\,. \nonumber
\end{eqnarray}
Note that the last term on the right hand side of Eq.~\eqref{eq:U_2nd_me_XY} corresponds the last diagram in Fig.~\ref{Utilda_2_V} and it is of skeleton type. This is the second-order contribution to $\Sigma^\infty_{\alpha\beta}$ that would appear also in the self-consistent expansion.

As we discussed above, it is customary in several practical applications to assume for the model space the very same orbits that diagonalize the unperturbed Hamiltonian, $\hat{H}_0$, and define the reference state. In this case the  amplitudes~(\ref{tran_ampl_ref}) become diagonal in the two indexes $i$ and $\alpha$.  Similarly, one may chose an HF reference state and in this case the term $\widetilde{U}^{(1)}$ vanishes because of the specific definition of the HF potential (as it is obvious from Eq.~\eqref{eq:U_1st_me}).  Whether or not it is convenient to take these assumptions--and in particular which is the best reference state to use--depends of the specific computational approach and on the specific system one needs to solve.
For completeness, we give the example of how Eq.~\eqref{eq:U_2nd_me_XY} would simplify if \emph{both} these assumptions are made:
\begin{align}
\label{eq:U_2nd_me_deltas}
  \widetilde U^{(2)}_{\alpha  \beta } ={}&  \left(
 \frac{1}{4} \sum_{\mu,  \nu \notin F} \; \sum_{\rho, \lambda \in F }  
   V_{\rho \lambda, \mu \nu}  \,  \frac{1}{ \left(\varepsilon_{\rho}^{-}+\varepsilon_{\lambda}^{-}-\varepsilon_{\mu}^{+}-\varepsilon_{\nu}^{+} \right) + i\eta  }
    \,  W_{\alpha \mu \nu, \beta \rho \lambda}
~+~
W_{\alpha \rho \lambda,\beta\mu \nu} \,  \frac{1}{ \left(\varepsilon_{\rho}^{-}+\varepsilon_{\lambda}^{-}-\varepsilon_{\mu}^{+}-\varepsilon_{\nu}^{+} \right) + i\eta  }
    \, V_{\mu \nu, \rho \lambda} 
 \right)  \, ,
\end{align}
where we have used the notations $\in F$ ($\notin F$) to restrict the sums over occupied (unoccupied) orbits.   

For the second-order term $\widetilde V^{(2)}_{\alpha \beta,  \gamma\delta} $ in the expansion of Eq.~(\ref{V_exp}), we have the three diagrams on the right-hand side of Fig.~\ref{Vtilda_2} with the following expressions:
\begin{eqnarray}
\label{eq:V_2nd_me}
  \widetilde V^{(2)}_{\alpha \beta,  \gamma\delta} &=& 
i\hbar \! \int  \! \frac{\mathrm d \omega}{2 \pi}  \sum_{\substack{\epsilon \eta \\ \lambda \mu }}W_{\alpha\beta \epsilon,\gamma\delta \eta} \, g^{(0)}_{\mu \epsilon}(\omega) g^{(0)}_{\eta \nu}(\omega) U_{\nu \mu}   - i\hbar  \! \int  \! \frac{\mathrm d \omega}{2 \pi}  \sum_{\substack{\epsilon \eta \nu \\ \lambda \mu \rho} }W_{\alpha\beta \epsilon,\gamma\delta \eta} \, g^{(0)}_{\mu \epsilon}(\omega) g^{(0)}_{\eta \nu}(\omega)  \rho^{(0)}_{\lambda \rho} \, V_{\nu \rho, \mu \lambda} \nonumber \\
&&-  \frac{i\hbar}{2} \! \int  \! \frac{\mathrm d \omega}{2 \pi}  \sum_{\substack{\epsilon \eta \nu \sigma \\ \lambda \mu \rho \tau}  }W_{\alpha \beta \epsilon,\gamma \delta  \eta} \, g^{(0)}_{ \mu \epsilon}(\omega) g^{(0)}_{\eta \nu }(\omega)   W_{ \nu \lambda \sigma, \mu \rho \tau}  \rho^{(0)}_{\rho \lambda}  \rho^{(0)}_{\tau \sigma} \nonumber \\
&= &- i\hbar \! \int \!  \! \frac{\mathrm d \omega}{2 \pi} \! \sum_{\substack{\epsilon \eta \\ \nu \mu }}W_{\alpha\beta \epsilon,\gamma\delta \eta} \, g^{(0)}_{\mu \epsilon}(\omega) g^{(0)}_{\eta \nu}(\omega)   \widetilde U^{(1)}_{\nu \mu}   \,,
\end{eqnarray}
where in the last equality we have written a more compact expression for $ \widetilde V^{(2)}$ by using the first term in the expansion of $ \widetilde U$.
\begin{figure}
  \centering
    \subfloat{\includegraphics[scale=0.65]{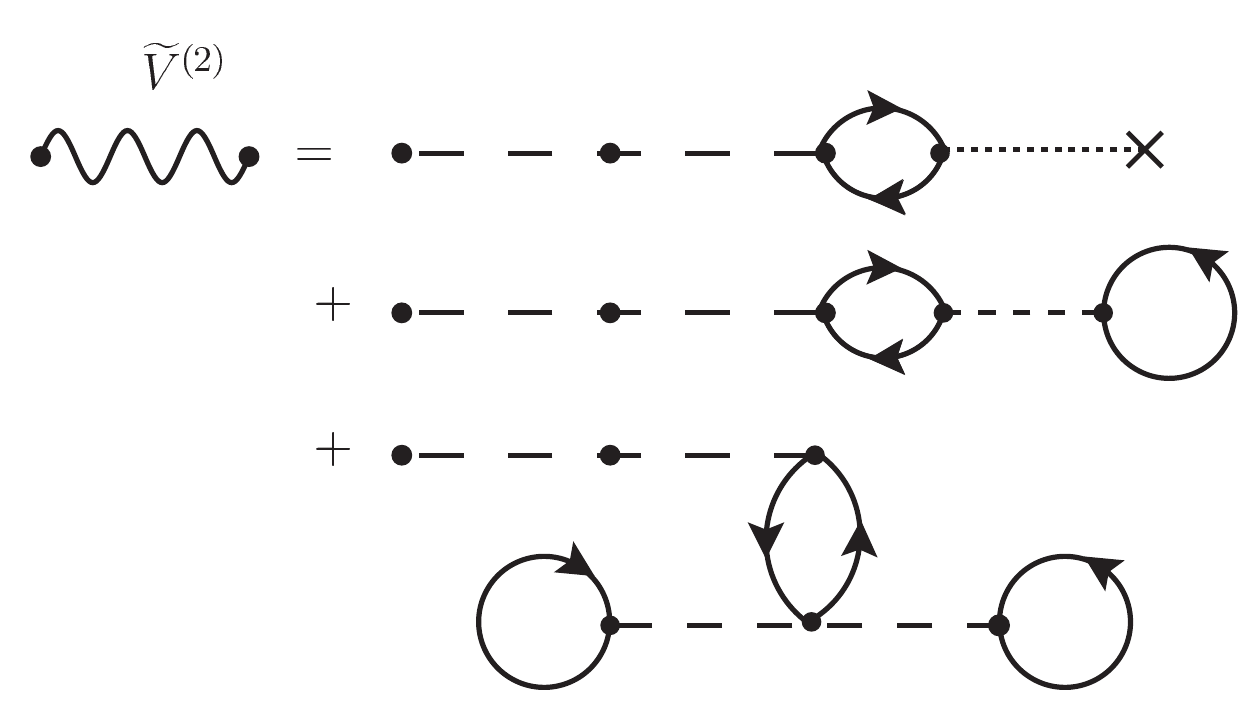}}
  \caption{ As described in the caption of Fig.~\ref{Vtilda_1} but for the second-order term $\widetilde V^{(2)}$.}
  \label{Vtilda_2}
\end{figure}
The integration over the frequency in Eq.~(\ref{eq:V_2nd_me}) gives the expression of  $\widetilde V^{(2)}$ in terms of the uncorrelated  transition amplitudes:
\begin{align}
\label{eq:V_2nd_me_RES_XY}
  \widetilde V^{(2)}_{\alpha \beta,  \gamma\delta} ={}& 
   \sum_{\substack{\epsilon \eta \\  \mu \nu }}
 W_{\alpha\beta \epsilon,\gamma\delta \eta} \, \left(\sum_{n_1 k_2 } \frac{ (X_{\mu}^{n_1} Y_{\nu}^{k_2})^* X_{\epsilon}^{n_1}  Y_{\eta}^{k_2}  }{- (\varepsilon_{n_1}^{+}-\varepsilon_{k_2}^{-}) + i\eta} -\sum_{k_1 n_2 } \frac{ Y_{\mu}^{k_1} X_{\nu}^{n_2} (Y_{\epsilon}^{k_1} X_{\eta}^{n_2})^*    }{- (\varepsilon_{k_1}^{-}-\varepsilon_{n_2}^{+}) - i\eta} \right) \,    \widetilde U^{(1)}_{\nu \mu}  \, ,
\end{align}
which is zero in the specific case of an HF reference state, due to $\widetilde U^{(1)}$ vanishing. 

\begin{figure*}
  \subfloat[(a)]{\label{U3_a}\includegraphics[scale=0.50]{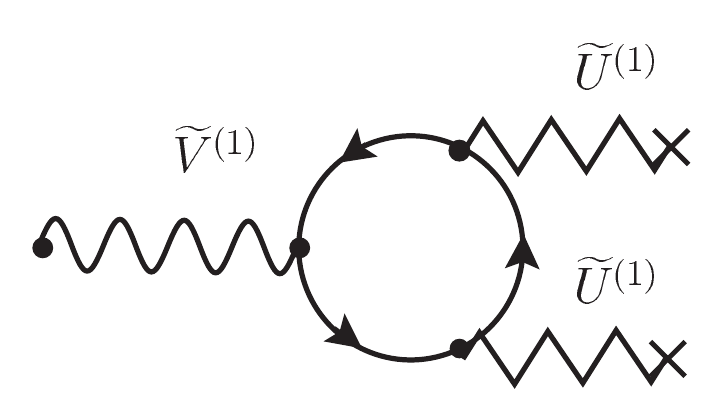}}
  \hspace{1.5cm}
  \subfloat[(b)]{\label{U3_b}\includegraphics[scale=0.50]{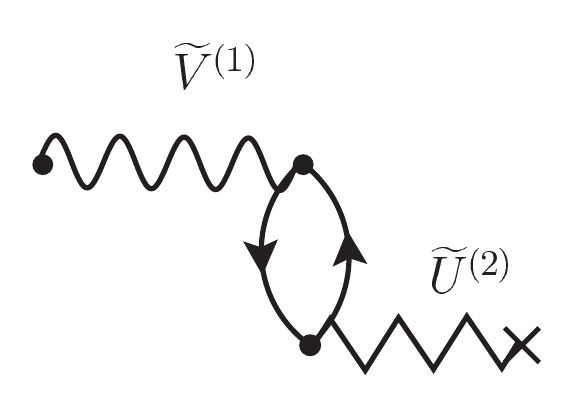}}
  \hspace{1.5cm}
   \subfloat[(c)]{\label{U3_c}\includegraphics[scale=0.50]{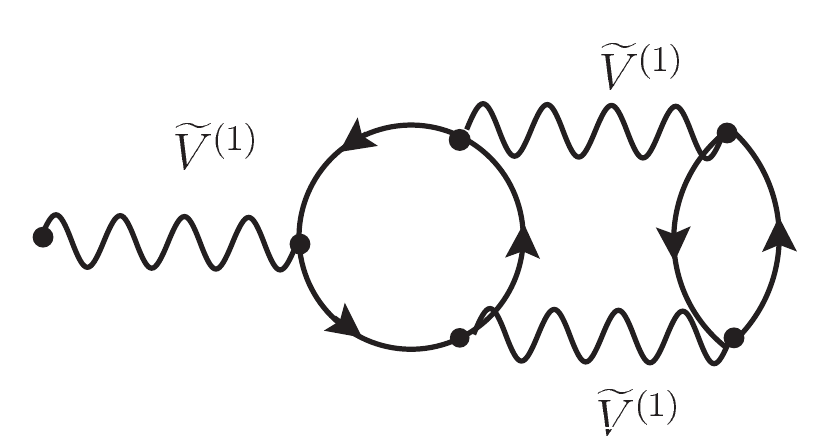}}
   \newline   \vskip .7cm
  \subfloat[(d)]{\label{U3_d}\includegraphics[scale=0.48]{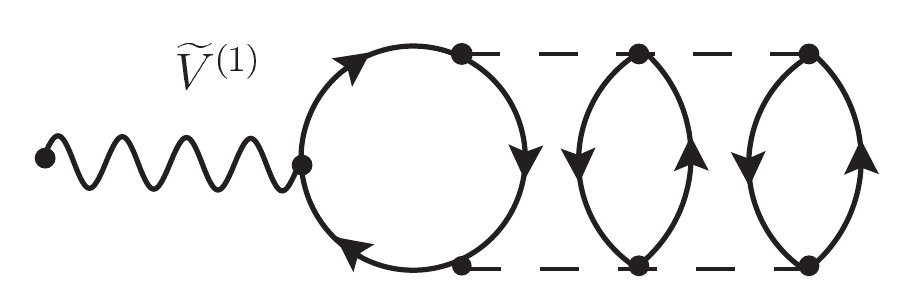}}
  \hspace{0.8cm}
  \subfloat[(e)]{\label{U3_V2}\includegraphics[scale=0.53]{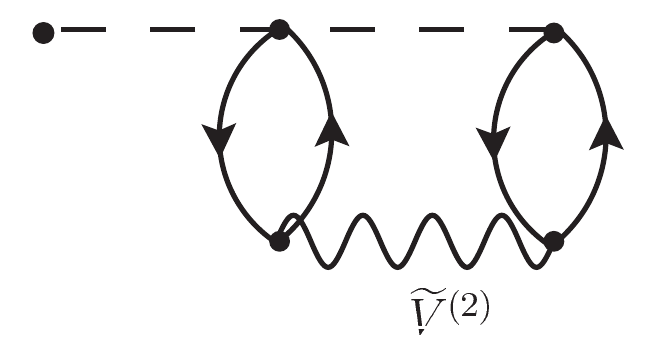}}
  \hspace{0.8cm}
  \subfloat[(f)]{\label{U3_e}\includegraphics[scale=0.48]{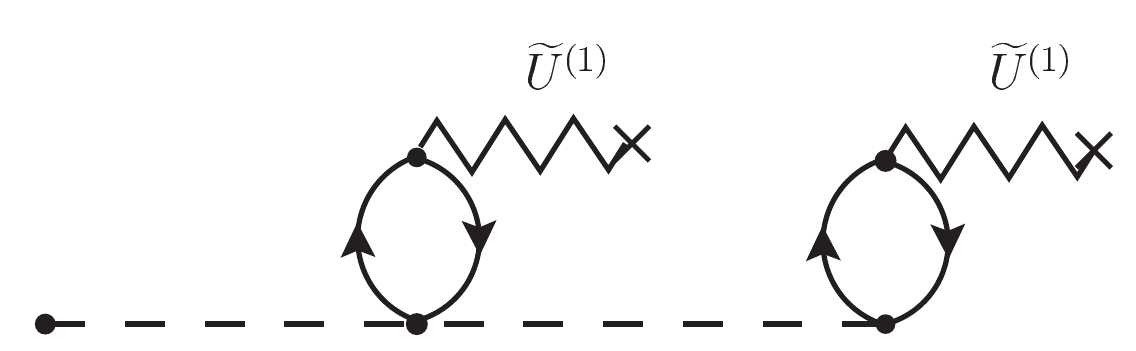}}  
       \newline 
\subfloat[(g)]{\label{U3_f}\includegraphics[scale=0.50]{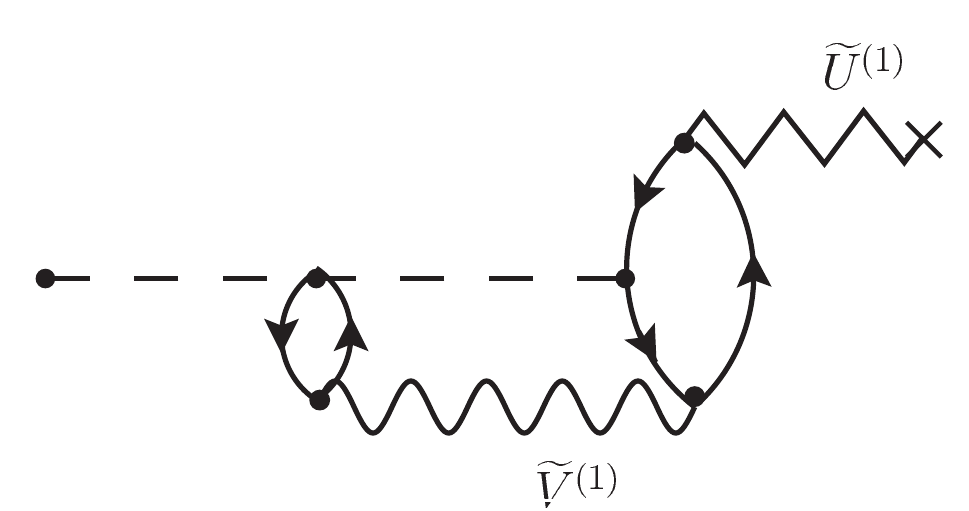}}
    \hspace{1.0cm}
       \subfloat[(h)]{\label{U3_g}\includegraphics[scale=0.50]{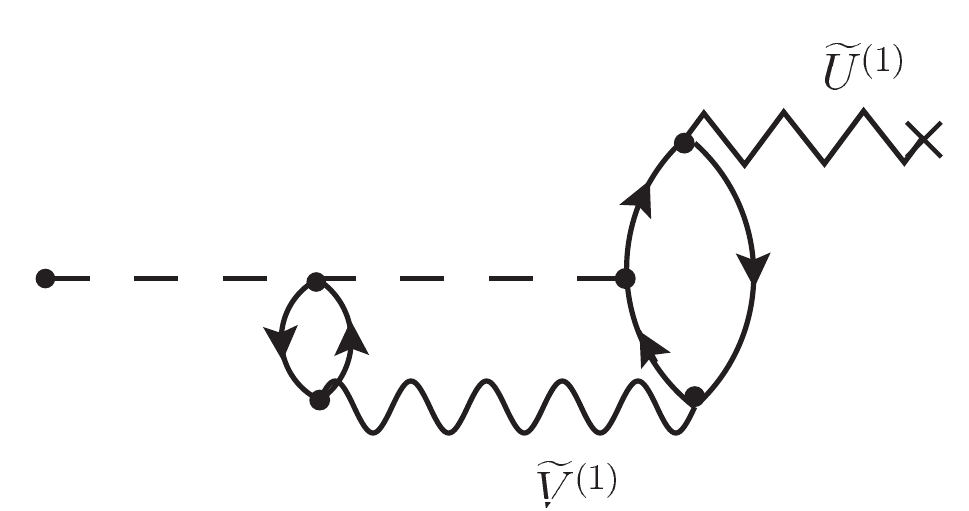}}
         \hspace{1.0cm}
  \subfloat[(i)]{\label{U3_h}\includegraphics[scale=0.50]{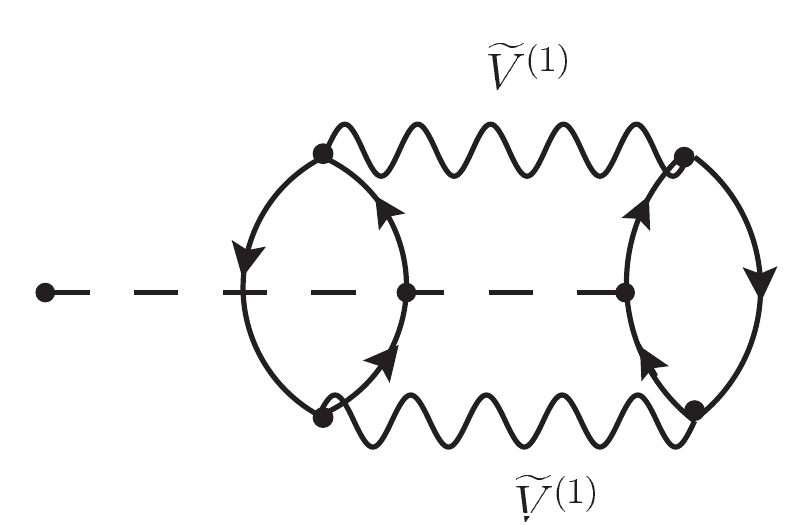}}
  \newline
  \subfloat[(j)]{\label{U3_i}\includegraphics[scale=0.48]{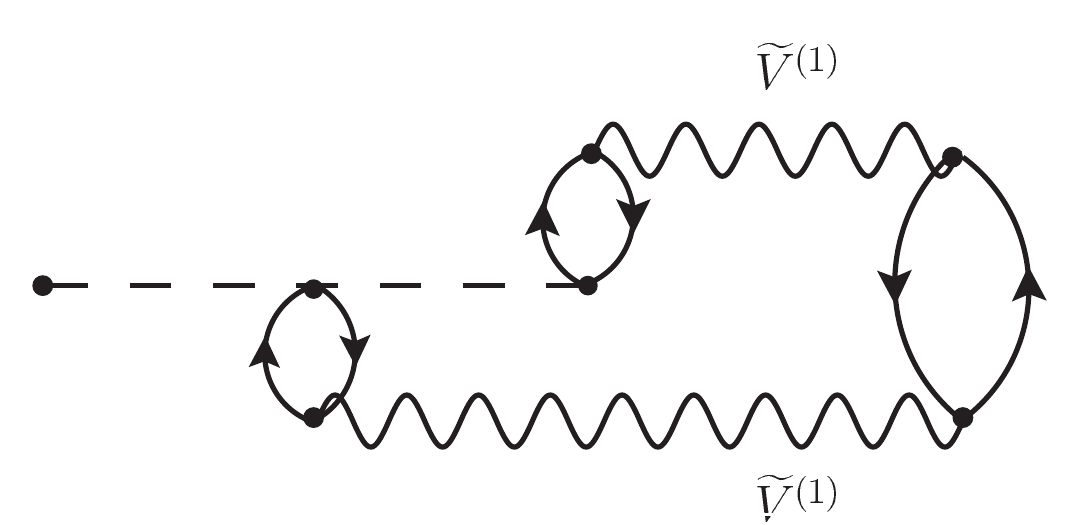}}  
    \hspace{0.6cm}
  \subfloat[(k)]{\label{U3_o}\includegraphics[scale=0.48]{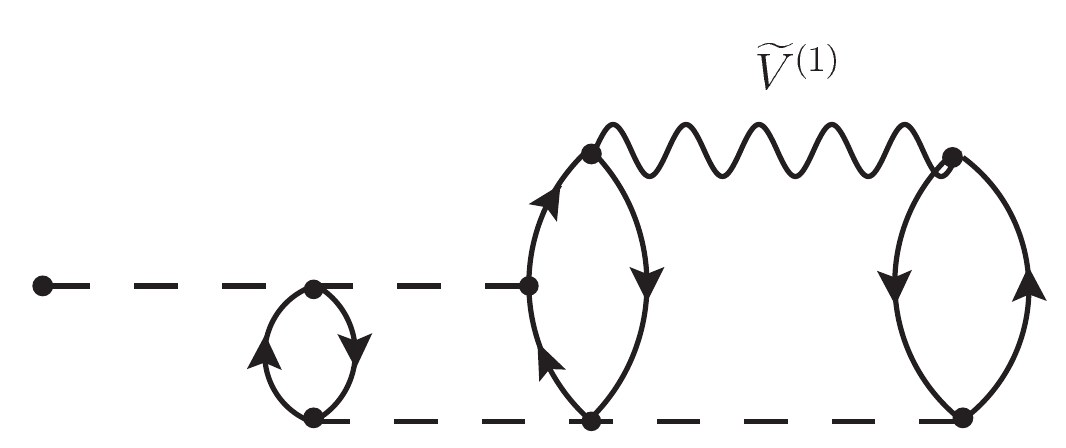}}
    \hspace{0.6cm}
  \subfloat[(l)]{\label{U3_p}\includegraphics[scale=0.48]{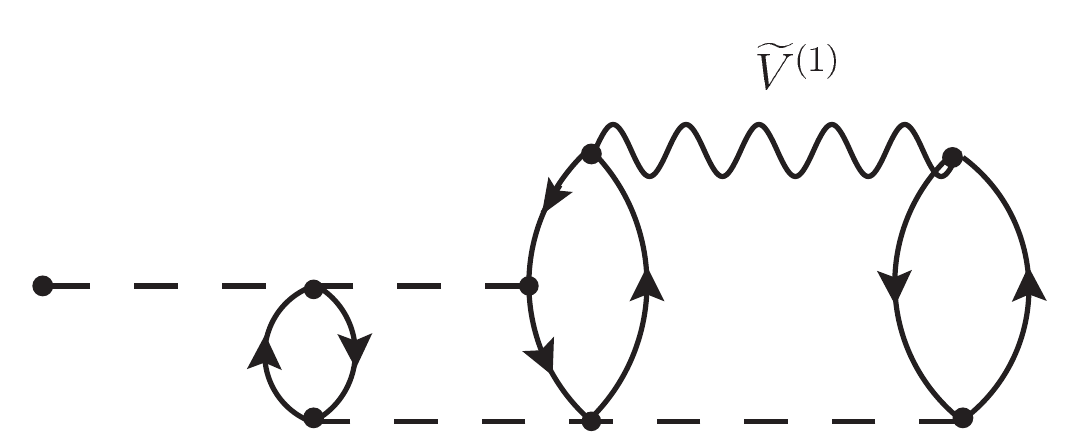}}
    \newline   \vskip .7cm
  \subfloat[(m)]{\label{U3_j}\includegraphics[scale=0.50]{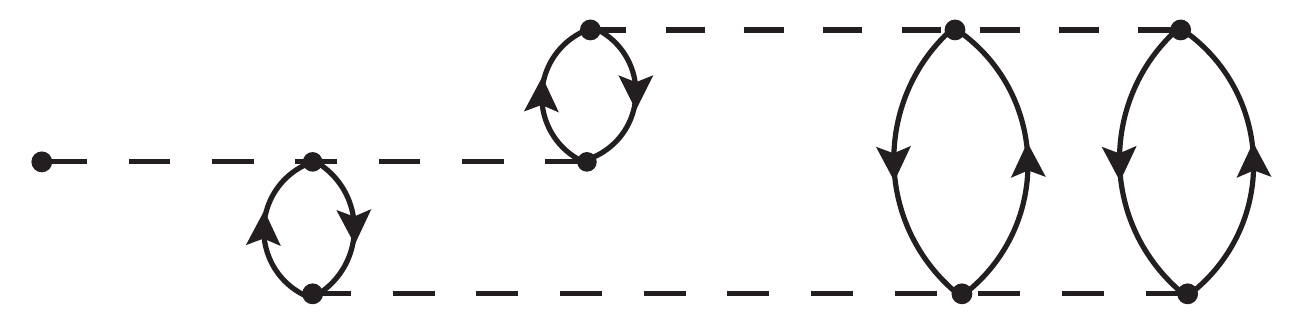}}
    \hspace{1.0cm}
  \subfloat[(n)]{\label{U3_k}\includegraphics[scale=0.50]{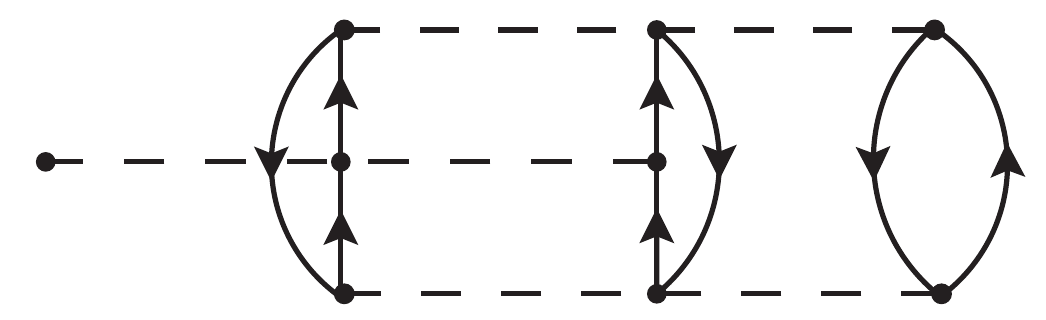}}
   \caption{As described in the caption of Fig.~\ref{Utilda_1} but for the third-order term $\widetilde U^{(3)}$.} 
  \label{Utilda_3}
\end{figure*} 

The expansion of $ \widetilde U$ in Eq.~(\ref{U_exp}) contains also the term $ \widetilde U^{(3)}_{\alpha \beta}$  composed by the 14 contributions shown in Fig.~\ref{Utilda_3}. By using the same Feynman rules applied for the terms at second and third order (see Appendix A of Ref.~\cite{CarAr13}), one can derive the expressions for those fourteen diagrams. Here we give the final equations after all  integrals over the frequencies have been performed.
Using the compact notation for $npnh$ coupled cluster amplitudes of Eqs.~(\ref{eq:t_1B})-(\ref{eq:t_3B}) and assuming Einstein's summing convention throughout, they are listed below according to the order of appearance in Fig.~\ref{Utilda_3}:
\begin{subequations}
\begin{eqnarray}
\label{eq:U_3rd_me_RES_XY_A}
  \widetilde U^{(3)}_{\alpha  \beta } (\text{\ref{U3_a}})&= &
   \widetilde V^{(1)}_{\alpha \gamma, \beta \delta}    (X_{\delta}^{n_1 })^*  X_{\gamma}^{n_2 } t^{n_1}_{k_3} (t^{n_2}_{k_3})^* -     \widetilde V^{(1)}_{\alpha \gamma, \beta \delta} 
( Y_{\gamma}^{k_2} )^*  Y_{\delta}^{k_1}  t^{n_3}_{k_2} (t^{n_3}_{k_1})^* \\
& &+  \widetilde V^{(1)}_{\alpha \gamma, \beta \delta}  \, \widetilde  U^{(1)}_{\epsilon \eta} \widetilde  U^{(1)}_{\mu \nu} \left( \frac{(X_{\nu}^{n_1 } X_{\eta}^{n_2 } Y_{\epsilon}^{k_3 })^*  X_{\gamma}^{n_1 } X_{\mu}^{n_2 } Y_{\delta}^{k_3 }}{(-(\varepsilon_{n_1}^{+}-\varepsilon_{k_3}^{-}) + i\eta)(-(\varepsilon_{n_2}^{+}-\varepsilon_{k_3}^{-}) + i\eta)} -     
 \frac{(Y_{\gamma}^{k_1} Y_{\mu}^{k_2} X_{\delta}^{n_3})^*  Y_{\nu}^{k_1} Y_{\eta}^{k_2} X_{\epsilon}^{n_3}}{(-(\varepsilon_{k_2}^{-}-\varepsilon_{n_3}^{+}) - i\eta)(-(\varepsilon_{k_1}^{-}-\varepsilon_{n_3}^{+}) - i\eta)} \right. \nonumber\\
& &\left.
 +  \frac{(X_{\delta}^{n_1 } X_{\eta}^{n_2 } Y_{\gamma}^{k_3 })^*  X_{\epsilon}^{n_1 } X_{\mu}^{n_2 } Y_{\nu}^{k_3 }}{(-(\varepsilon_{n_2}^{+}-\varepsilon_{k_3}^{-}) + i\eta)(-(\varepsilon_{n_1}^{+}-\varepsilon_{k_3}^{-}) + i\eta)} -     
 \frac{(Y_{\mu}^{k_1} Y_{\epsilon}^{k_2} X_{\nu}^{n_3})^*  Y_{\eta}^{k_1} Y_{\delta}^{k_2} X_{\gamma}^{n_3}}{(-(\varepsilon_{k_1}^{-}-\varepsilon_{n_3}^{+}) - i\eta)(-(\varepsilon_{k_2}^{-}-\varepsilon_{n_3}^{+}) - i\eta)} 
\right)\nonumber
\, \,;
\end{eqnarray}
\begin{eqnarray}
\label{eq:U_3rd_me_RES_XY_B}
  \widetilde U^{(3)}_{\alpha  \beta } (\text{\ref{U3_b}})&= &
  \! \widetilde V^{(1)}_{\alpha \gamma, \beta \delta}  \, \left((X_{\delta}^{n_1 } Y_{\gamma}^{k_2 })^* t^{n_1}_{k_2}
-  (t^{n_2}_{k_1})^* X_{\gamma}^{n_2} Y_{\delta}^{k_1}\right) 
\, \,;
\end{eqnarray}
\begin{eqnarray}
\label{eq:U_3rd_me_RES_XY_C}
  \widetilde U^{(3)}_{\alpha  \beta } (\text{\ref{U3_c}})&= &
 -  \frac{1}{2} \!  \widetilde V^{(1)}_{\alpha \gamma, \beta \delta}  \left(  (X_{\delta}^{n_2} )^* X_{\gamma}^{n_1}(t^{n_1 n_3}_{k_4 k_5})^* t^{n_2 n_3}_{k_4 k_5}+(Y_{\gamma}^{k_2} )^* Y_{\delta}^{k_1} (t^{n_4 n_5}_{k_1 k_3})^* t^{n_4 n_5}_{k_2 k_3} \right) \nonumber \\
&  & -  \frac{1}{2} \!  \widetilde V^{(1)}_{\alpha \gamma, \beta \delta}  \, V^{(1)}_{\mu \tau, \nu \zeta} \left(   \frac{( X_{\nu}^{n_2} )^* X_{\mu}^{n_1} X_{\gamma}^{n_2} X_{\tau}^{n_3}  Y_{\zeta}^{k_4} Y_{\delta}^{k_5}}{(-(\varepsilon_{k_5}^{-}-\varepsilon_{n_2}^{+}) - i\eta)} (t^{n_1 n_3}_{k_4 k_5})^*  +   \frac{(Y_{\gamma}^{k_1} Y_{\mu}^{k_2} Y_{\tau}^{k_3}  X_{\delta}^{n_4} X_{\zeta}^{n_5})^*           Y_{\nu}^{k_1}}{(-(\varepsilon_{k_1}^{-}-\varepsilon_{n_4}^{+}) - i\eta)}  t^{n_4 n_5}_{k_2 k_3}  \right) \nonumber \\
&  & -  \frac{1}{2} \!  \widetilde V^{(1)}_{\alpha \gamma, \beta \delta}  \,\widetilde V^{(1)}_{\epsilon \sigma, \eta \rho } \left(  \frac{(X_{\delta}^{n_1} X_{\eta}^{n_2} X_{\rho}^{n_3}  Y_{\gamma}^{k_4} Y_{\sigma}^{k_5})^* X_{\epsilon}^{n_1} }{(-(\varepsilon_{k_4}^{-}-\varepsilon_{n_1}^{+}) - i\eta)}  t^{n_2 n_3}_{k_4 k_5}  + \frac{(Y_{\epsilon}^{k_1} )^* Y_{\delta}^{k_1} Y_{\eta}^{k_2} Y_{\rho}^{k_3}  X_{\gamma}^{n_4} X_{\sigma}^{n_5}}{(-(\varepsilon_{k_1}^{-}-\varepsilon_{n_4}^{+}) - i\eta)}  (t^{n_4 n_5}_{k_2 k_3} )^* \right) 
;
\end{eqnarray}
\begin{eqnarray}
\label{eq:U_3rd_me_RES_XY_D}
  \widetilde U^{(3)}_{\alpha  \beta } (\text{\ref{U3_d}})&= &
  \frac{1}{12} \! \widetilde V^{(1)}_{\alpha \gamma, \beta \delta}   \left(-  (X_{\delta}^{n_2} )^*  X_{\gamma}^{n_1} (t^{n_1 n_3 n_4}_{k_5 k_6 k_7} )^* t^{n_2 n_3 n_4}_{k_5 k_6 k_7}   - (Y_{\gamma}^{k_1} )^*  Y_{\delta}^{k_2}  (t^{n_5 n_6 n_7}_{k_2 k_3 k_4} )^* t^{n_5 n_6 n_7}_{k_1 k_3 k_4}\right) \nonumber \\
&& +  \frac{1}{12} \! \widetilde V^{(1)}_{\alpha \gamma, \beta \delta}  \, W_{\chi \eta \epsilon, \lambda \nu \mu} \left(  \frac{(X_{\lambda}^{n_1})^*  X_{\gamma}^{n_1} X_{\chi}^{n_2} X_{\eta}^{n_3} X_{\epsilon}^{n_4} Y_{\delta}^{k_5}  Y_{\nu}^{k_6}  Y_{\mu}^{k_7}}{(-(\varepsilon_{n_1}^{+}-\varepsilon_{k_5}^{-}) + i\eta)}  (t^{n_2 n_3 n_4}_{k_5 k_6 k_7} )^*  \right. \nonumber \\
&& \left. + \frac{(Y_{\gamma}^{k_1} Y_{\chi}^{k_2} Y_{\eta}^{k_3} Y_{\epsilon}^{k_4} X_{\delta}^{n_5}  X_{\nu}^{n_6}  X_{\mu}^{n_7})^*  Y_{\lambda}^{k_1} }{(-(\varepsilon_{n_5}^{+}-\varepsilon_{k_1}^{-}) + i\eta)} t^{n_5 n_6 n_7}_{k_2 k_3 k_4} \right) \nonumber \\
&& + \frac{1}{12} \! \widetilde V^{(1)}_{\alpha \gamma, \beta \delta}  \,  W_{\sigma \zeta \theta , \tau \rho \kappa} \left(  \frac{(X_{\delta}^{n_1} X_{\tau}^{n_2} X_{\rho}^{n_3} X_{\kappa}^{n_4} Y_{\gamma}^{k_5}  Y_{\zeta}^{k_6}  Y_{\theta}^{k_7})^*  X_{\sigma}^{n_1} }{(-(\varepsilon_{n_1}^{+}-\varepsilon_{k_5}^{-}) + i\eta)} t^{n_2 n_3 n_4}_{k_5 k_6 k_7} \right. \nonumber \\
&& \left. + \frac{(Y_{\sigma}^{k_1} )^*  Y_{\delta}^{k_1} Y_{\tau}^{k_2} Y_{\rho}^{k_3} Y_{\kappa}^{k_4} X_{\gamma}^{n_5}  X_{\zeta}^{n_6}  X_{\theta}^{n_7}}{(-(\varepsilon_{n_5}^{+}-\varepsilon_{k_1}^{-}) + i\eta)}  (t^{n_5 n_6 n_7}_{k_2 k_3 k_4} )^* \right) 
\, \,;
\end{eqnarray}
\begin{eqnarray}
\label{eq:U_3rd_me_RES_XY_E_true}
  \widetilde U^{(3)}_{\alpha  \beta } (\text{\ref{U3_V2}})&= &\frac{1}{4} \sum_{\substack{\epsilon \eta \nu \delta \\ \gamma \lambda \mu \rho} } \! W_{\alpha\gamma \epsilon,\beta\delta \eta} \,     \left(
\sum_{\substack{n_1 n_2 \\ k_3 k_4}} \frac{(X_{\nu}^{n_1} X_{\lambda}^{n_2} Y_{\rho}^{k_3} Y_{\mu}^{k_4})^* X_{\gamma}^{n_1} X_{\epsilon}^{n_2}  Y_{\delta}^{k_3}  Y_{\eta}^{k_4}}{ -\left(\varepsilon_{n_1}^{+}+\varepsilon_{n_2}^{+}-\varepsilon_{k_3}^{-}-\varepsilon_{k_4}^{-} \right) + i\eta  } 
\right. \nonumber \\
&& \left. 
- \! \sum_{ \substack{k_1 k_2 \\ n_3 n_4}} \! \frac{ Y_{\nu}^{k_1} Y_{\lambda}^{k_2} X_{\rho}^{n_3} X_{\mu}^{n_4} 
(Y_{\gamma}^{k_1} Y_{\epsilon}^{k_2}  X_{\delta}^{n_3}  X_{\eta}^{n_4})^*
 }{- (\varepsilon_{k_1}^{-}+\varepsilon_{k_2}^{-}-\varepsilon_{n_3}^{+} -\varepsilon_{n_4}^{+} ) - i \eta }  \right) 
 \widetilde V^{(2)}_{\rho \mu, \nu \lambda}
\, \,;
\end{eqnarray}
\begin{eqnarray}
\label{eq:U_3rd_me_RES_XY_E}
  \widetilde U^{(3)}_{\alpha  \beta } (\text{\ref{U3_e}})&= &
    \! W_{\alpha\gamma \epsilon,\beta\delta \eta} \, \left((X_{\delta}^{n_1} X_{\eta}^{n_2} Y_{\gamma}^{k_3} Y_{\epsilon}^{k_4})^* t^{n_1}_{k_3}  t^{n_2}_{k_4}     
+(X_{\delta}^{n_1} Y_{\gamma}^{k_3} )^* X_{\epsilon}^{n_2}  Y_{\eta}^{k_4}  t^{n_1}_{k_3}  (t^{n_2}_{k_4})^*  
 \right. \nonumber\\ 
 &    &\left. +(Y_{\epsilon}^{k_2}  X_{\eta}^{n_4})^* Y_{\delta}^{k_1}  X_{\gamma}^{n_3}  t^{n_4}_{k_2}  (t^{n_3}_{k_1})^*
 +Y_{\delta}^{k_1} Y_{\eta}^{k_2} X_{\gamma}^{n_3} X_{\epsilon}^{n_4}      (t^{n_4}_{k_2})^*  (t^{n_3}_{k_1})^*
 \right)  
\, \,;
\end{eqnarray}
\begin{eqnarray}
\label{eq:U_3rd_me_RES_XY_F}
  \widetilde U^{(3)}_{\alpha  \beta } (\text{\ref{U3_f}})&= &
  -\frac{1}{2} \!   W_{\alpha\gamma \epsilon,\beta\delta \eta} \,     \left( -  (X_{\eta}^{n_1})^*  X_{\epsilon}^{n_2} X_{\gamma}^{n_3}  Y_{\delta}^{k_5}   t^{n_1}_{k_4}  (t^{n_3 n_2}_{k_5 k_4})^* + (Y_{\epsilon}^{k_2} Y_{\gamma}^{k_3} X_{\delta}^{n_5} )^*  Y_{\eta}^{k_1} (t^{n_4}_{k_1})^*  t^{n_5 n_4}_{k_3 k_2} \right) \nonumber \\
&   &-\frac{1}{2} \!   W_{\alpha\gamma \epsilon,\beta\delta \eta} \,    U^{(1)}_{\sigma \chi}  \left( -  \frac{(X_{\eta}^{n_1} X_{\chi}^{n_2} X_{\delta}^{n_3} Y_{\epsilon}^{k_4} Y_{\gamma}^{k_5})^* X_{\sigma}^{n_1} }{(-(\varepsilon_{n_1}^{+}+\varepsilon_{n_3}^{+}-\varepsilon_{k_4}^{-}-\varepsilon_{k_5}^{-}) + i\eta)} t^{n_3 n_2}_{k_5 k_4}  +  \frac{(Y_{\sigma}^{k_1})^*  Y_{\eta}^{k_1} Y_{\chi}^{k_2} Y_{\delta}^{k_3} X_{\gamma}^{n_4} X_{\epsilon}^{n_5}}{\varepsilon_{k_1}^{-}+\varepsilon_{k_3}^{-}-\varepsilon_{n_4}^{+}-\varepsilon_{n_5}^{+} + i\eta}  (t^{n_4 n_5}_{k_3 k_2})^* \right) \nonumber \\  
 &  & -\frac{1}{2} \!   W_{\alpha\gamma \epsilon,\beta\delta \eta} \,   \widetilde V^{(1)}_{\mu \lambda, \nu \rho}\left( -  \frac{(X_{\rho}^{n_1}  X_{\nu}^{n_3}  Y_{\mu}^{k_5})^* X_{\epsilon}^{n_1} X_{\lambda}^{n_2} X_{\gamma}^{n_3} Y_{\eta}^{k_4} Y_{\delta}^{k_5}}{(-(\varepsilon_{n_1}^{+}+\varepsilon_{n_3}^{+}-\varepsilon_{k_4}^{-}-\varepsilon_{k_5}^{-}) + i\eta)} (t^{n_2}_{k_4})^*  \right. \nonumber \\
 &   &\left. + \frac{(Y_{\epsilon}^{k_1} Y_{\lambda}^{k_2} Y_{\gamma}^{k_3} X_{\eta}^{n_4} X_{\delta}^{n_5} )^*  Y_{\rho}^{k_1}  Y_{\nu}^{k_3}  X_{\mu}^{n_5}}{\varepsilon_{k_1}^{-}+\varepsilon_{k_3}^{-}-\varepsilon_{n_4}^{+}-\varepsilon_{n_5}^{+} + i\eta} t^{n_4}_{k_2} \right)
\, \,;
\end{eqnarray}
\begin{eqnarray}
\label{eq:U_3rd_me_RES_XY_G}
  \widetilde U^{(3)}_{\alpha  \beta } (\text{\ref{U3_g}})&= &
  -\frac{1}{2} \!  W_{\alpha\gamma \epsilon,\beta\delta \eta} \,    \left( -  (X_{\eta}^{n_1} X_{\delta}^{n_3}  Y_{\gamma}^{k_5})^*X_{\epsilon}^{n_2}   (t^{n_2}_{k_4})^*  t^{n_3 n_1}_{k_5 k_4}  + (Y_{\epsilon}^{k_2})^*  Y_{\eta}^{k_1} Y_{\delta}^{k_3}  X_{\gamma}^{n_5}   t^{n_4}_{k_2}  
  (t^{n_5 n_4}_{k_3 k_1})^* \right) \nonumber \\        
&   &  -\frac{1}{2} \!  W_{\alpha\gamma \epsilon,\beta\delta \eta} \,  \widetilde U^{(1)}_{\sigma \chi}   \left( -  \frac{(X_{\chi}^{n_2})^* X_{\sigma}^{n_1} X_{\epsilon}^{n_2} X_{\gamma}^{n_3} Y_{\eta}^{k_4} Y_{\delta}^{k_5}}{(-(\varepsilon_{n_2}^{+}+\varepsilon_{n_3}^{+}-\varepsilon_{k_4}^{-}-\varepsilon_{k_5}^{-}) + i\eta)} (t^{n_3 n_1}_{k_5 k_4})^* \right. \nonumber \\
&    &\left. +  \frac{(Y_{\epsilon}^{k_1} Y_{\sigma}^{k_2} Y_{\gamma}^{k_3} X_{\eta}^{n_4} X_{\delta}^{n_5} )^*  Y_{\chi}^{k_1} }{\varepsilon_{k_1}^{-}+\varepsilon_{k_3}^{-}-\varepsilon_{n_4}^{+}-\varepsilon_{n_5}^{+} + i\eta} t^{n_5 n_4}_{k_3 k_2} \right) \nonumber \\    
&   &  -\frac{1}{2} \!  W_{\alpha\gamma \epsilon,\beta\delta \eta} \,   \widetilde V^{(1)}_{\mu \lambda, \nu \rho}   \left( -  \frac{(X_{\eta}^{n_1} X_{\rho}^{n_2} X_{\delta}^{n_3} Y_{\epsilon}^{k_4} Y_{\gamma}^{k_5})^* X_{\lambda}^{n_1}  X_{\mu}^{n_3}  Y_{\nu}^{k_5}}{(-(\varepsilon_{n_1}^{+}+\varepsilon_{n_3}^{+}-\varepsilon_{k_4}^{-}-\varepsilon_{k_5}^{-}) + i\eta)} t^{n_2}_{k_4} \right. \nonumber \\
&   &\left. +   \frac{(Y_{\lambda}^{k_2} Y_{\mu}^{k_3} X_{\nu}^{n_5})^*  Y_{\rho}^{k_1} Y_{\eta}^{k_2} Y_{\delta}^{k_3} X_{\epsilon}^{n_4} X_{\gamma}^{n_5}}{\varepsilon_{k_2}^{-}+\varepsilon_{k_3}^{-}-\varepsilon_{n_4}^{+}-\varepsilon_{n_5}^{+} + i\eta} (t^{n_4}_{k_1})^*    \right)  
\, \,;
\end{eqnarray}
\begin{eqnarray}
\label{eq:U_3rd_me_RES_XY_H}
  \widetilde U^{(3)}_{\alpha  \beta } (\text{\ref{U3_h}})&= &
  \frac{1}{8}  W_{\alpha\gamma \epsilon,\beta\delta \eta} \,   \left(  (X_{\delta}^{n_2}  X_{\eta}^{n_4})^* \,  X_{\gamma}^{n_1}   X_{\epsilon}^{n_3}   (t^{n_1 n_3}_{k_5 k_6})^* t^{n_2 n_4}_{k_5 k_6} + (Y_{\gamma}^{k_1}  Y_{\epsilon}^{k_3})^* \,   Y_{\delta}^{k_2}  Y_{\eta}^{k_4}   (t^{n_5 n_6}_{k_2 k_4})^* t^{n_5 n_6}_{k_1 k_3}  \right) \nonumber \\
&  & + \frac{1}{8}  W_{\alpha\gamma \epsilon,\beta\delta \eta} \,  \widetilde V^{(1)}_{\tau \sigma, \zeta \chi } \left( + \frac{  (X_{\zeta}^{n_1} X_{\chi}^{n_3})^* \,  X_{\gamma}^{n_1}  X_{\tau}^{n_2} X_{\epsilon}^{n_3} X_{\sigma}^{n_4} Y_{\delta}^{k_5}  Y_{\eta}^{k_6}  }{(-(\varepsilon_{n_1}^{+}+\varepsilon_{n_3}^{+}-\varepsilon_{k_5}^{-}-\varepsilon_{k_6}^{-}) + i\eta)} (t^{n_2 n_4}_{k_5 k_6})^* \right. \nonumber \\   
&    &\left. +  \frac{ (Y_{\gamma}^{k_1}  Y_{\tau}^{k_2} Y_{\epsilon}^{k_3}  Y_{\sigma}^{k_4} X_{\delta}^{n_5}  X_{\eta}^{n_6}  )^* \,  Y_{\zeta}^{k_1}  Y_{\chi}^{k_3}   }{\varepsilon_{k_1}^{-}+\varepsilon_{k_3}^{-}-\varepsilon_{n_5}^{+}-\varepsilon_{n_6}^{+}+ i\eta} t^{n_5 n_6}_{k_2 k_4} \right) \nonumber \\   
&    & + \frac{1}{8}  W_{\alpha\gamma \epsilon,\beta\delta \eta} \,  \widetilde V^{(1)}_{\mu \lambda, \nu \rho}  \left( + \frac{   (X_{\nu}^{n_1}  X_{\delta}^{n_2} X_{\rho}^{n_3} X_{\eta}^{n_4} Y_{\gamma}^{k_5}  Y_{\epsilon}^{k_6} )^* \,   X_{\mu}^{n_2}  X_{\lambda}^{n_4} }{(-(\varepsilon_{n_2}^{+}+\varepsilon_{n_4}^{+}-\varepsilon_{k_5}^{-}-\varepsilon_{k_6}^{-}) + i\eta)}  t^{n_1 n_3}_{k_5 k_6}\right. \nonumber \\    
&    &\left. + \frac{  (Y_{\mu}^{k_2}  Y_{\lambda}^{k_4})^* \,  Y_{\nu}^{k_1}  Y_{\delta}^{k_2} Y_{\rho}^{k_3}  Y_{\eta}^{k_4} X_{\gamma}^{n_5}  X_{\epsilon}^{n_6}   }{\varepsilon_{k_2}^{-}+\varepsilon_{k_4}^{-}-\varepsilon_{n_5}^{+}-\varepsilon_{n_6}^{+} + i\eta}  (t^{n_5 n_6}_{k_1 k_3})^*  \right)
\, \,;
\end{eqnarray}
\begin{eqnarray}
\label{eq:U_3rd_me_RES_XY_I}
  \widetilde U^{(3)}_{\alpha  \beta } (\text{\ref{U3_i}})&= &
  \frac{1}{2} W_{\alpha\gamma \epsilon,\beta\delta \eta} \,  \left(  (X_{\delta}^{n_1}   Y_{\gamma}^{k_4})^* \, X_{\epsilon}^{n_2} Y_{\eta}^{k_5} (t^{n_2 n_3}_{k_5 k_6})^* t^{n_1 n_3}_{k_4 k_6}  + (Y_{\epsilon}^{k_2}  X_{\eta}^{n_5})^* \, Y_{\delta}^{k_1} X_{\gamma}^{n_4} (t^{n_4 n_6}_{k_1 k_3})^* t^{n_5 n_6}_{k_2 k_3}   \right) \nonumber \\
&  &  + \frac{1}{2} W_{\alpha\gamma \epsilon,\beta\delta \eta} \,  \widetilde V^{(1)}_{\mu \lambda,  \nu \rho}  \left(  \frac{ (X_{\nu}^{n_1}  Y_{\mu}^{k_4})^* \, X_{\gamma}^{n_1}  X_{\lambda}^{n_2} X_{\epsilon}^{n_3}  Y_{\delta}^{k_4}   Y_{\rho}^{k_5}  Y_{\eta}^{k_6} }{(-(\varepsilon_{n_1}^{+}+\varepsilon_{n_3}^{+}-\varepsilon_{k_4}^{-}-\varepsilon_{k_6}^{-}) + i\eta)} (t^{n_3 n_2}_{k_6 k_5})^*  \right. \nonumber \\ 
&    &\left. +  \frac{ ( Y_{\gamma}^{k_1}   Y_{\lambda}^{k_2}  Y_{\epsilon}^{k_3} X_{\delta}^{n_4}  X_{\rho}^{n_5} X_{\eta}^{n_6}   )^* \, Y_{\nu}^{k_1}    X_{\mu}^{n_4}   }{\varepsilon_{k_2}^{-}+\varepsilon_{k_3}^{-}-\varepsilon_{n_5}^{+}-\varepsilon_{n_6}^{+} + i\eta} t^{n_6 n_5}_{k_3 k_2}  \right) \nonumber \\
&   &+ \frac{1}{2} W_{\alpha\gamma \epsilon,\beta\delta \eta} \,   \widetilde V^{(1)}_{\tau \sigma, \zeta \chi} \left(  \frac{ (X_{\delta}^{n_1}  X_{\eta}^{n_2} X_{\chi}^{n_3}  Y_{\gamma}^{k_4}   Y_{\epsilon}^{k_5}  Y_{\sigma}^{k_6} )^* \,  X_{\tau}^{n_2}   Y_{\zeta}^{k_5}  }{(-(\varepsilon_{n_1}^{+}+\varepsilon_{n_2}^{+}-\varepsilon_{k_4}^{-}-\varepsilon_{k_5}^{-}) + i\eta)} t^{n_1 n_3}_{k_4 k_6}  \right. \nonumber \\ 
&  &\left. +   \frac{ (Y_{\tau}^{k_2} X_{\zeta}^{n_5})^* \, Y_{\delta}^{k_1}   Y_{\eta}^{k_2}  Y_{\chi}^{k_3} X_{\gamma}^{n_4}  X_{\epsilon}^{n_5} X_{\sigma}^{n_6} }{\varepsilon_{k_1}^{-}+\varepsilon_{k_2}^{-}-\varepsilon_{n_4}^{+}-\varepsilon_{n_5}^{+} + i\eta} (t^{n_4 n_6}_{k_1 k_3})^* \right) 
\, \,;
\end{eqnarray}

\begin{eqnarray}
\label{eq:U_3rd_me_RES_XY_Ibis}
  \widetilde U^{(3)}_{\alpha  \beta } (\text{\ref{U3_o}})&= &
  \frac{1}{4}   W_{\alpha\gamma \epsilon,\beta\delta \eta} \,  \left(  \frac{Y_{\delta}^{k_1} Y_{\eta}^{k_5}  X_{\gamma}^{n_2} X_{\epsilon}^{n_7}}{\varepsilon_{n_2}^{+}+\varepsilon_{n_7}^{+}-\varepsilon_{k_1}^{-}-\varepsilon_{k_5}^{-} - i\eta}   (Y_{\nu}^{k_1} Y_{\kappa}^{k_5}  X_{\rho}^{n_2})^*   W_{\nu\kappa \pi,\rho\theta \phi} X_{\pi}^{n_3} Y_{\theta}^{k_6}  Y_{\phi}^{k_4} (t^{n_3 n_7}_{ k_4 k_6})^* \right.
 \nonumber \\  
&     & -   \left. (Y_{\gamma}^{k_2} X_{\delta}^{n_1}  X_{\eta}^{n_5})^*  X_{\epsilon}^{n_7} t^{n_1 n_3 n_5}_{k_2 k_4 k_6} (t^{n_3 n_7}_{ k_4 k_6})^*
+ X_{\gamma}^{n_2} (Y_{\epsilon}^{k_7})^* Y_{\delta}^{k_1}   Y_{\eta}^{k_5}  (t^{n_2 n_4 n_6}_{k_1 k_3 k_5})^* t^{n_4 n_6}_{ k_3 k_7} \right. \nonumber \\  
&     & +  \left.  X_{\gamma}^{n_2} X_{\epsilon}^{n_7}  Y_{\delta}^{k_1} Y_{\eta}^{k_5}
(t^{n_2 n_4 n_6}_{k_1 k_3 k_5})^* \frac{X_{\mu}^{n_6} X_{\tau}^{n_7}  Y_{\sigma}^{k_3} (X_{\lambda}^{n_7})^*}{\varepsilon_{n_2}^{+}+\varepsilon_{n_7}^{+}-\varepsilon_{k_1}^{-}-\varepsilon_{k_5}^{-} - i\eta}  \widetilde V^{(1)}_{\mu \tau, \lambda \sigma} \right. \nonumber \\ 
&     & +  \left. ( Y_{\gamma}^{k_2}  Y_{\epsilon}^{k_7} X_{\delta}^{n_1} X_{\eta}^{n_5})^*
  \frac{X_{\nu}^{n_1} X_{\kappa}^{n_5} (Y_{\pi}^{k_3})^* Y_{\rho}^{k_2} (X_{\theta}^{n_6} X_{\phi}^{n_4})^*}{\varepsilon_{n_1}^{+}+\varepsilon_{n_5}^{+}-\varepsilon_{k_2}^{-}-\varepsilon_{k_7}^{-} - i\eta}    W_{\nu\kappa \pi,\rho\theta \phi} t^{n_4 n_6}_{ k_3 k_7} \right. \nonumber \\ 
&     &  - \left.   \frac{(Y_{\gamma}^{k_2} Y_{\epsilon}^{k_7} X_{\delta}^{n_1} X_{\eta}^{n_5})^* }{\varepsilon_{k_2}^{-}+\varepsilon_{k_7}^{-}-\varepsilon_{n_1}^{+}-\varepsilon_{n_5}^{+} + i\eta}  t^{n_1 n_3 n_5}_{k_2 k_4 k_6} (Y_{\mu}^{k_6} Y_{\tau}^{k_4})^* \widetilde V^{(1)}_{\mu \tau, \lambda \sigma}  Y_{\lambda}^{k_7} (X_{\sigma}^{n_3})^* \right)
\, \,;
\end{eqnarray}

\begin{eqnarray}
\label{eq:U_3rd_me_RES_XY_Itris}
  \widetilde U^{(3)}_{\alpha  \beta } (\text{\ref{U3_p}})&= &
  \frac{1}{4}   W_{\alpha\gamma \epsilon,\beta\delta \eta} \,   \left(\frac{X_{\gamma}^{n_2} X_{\epsilon}^{n_5}  Y_{\delta}^{k_1} Y_{\eta}^{k_7}}{\varepsilon_{n_2}^{+}+\varepsilon_{n_5}^{+}-\varepsilon_{k_1}^{-}-\varepsilon_{k_7}^{-} + i\eta}
 (t^{n_2 n_4 n_5}_{k_1 k_3 k_6})^* (Y_{\mu}^{k_7})^* X_{\tau}^{n_4} \widetilde V^{(1)}_{\mu \tau, \lambda \sigma}  Y_{\lambda}^{k_6} Y_{\sigma}^{k_3} \right. \nonumber \\ 
  & & + \left.     \frac{(Y_{\gamma}^{k_2} Y_{\epsilon}^{k_5}  X_{\delta}^{n_1} X_{\eta}^{n_7})^*}{\varepsilon_{k_2}^{-}+\varepsilon_{k_5}^{-}-\varepsilon_{n_1}^{+}-\varepsilon_{n_7}^{+} + i\eta}   X_{\nu}^{n_1} (Y_{\kappa}^{k_6}  Y_{\pi}^{k_3})^*   W_{\nu\kappa \pi,\rho\theta \phi} Y_{\rho}^{k_2} Y_{\theta}^{k_5}  (X_{\phi}^{n_4})^* t^{n_4 n_7}_{ k_3 k_6} \right.
 \nonumber \\  
 &     & \left. -   X_{\gamma}^{n_2} X_{\epsilon}^{n_5} Y_{\delta}^{k_1}  (X_{\eta}^{n_7})^* (t^{n_2 n_4 n_5}_{k_1 k_3 k_6})^* t^{n_4 n_7}_{ k_3 k_6}   +  (Y_{\gamma}^{k_2} X_{\delta}^{n_1}  Y_{\epsilon}^{k_5})^*  Y_{\eta}^{k_7} t^{n_1 n_3 n_6}_{k_2 k_4 k_5} (t^{n_3 n_6}_{ k_4 k_7})^*
 \right. \nonumber \\  
 &     & \left.  -  \frac{X_{\gamma}^{n_2} X_{\epsilon}^{n_5}  Y_{\delta}^{k_1} Y_{\eta}^{k_7}}{\varepsilon_{n_2}^{+}+\varepsilon_{n_5}^{+}-\varepsilon_{k_1}^{-}-\varepsilon_{k_7}^{-} + i\eta}
  (Y_{\nu}^{k_1})^* X_{\kappa}^{n_6}  X_{\pi}^{n_3}   W_{\nu\kappa \pi,\rho\theta \phi} (X_{\rho}^{n_2} X_{\theta}^{n_5})^*  Y_{\phi}^{k_4} (t^{n_3 n_6}_{ k_4 k_7})^* \right. \nonumber \\ 
& & \left.  - \frac{(Y_{\gamma}^{k_2} Y_{\epsilon}^{k_5} X_{\delta}^{n_1} X_{\eta}^{n_7})^* }{\varepsilon_{k_2}^{-}+\varepsilon_{k_5}^{-}-\varepsilon_{n_1}^{+}-\varepsilon_{n_7}^{+} + i\eta}  t^{n_1 n_3 n_6}_{k_2 k_4 k_5} X_{\mu}^{n_7} (Y_{\tau}^{k_4})^* \widetilde V^{(1)}_{\mu \tau, \lambda \sigma}  (X_{\lambda}^{n_6} X_{\sigma}^{n_3})^* \right) 
\, \,;
\end{eqnarray}
\begin{eqnarray}
\label{eq:U_3rd_me_RES_XY_J}
  \widetilde U^{(3)}_{\alpha  \beta } (\text{\ref{U3_j}})&= &
  \frac{1}{4}   W_{\alpha\gamma \epsilon,\beta\delta \eta} \,  \left(  (X_{\delta}^{n_1} Y_{\gamma}^{k_5})^* \, X_{\epsilon}^{n_2} Y_{\eta}^{k_6}  (t^{n_2 n_3 n_4}_{k_6 k_7 k_8})^* t^{n_1 n_3 n_4}_{k_5 k_7 k_8} +(Y_{\epsilon}^{k_2}  X_{\eta}^{n_6})^* \, Y_{\delta}^{k_1}  X_{\gamma}^{n_5}  (t^{n_5 n_7 n_8}_{k_1 k_3 k_4})^* t^{n_6 n_7 n_8}_{k_2 k_3 k_4} \right) \nonumber \\  
&     & +   \frac{1}{4}   W_{\alpha\gamma \epsilon,\beta\delta \eta} \,W_{\nu \kappa \pi, \rho \theta \phi}  \, \left(  \frac{  (X_{\rho}^{n_2}  Y_{\nu}^{k_6})^* X_{\epsilon}^{n_1} X_{\gamma}^{n_2} X_{\kappa}^{n_3} X_{\pi}^{n_4} Y_{\eta}^{k_5} Y_{\delta}^{k_6} Y_{\theta}^{k_7} Y_{\phi}^{k_8}   }{(-(\varepsilon_{n_1}^{+}+\varepsilon_{n_2}^{+}-\varepsilon_{k_5}^{-}-\varepsilon_{k_6}^{-}) + i\eta)}  (t^{n_1 n_3 n_4}_{k_5 k_7 k_8})^*\right. \nonumber \\         
 &  &\left. +  \frac{ ( Y_{\epsilon}^{k_1} Y_{\gamma}^{k_2} Y_{\kappa}^{k_3} Y_{\pi}^{k_4}  X_{\eta}^{n_5} X_{\delta}^{n_6} X_{\theta}^{n_7} X_{\phi}^{n_8} )^* \,  Y_{\rho}^{k_2}  X_{\nu}^{n_6} }{\varepsilon_{k_1}^{-}+\varepsilon_{k_2}^{-}-\varepsilon_{n_5}^{+}-\varepsilon_{n_6}^{+} + i\eta} t^{n_5 n_7 n_8}_{k_1 k_3 k_4}  \right) \nonumber \\           
&    &  + \frac{1}{4}    W_{\alpha\gamma \epsilon,\beta\delta \eta} \, W_{\mu \tau \chi, \lambda \sigma \zeta} \left(\frac{  ( X_{\delta}^{n_1} X_{\eta}^{n_2} X_{\sigma}^{n_3} X_{\zeta}^{n_4} Y_{\gamma}^{k_5} Y_{\epsilon}^{k_6} Y_{\tau}^{k_7} Y_{\chi}^{k_8} )^* \,    X_{\mu}^{n_2}  Y_{\lambda}^{k_6}   }{(-(\varepsilon_{n_1}^{+}+\varepsilon_{n_2}^{+}-\varepsilon_{k_5}^{-}-\varepsilon_{k_6}^{-}) + i\eta)}  t^{n_1 n_3 n_4}_{k_5 k_7 k_8} \right. \nonumber \\         
 &     &\left. + \frac{ (Y_{\mu}^{k_2}  X_{\lambda}^{n_6})^* \, Y_{\delta}^{k_1} Y_{\eta}^{k_2} Y_{\sigma}^{k_3} Y_\zeta{}^{k_4}  X_{\gamma}^{n_5} X_{\epsilon}^{n_6} X_{\tau}^{n_7} X_{\chi}^{n_8}    }{\varepsilon_{k_1}^{-}+\varepsilon_{k_2}^{-}-\varepsilon_{n_5}^{+}-\varepsilon_{n_6}^{+} + i\eta}  (t^{n_5 n_7 n_8}_{k_1 k_3 k_4})^* \right)  
\, \,;
\end{eqnarray}
\begin{eqnarray}
\label{eq:U_3rd_me_RES_XY_K}
  \widetilde U^{(3)}_{\alpha  \beta } (\text{\ref{U3_k}})&= &
  \frac{1}{24}  W_{\alpha\gamma \epsilon,\beta\delta \eta} \, \left((X_{\delta}^{n_2}  X_{\eta}^{n_4})^* \, X_{\gamma}^{n_1}  X_{\epsilon}^{n_3}  (t^{n_1 n_3 n_5}_{k_6 k_7 k_8})^* t^{n_2 n_4 n_5}_{k_6 k_7 k_8}  +(Y_{\gamma}^{k_2}   Y_{\epsilon}^{k_4})^* \, Y_{\delta}^{k_1}   Y_{\eta}^{k_3} (t^{n_6 n_7 n_8}_{k_1 k_3 k_5})^* t^{n_6 n_7 n_8}_{k_2 k_4 k_5} \right) \nonumber \\           
 &   &+   \frac{1}{24}  W_{\alpha\gamma \epsilon,\beta\delta \eta} \,  W_{\nu \kappa \pi, \rho \theta \phi} \left( \frac{   ( X_{\delta}^{n_1}  X_{\rho}^{n_2} X_{\eta}^{n_3} X_{\theta}^{n_4} X_{\phi}^{n_5} Y_{\gamma}^{k_6}  Y_{\epsilon}^{k_7}  Y_{\pi}^{k_8} )^* \, X_{\nu}^{n_1}   X_{\kappa}^{n_3}    }{(-(\varepsilon_{n_1}^{+}+\varepsilon_{n_3}^{+}-\varepsilon_{k_6}^{-}-\varepsilon_{k_7}^{-}) + i\eta)}  t^{n_2 n_4 n_5}_{k_6 k_7 k_8}  \right. \nonumber \\  
&   &\left. + \frac{   (Y_{\nu}^{k_2}   Y_{\kappa}^{k_4})^* \, Y_{\rho}^{k_1}  Y_{\delta}^{k_2}  Y_{\theta}^{k_3} Y_{\eta}^{k_4}  Y_{\phi}^{k_5} X_{\gamma}^{n_6}  X_{\epsilon}^{n_7} X_{\pi}^{n_8}    }{\varepsilon_{k_2}^{-}+\varepsilon_{k_4}^{-}-\varepsilon_{n_6}^{+}-\varepsilon_{n_7}^{+} + i\eta} (t^{n_6 n_7 n_8}_{k_1 k_3 k_5})^* \right) \nonumber \\       
 &    &+  \frac{1}{24}  W_{\alpha\gamma \epsilon,\beta\delta \eta} \,  W_{\mu \tau \chi, \lambda \sigma \zeta} \left( \frac{   (X_{\lambda}^{n_2} X_{\sigma}^{n_4})^* \, X_{\mu}^{n_1}  X_{\gamma}^{n_2} X_{\tau}^{n_3} X_{\epsilon}^{n_4} X_{\chi}^{n_5} Y_{\delta}^{k_6}  Y_{\eta}^{k_7}  Y_{\zeta}^{k_8}   }{(-(\varepsilon_{n_2}^{+}+\varepsilon_{n_4}^{+}-\varepsilon_{k_6}^{-}-\varepsilon_{k_7}^{-}) + i\eta)} (t^{n_1 n_3 n_5}_{k_6 k_7 k_8})^* \right. \nonumber \\          
&      &\left. +  \frac{   ( Y_{\gamma}^{k_1}  Y_{\mu}^{k_2}  Y_{\epsilon}^{k_3} Y_{\tau}^{k_4}  Y_{\chi}^{k_5} X_{\delta}^{n_6}  X_{\eta}^{n_7} X_{\zeta}^{n_8}   )^* \, Y_{\lambda}^{k_1}   Y_{\sigma}^{k_3}  }{\varepsilon_{k_1}^{-}+\varepsilon_{k_3}^{-}-\varepsilon_{n_6}^{+}-\varepsilon_{n_7}^{+} + i\eta}  t^{n_6 n_7 n_8}_{k_2 k_4 k_5}  \right)       
\, \,.
\end{eqnarray}
\end{subequations}

Together with the last term of Eq.~\eqref{eq:U_2nd_me_XY}, the third-order diagrams in Figs.~\ref{U3_V2},~\ref{U3_h},~\ref{U3_i},~\ref{U3_o},~\ref{U3_p},~\ref{U3_j} and~\ref{U3_k} are skeleton and therefore they would need to be included in a fully self-consistent ADC(3) formulation.  Note, however, that Eq.~\eqref{eq:U_2nd_me_XY} and the diagram~\ref{U3_V2} reduce to a single contribution if the full $\widetilde{V}$, from Eq.~\eqref{veff}, is used.
Again, choosing an HF state as the unperturbed reference would force the diagrams of Figs.~\ref{U3_a},~\ref{U3_e},~\ref{U3_f} and~\ref{U3_g}  to vanish.

\end{widetext}

\subsection{\label{Dynamic_SE} Dynamic self-energy}
When a self-consistent formulation is possible, some of the correlation effects beyond mean field are already included through the use of dressed reference propagators.
However, for a  reference state that is not dressed, additional nonskeleton diagrams contribute to the energy-dependent self-energy at
third and higher orders. Hence, their contributions should be added to the ADC($n$) equations.  

\begin{figure}[t]
\setcounter{subfigure}{17}
  \centering
  \subfloat[(r)]{\label{3ord_U1_a}\includegraphics[scale=0.45]{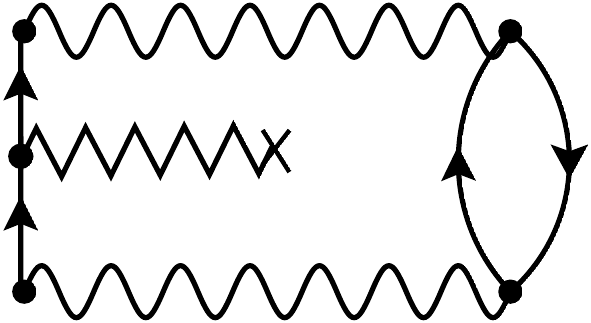}}
   \hspace{1.35cm}
  \subfloat[(s)]{\label{3ord_U1_b}\includegraphics[scale=0.45]{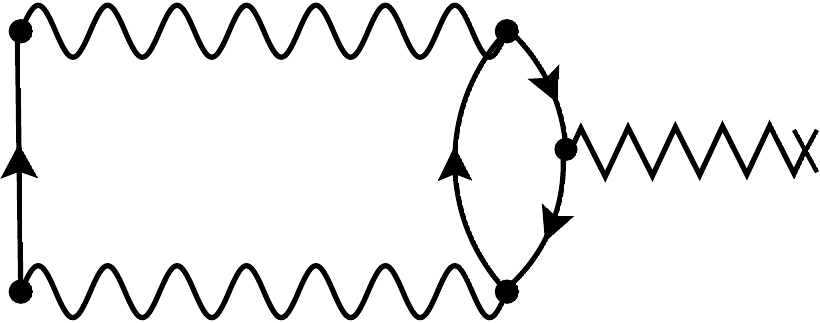}}
    \newline   \vskip .7cm
 \subfloat[(t)]{\label{3ord_U1_c}\includegraphics[scale=0.45]{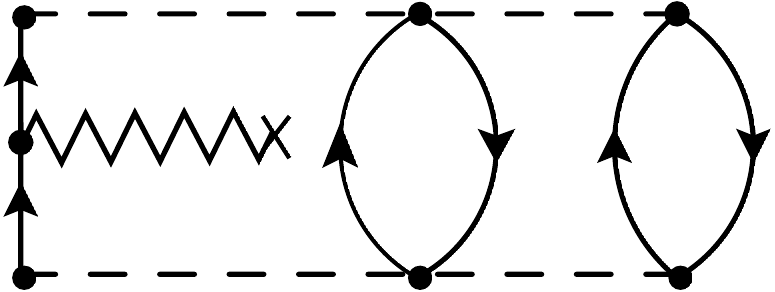}}
 \hspace{0.35cm}
   \subfloat[(u)]{\label{3ord_U1_d}\includegraphics[scale=0.45]{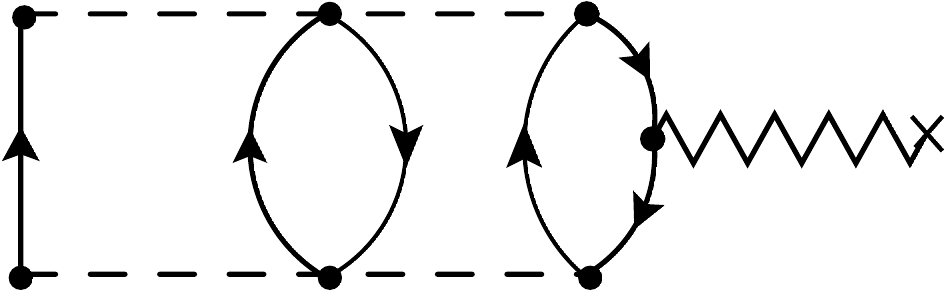}}
  \caption{ Third-order nonskeleton diagrams of the dynamic self-energy, complementing the ones in Figs.~\ref{3ord_a_b_c} and \ref{3ord_remn}, for the general ADC(3), not based on a self-consistent reference propagator. Zigzag lines represent the effective 1B interaction of Eq.~(\ref{eq:U_1st_me}). }
  \label{3ord_U1}
\end{figure}

Specifically, at the ADC(3) level, the  one-particle irreducible and interaction-irreducible diagrams considered for the energy-dependent self-energy in Sec.~\ref{ADC(N)} and Appendix~\ref{M_N_matrices}  must be complemented with the four Feynman diagrams of Fig.~\ref{3ord_U1}. Diagrammatically, they are obtained by inserting the first-order 1B effective interaction $\widetilde U^{(1)}$ of Fig.~\ref{Utilda_1} into the second-order diagrams of Fig.~\ref{2ord}. Since we are not considering four and higher orders here, only the $\widetilde U^{(1)}$ needs to be included for ADC(3).

In this case, the ADC(3) expansion of the self-energy matrices that appear  in Eq.~(\ref{eq:Dy_residual_5}) is enriched with additional coupling corrections. These are already listed in Eqs.~(\ref{ADC3_somm_final}) and~(\ref{ADC3_somm_final_N}), respectively for the forward-in-time and backward-in-time cases, and they are repeated here for completeness:
\begin{align}
\label{ADC3_somm_dynamic}
\textbf{M}^{(\textrm{II})}_{j\alpha}  ={}& 
\begin{cases}
  \textbf{M}^{(\textrm{IIr})}_{r \alpha } + \textbf{M}^{(\textrm{IIs})}_{r \alpha }  &\hbox{if $j$=$r$ ($2p1h$),} \\  ~ \\
    \textbf{M}^{(\textrm{IIt})}_{q \alpha }  +  \textbf{M}^{(\textrm{IIu})}_{q \alpha }  &\hbox{if $j$=$q$ ($3p2h$),}
\end{cases} 
\\ \nonumber   \\
\label{ADC3_somm_dynamic_N}\textbf{N}^{(\textrm{II})}_{\alpha k }  ={}& 
\begin{cases}
\textbf{N}^{(\textrm{IIr})}_{\alpha s} + \textbf{N}^{(\textrm{IIs})}_{\alpha s}  &\hbox{if $k$=$s$ ($2h1p$),} \\ ~\\
\textbf{N}^{(\textrm{IIt})}_{\alpha u}   + \textbf{N}^{(\textrm{IIu})}_{\alpha u}   &\hbox{if $k$=$u$ ($3h2p$).}
\end{cases}
\end{align}
Also for the interaction matrices, the nonskeleton expansion is enriched by additional $\textbf{C}_{j j'}$ and $\textbf{D}_{k k'} $ matrices.  These terms are included in Eqs.~(\ref{C_final}-\ref{D_final}) and are:
\begin{align}
\label{C_final_dynamic}
\textbf{C}_{j j'} ={}&
\begin{cases}
    \textbf{C}^{\widetilde{U} p}_{r r'} + \textbf{C}^{\widetilde{U} h}_{r r'}    &\hbox{if $j$=$r$ and $j'$=$r'$ ($2p1h$),} \\ ~ \\
     \textbf{C}^{\widetilde{U} p}_{q q'} + \textbf{C}^{\widetilde{U} h}_{q q'} &\hbox{if $j$=$q$ and $j'$=$q'$ ($3p2h$),} \\ ~ \\
     0  & \hbox{otherwise,}
\end{cases}
\\ \nonumber \\
\label{D_final_dynamic}
\textbf{D}_{k k'}  ={}&
\begin{cases}
 \textbf{D}^{\widetilde{U} h}_{s s'} + \textbf{D}^{\widetilde{U} p}_{s s'}     &\hbox{if $k$=$s$ and $k'$=$s'$ ($2h1p$),} \\ ~ \\
   \textbf{D}^{\widetilde{U} p}_{u u'} + \textbf{D}^{\widetilde{U} h}_{u u'}  &\hbox{if $k$=$u$ and $k'$=$u'$ ($3h2p$),} \\ ~ \\
     0  & \hbox{otherwise.}
\end{cases}
\end{align}

\subsubsection{\label{2_1_U1} ADC(3) terms with $2p1h$ and $2h1p$ ISCs}

By following the same procedure used to find the expressions of the coupling matrices containing 2NFs and/or 3NFs, we can derive the analogous expressions of $\textbf{M}_{j \alpha } $ and $\textbf{N}_{\alpha k}$ containing one $\widetilde U$ insertion, and expressed in terms of the uncorrelated transition amplitudes of Eq.~(\ref{tran_ampl_ref}).

From the Goldstone-Feynman diagrams  in Figs.~\ref{3ord_U1_a} and~\ref{3ord_U1_b}  we find,
\begin{equation}
\label{eq:M_2N_U1_a}
\textbf{M}^{(\textrm{IIr})}_{r \alpha } \equiv \frac{-1}{\sqrt{2}} 
 \mathcal{A}_{1 2}   
\frac{X_{\gamma}^{n_1}  \widetilde{U}^{(1)}_{\gamma \delta}  Y_{\delta}^{k_4} }{\varepsilon_{k_4}^{-}-\varepsilon_{n_1}^{+}} 
(Y_{\mu}^{k_4})^*  X_{\nu}^{n_2}  Y_{\lambda}^{k_3} \, \widetilde{V}_{\mu\nu,\alpha \lambda} \,,
\end{equation}
and 
\begin{equation}
\label{eq:M_2N_U1_b}
\textbf{M}^{(\textrm{IIs})}_{r \alpha} \equiv \frac{1}{\sqrt{2}}  
\frac{X_{\gamma}^{n_4}  \widetilde{U}^{(1)}_{\gamma \delta}  Y_{\delta}^{k_3}  }{\varepsilon_{k_3}^{-}-\varepsilon_{n_4}^{+}}
X_{\mu}^{n_1}  X_{\nu}^{n_2}  (X_{\lambda}^{n_4})^*\
\widetilde{V}_{\mu\nu,\alpha \lambda}   \, ,
\end{equation}
respectively. 

The $2p1h$ interaction matrices  in Eq.~(\ref{C_final_dynamic}) are
\begin{equation}
 \label{C_123_456_U1_pp}
\textbf{C}^{\widetilde{U} p}_{r r'} = \frac{1}{2}    \mathcal{A}_{1 2}  \mathcal{A}_{4 5}  \,
 X_{\gamma}^{n_1} \widetilde{U}^{(1)}_{\gamma \delta}  (X_{\delta}^{n_4})^* 
 \delta_{n_2 n_5}  \delta_{k_3 k_6} \, ,
 \end{equation}
 and 
 \begin{equation}
 \label{C_123_456_U1_hh}
 \textbf{C}^{\widetilde{U} h}_{r r'} = \frac{-1}{2}   \mathcal{A}_{1 2} 
   Y_{\delta}^{k_3}   \widetilde{U}^{(1)}_{\gamma \delta} (Y_{\gamma}^{k_6})^* \,  \delta_{n_1 n_4} \delta_{n_2 n_5} \,  .
 \end{equation}

In the backward-in-time Goldstone diagrams of Figs.~\ref{3ord_U1_a} and~\ref{3ord_U1_b} one finds the coupling matrices,
\begin{equation}
\label{eq:N_2N_U1_a}
\textbf{N}^{(\textrm{IIr})}_{\alpha s} \equiv \frac{1}{\sqrt{2}}  \widetilde{V}_{\alpha \lambda , \mu \nu} \mathcal{A}_{1 2}  
 \frac{ X_{\gamma}^{n_4} \widetilde{U}^{(1)}_{\gamma \delta}  Y_{\delta}^{k_1} }{\varepsilon_{k_1}^{-}-\varepsilon_{n_4}^{+}} \, 
(X_{\mu}^{n_4})^*  Y_{\nu}^{k_2}  X_{\lambda}^{n_3}  
\end{equation}
and 
\begin{equation}
\label{eq:N_2N_U1_b}
\textbf{N}^{(\textrm{IIs})}_{\alpha s} \equiv \frac{-1}{\sqrt{2}} \widetilde{V}_{\alpha \lambda, \mu\nu}
 \frac{ X_{\gamma}^{n_3} \widetilde{U}^{(1)}_{\gamma \delta}  Y_{\delta}^{k_4} }{\varepsilon_{k_4}^{-}-\varepsilon_{n_3}^{+}} \, 
 Y_{\mu}^{k_1}  Y_{\nu}^{k_2} (Y_{\lambda}^{k_4})^*  \, ,
\end{equation}
while the corresponding interaction matrices  in Eq.~(\ref{D_final_dynamic}) 
connecting $2h1p$ ISCs, are
 \begin{equation}
 \label{D_123_456_U1_hh}
\textbf{D}^{\widetilde{U} h}_{s s'} = \frac{1}{2}   \mathcal{A}_{1 2}  \mathcal{A}_{4 5}  \,
 (Y_{\gamma}^{k_1})^* \widetilde{U}^{(1)}_{\gamma \delta}  Y_{\delta}^{k_4} \,
\delta_{k_2 k_5}  \delta_{n_3 n_6} 
 \end{equation}
 and 
 \begin{equation}
 \label{D_123_456_U1_pp}
 \textbf{D}^{\widetilde{U} p}_{s s'} = \frac{-1}{2}   \mathcal{A}_{1 2}  \,   
 (X_{\delta}^{n_3})^* \widetilde{U}^{(1)}_{\gamma \delta} X_{\gamma}^{n_3} \,
 \delta_{k_1 k_4} \delta_{k_2 k_5} \,  .
 \end{equation}
 
Equations~(\ref{eq:M_2N_U1_a})-(\ref{D_123_456_U1_pp}) above are third-order terms composed by effective 1B interaction and effective 2NFs. In the next section we proceed by introducing the set of expressions with effective 1B interaction and 3NFs, which connect $3p2h$ and $3h2p$ ISCs.

\pagebreak
  \begin{widetext}

 \subsubsection{\label{3_1_U1} ADC(3) terms with $3p2h$ and $3h2p$ ISCs}
 
 The Goldstone-Feynman diagrams of Figs.~\ref{3ord_U1_c} and~\ref{3ord_U1_d} involve $3p2h$ and $3h2p$ ISCs. They contain the coupling matrices that complete the expressions for $\textbf{M}_{j \alpha } $ and $\textbf{N}_{\alpha k} $, when the reference state adopted has not be calculated self-consistently.

The working equations for the forward-in-time coupling matrices are,
 \begin{eqnarray}
\label{eq:M_3N_U1_d}
\textbf{M}^{(\textrm{IIt})}_{q \alpha } \equiv \frac{-1}{\sqrt{12}}  \mathcal{P}_{1 2 3} \,
 \frac{X_{\gamma}^{n_1}  \widetilde{U}^{(1)}_{\gamma \delta}  Y_{\delta}^{k_6} }{\varepsilon_{k_6}^{-} - \varepsilon_{n_1}^{+}} \, 
 (Y_{\mu}^{k_6})^* X_{\nu}^{n_2}  X_{\rho}^{n_3}  Y_{\lambda}^{k_4} Y_{\eta}^{k_5} \,
 W_{\mu\nu \rho,\alpha \lambda \eta} \, ,
\end{eqnarray}
and
 \begin{eqnarray}
\label{eq:M_3N_U1_c}
\textbf{M}^{(\textrm{IIu})}_{q \alpha } \equiv  \frac{1}{\sqrt{12}}    \mathcal{A}_{4 5} 
  \frac{X_{\gamma}^{n_6}  \widetilde{U}^{(1)}_{\gamma \delta}  Y_{\delta}^{k_5}  }{\varepsilon_{k_5}^{-}-\varepsilon_{n_6}^{+}} \,
 X_{\mu}^{n_1} X_{\nu}^{n_2}  X_{\rho}^{n_3}  Y_{\lambda}^{k_4} (X_{\eta}^{n_6} )^* \,
W_{\mu\nu \rho,\alpha \lambda \eta} \, , 
\end{eqnarray}
while the interaction matrices in Eq.~(\ref{C_final_dynamic}) connecting two $3p2h$ ISCs are,
   \begin{equation}
 \label{C_12345_678910_U1_pp}
\textbf{C}^{\widetilde{U} p}_{q q'}  = \frac{1}{12}  \mathcal{A}_{1 2 3}  \mathcal{A}_{4 5}  \mathcal{P}_{6 7 8}  \,  
X_{\gamma}^{n_1}  \widetilde{U}^{(1)}_{\gamma \delta}   (X_{\delta}^{n_6})^*  \,
 \delta_{n_2 n_7} \delta_{n_3 n_8} \delta_{k_4 k_9} \delta_{k_5 k_{10}} \, ,
 \end{equation}
 and
   \begin{equation}
 \label{C_12345_678910_U1_hh}
 \textbf{C}^{\widetilde{U} h}_{q q'} = \frac{-1}{12}  \mathcal{A}_{4 5}  \mathcal{A}_{9  \, 10} \mathcal{A}_{1 2 3}  \,  
 Y_{\delta}^{k_5} \widetilde{U}^{(1)}_{\gamma \delta}  (Y_{\gamma}^{k_{10}})^*  \,
  \delta_{n_1 n_6}  \delta_{n_2 n_7}  \delta_{n_3 n_8} \delta_{k_4 k_9} \, .
 \end{equation}

Expressions for the backward-in-time coupling matrices, containing one effective 1B interaction and one interaction-irreducible 3NF insertion, are
 \begin{eqnarray}
\label{eq:N_3N_U1_d}
\textbf{N}^{(\textrm{IIt})}_{\alpha u} \equiv \frac{1}{\sqrt{12}}  W_{\alpha \lambda \eta, \mu\nu \rho} \mathcal{P}_{1 2 3} 
 \frac{ X_{\gamma}^{n_6}   \widetilde{U}^{(1)}_{\gamma \delta} Y_{\delta}^{k_1} }{\varepsilon_{k_1}^{-}-\varepsilon_{n_6}^{+}} \, 
 (X_{\mu}^{n_6})^*   Y_{\nu}^{k_2}  Y_{\rho}^{k_3}  X_{\lambda}^{n_4} X_{\eta}^{n_5}   \, ,
\end{eqnarray}
and
\begin{eqnarray}
\label{eq:N_3N_U1_c}
\textbf{N}^{(\textrm{IIu})}_{\alpha u} \equiv \frac{-1}{\sqrt{12}}
  W_{\alpha \lambda \eta, \mu\nu \rho} \mathcal{A}_{4 5}   
  \frac{  X_{\gamma}^{n_5} \widetilde{U}^{(1)}_{\gamma \delta}  Y_{\delta}^{k_6} }{ \varepsilon_{k_6}^{-} - \varepsilon_{n_5}^{+} } \, 
Y_{\mu}^{k_1} Y_{\nu}^{k_2}  Y_{\rho}^{k_3}  X_{\lambda}^{n_4}  ( Y_{\eta}^{k_6})^* \, ,
\end{eqnarray}
while the interaction matrices  in Eq.~(\ref{D_final_dynamic}) connecting two $3h2p$ ISCs are
\begin{equation}
 \label{D_12345_678910_U1_pp}
 \textbf{D}^{\widetilde{U} p}_{u u'} = -\frac{1}{12}   \mathcal{A}_{4 5}  \mathcal{A}_{9  \, 10} \mathcal{A}_{1 2 3} \,  
  (X_{\delta}^{n_{5}})^* \,    \widetilde{U}^{(1)}_{\gamma \delta} \, X_{\gamma}^{n_{10}} \, 
  \delta_{k_1 k_6} \delta_{k_2 k_7} \delta_{k_3 k_8} \delta_{n_4 n_9}  \, ,
 \end{equation}
and
   \begin{equation}
 \label{D_12345_678910_U1_hh}
\textbf{D}^{\widetilde{U} h}_{u u'}  = \frac{1}{12}   \mathcal{A}_{1 2 3}  \mathcal{A}_{4 5}  \mathcal{P}_{6 7 8} \,  
(Y_{\gamma}^{k_1})^*   \, \widetilde{U}^{(1)}_{\gamma \delta} \,   Y_{\delta}^{k_6} \,
 \delta_{k_2 k_7}  \delta_{k_3 k_8} \delta_{n_4 n_9} \delta_{n_5 n_{10}}   \, .
 \end{equation}

\end{widetext}

\bibliography{ADC3_paper_22}

\end{document}